\documentclass[mph,copyrightpage]{mqthesis}
\usepackage{pstricks}
\usepackage{pst-3dplot}
\usepackage{graphicx}
\usepackage[round]{natbib}
\usepackage{layout} 
\usepackage{rotating}
\usepackage{longtable}
\usepackage{supertabular}
\usepackage[]{lscape}
\usepackage{afterpage}
\usepackage{dcolumn}
\usepackage{threeparttable}
\SpecialCoor
\begin{document}

\frontmatter

%%%%% Acknowledgements, titlepage, abstract, list of publications
\title{An Analysis of HCN Observations of
\\The Galactic Centre's Circumnuclear Disk}
%\\A Study of its Physical Properties}
\author{Ian Lindsay Smith}
\department{Physics and Astronomy} 
\submitdate{April 2012}
% honours students will want to use the keyword honours instead of phd :)
% or override the \degreetext variable with
% something appropriate like:
%\renewcommand{\degreetext}%
%{in fulfilment of the Degree of Master of Philosophy}
\copyrightyear{2012}
\titlepage

\chapter{Acknowledgements}

I would like to thank the following people, \\

Mark Wardle for his patient guidance and encouragement as my principal supervisor. \\

Maria Montero-Casta{\~n}o for generously providing the co-ordinates for the HCN(4-3) cores that she convolved with HCN(1-0) cores and Fahard Yusef-Zadeh for providing the co-ordinates of methanol and water masers detected in the CND.\\ 

Jackie Chapman for introducing me to the Molex and IDL software programmes that she modified to present analytical results. Catherine Braiding for proof reading, suggestions and general help with computer software problems. Ross Moore for his expert advice on \LaTeX \, packages and code. \\

Quentin Parker, my assistant supervisor, for his encouragement.\\ 

Carol McNaught for her guidance through the administrative maze.\\

My wife Anne for proof reading my thesis and patiently supporting me throughout this study.\\

\chapter{Abstract}
\paragraph*{}
The Circumnuclear Disk (CND) is a torus of molecular dust and gas rotating about the galactic centre and extending from 1.6pc to 7pc from the central massive black hole SgrA$^{*}$. Observations of the CND in a number of transitions of HCN have shown the gas to be clumpy. The HCN(1-0) transition has been interpreted as being optically thick with molecular hydrogen  number densities $\simeq10^{7}$ cm$^{-3}$ implying  that the cores are tidally stable. Given this stability a predicted life for the disk of millions of years would allow star formation to occur through core condensation.

\paragraph*{}
Large Velocity Gradient modelling of the intensity lines of a number of selected HCN transitions  is used to infer hydrogen density and  HCN optical depth. The selection of HCN cores for (LVG) modelling  requires identification of three transitions that share common locations and velocity spaces for valid comparisons and predictions of relevant parameter values. The geometry of the CND is explored as the first step in the core selection process. The projected co-ordinates and deprojected distances from SgrA$^{*}$ listed in \citet{Chris2005} are used to establish the disk's attitude relative to the plane of the sky, and deprojected co-ordinates that when plotted reveal a circular pattern of cores about a central cavity. A flat rotational velocity model compares modelled with observed HCN(1-0) core radial velocities that indicate eighteen out of twenty-six cores could be considered part of the CND.    

\paragraph*{}
Previous studies suggest that HCN(1-0) is optically thick (\large{$\tau$}\normalsize = 4) whereas the LVG modelling in this study suggests that the HCN(1-0) and H$^{13}$CN(1-0) emission is optically thin with weakly inverted populations and  \large{$\tau$}\normalsize(H$^{12}$CN) $\sim$ -0.2. The excitation temperatures for H$^{12}$CN and H$^{13}$CN are markedly different, undermining earlier arguments for optically thick HCN(1-0). The molecular hydrogen density is found to range from 0.1 to 2 $\times$ 10$^{6}$ cm$^{-3}$, about an order of magnitude less than the previous estimate. This implies that the cores are tidally unstable and that the total mass of the disk is about  10$^{5} M_{\odot}$ which is an order of magnitude lower than previous estimates  based on HCN data and consistent with thermal emission from dust and dynamical arguments. Star formation within the disk therefore is not expected to occur without some significant ``triggering'' event.

\tableofcontents
% comment out these as required for your discipline
\listoffigures
\listoftables

\mainmatter
\newcolumntype{d}[1]{D{.}{.}{2}}
%%%%% Introduction
%\input{chap_intro}
%\input{chapter_1}
%\input{chapter_2}
%\input{chapter_3}
%\input{chapter_4}
%\input{chapter_5}
%\input{chapter_6}
\chapter{The Circumnuclear Disk} \label{CNDintro}
\section{Introduction}
\paragraph*{}
The Circumnuclear Disk, CND, is a ring of gas and dust located close to and around the Milky Way's galactic centre. The gas is mainly molecular and atomic gas heated by the stellar group in the central cavity. The dust is primarily composed of  silicon and carbon that is the source of infra-red  radiation, IR, which is the result of heating from radiation sources inside the CND's inner radius.

\begin{figure}[ht]
\centering 
\includegraphics[bb= -5 -5 205 205,clip=true,scale=1.75]
{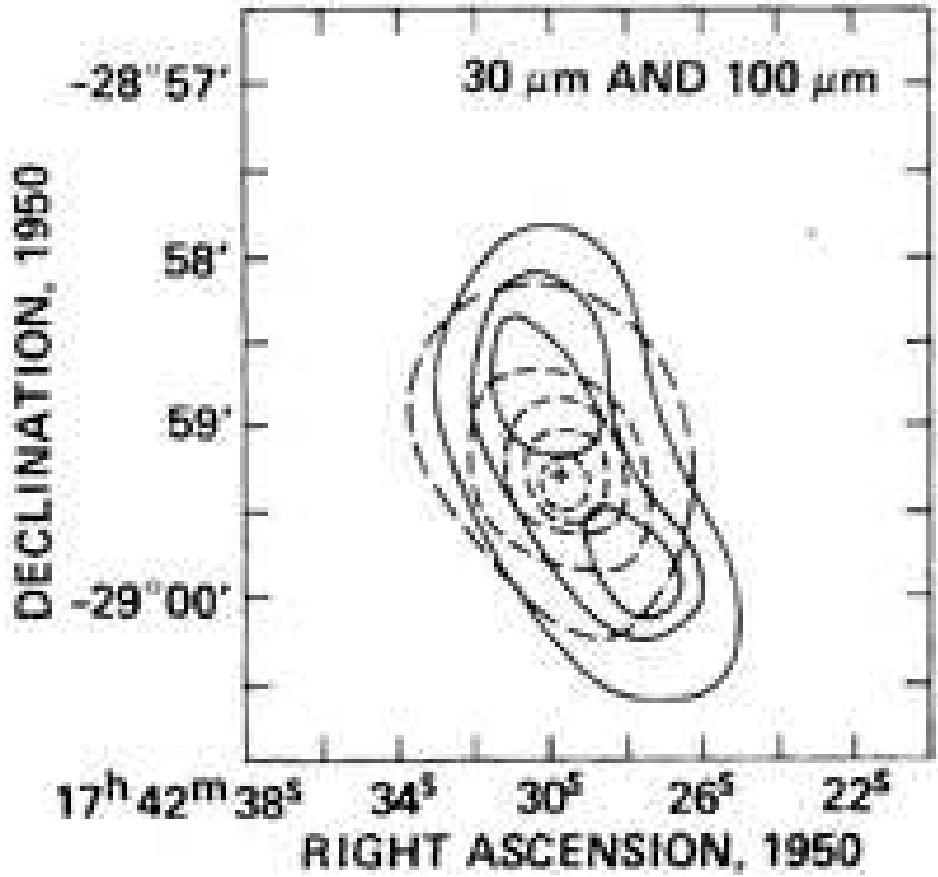} \fontsize{9} {9} \caption[Early CND IR Image]{\label{CND IR} High angular resolution far IR continuum observations of the galactic centre \citep{Becklin1982}. The 30$\mu$m map (dashed) superimposed on 100$\mu$m (solid) contours that outline the CND's lobes.}
\end{figure}

\paragraph*{}
The CND was first observed in IR continuum emission from dust at 30, 50 and 100$\mu$m using the 1 metre diameter Kuiper airborne observatory. The CND appeared as two lobes that were symmetrically located to the NE and SW about a relatively dust-free central cavity in 100$\mu$m emission with some 30$\mu$m emission close to the galactic centre (see Fig.\ \ref{CND IR}). The symmetry and orientation of the lobes suggested a ring like structure with a major axis approximately aligned with the galactic plane \citep{Becklin1982}.

\paragraph*{}
This chapter outlines the currently known information about the CND from observations spanning some 25 years including two reviews, the first summarises earlier work up to 1989 \citep{Genzel1989}, the second written last year covers more recent research \citep{Genzel2010}. It closes with discussion of questions posed by its existence and an outline of this thesis' structure.

\section{Current Status}
\paragraph*{}
Observations of CO, CS and HCN  subsequently led to the discovery of the CND rotating about the galactic centre \citep{Serabyn1985,SGW1986,Guesten1987}. The disk was found to have an inner radius of 1.5 to 1.7pc and extend to 5pc in HCN and $>$7pc in CO \citep{Guesten1987}, and more recent observations have detected HCN out to 7pc and CO up to 9pc from the galactic centre. \citep{Chris2005,Oka2007,Oka2011}.

\paragraph*{}
The disk is composed of clumps (cores), with diameters from 0.14 to 0.43 pc, rotating in a number of kinematically distinct streamers about the galactic centre \citep{Guesten1987,Jacks1993}. The disk's major axis is aligned to a position angle of $\sim$ 25$^{\circ}$ and inclination of $\sim$ 70$^{\circ}$ to the plane of the sky. \citep{SGW1986,Jacks1993,Marshall1995}. It was noted that the rotation was perturbed in several ways with a large local velocity dispersed throughout the disk. The position angle changed with radius and in inclination with azimuthal angle, i.e. the disk was warped. The perturbations together with the disk's clumpiness indicated a non-equilibrium configuration with a short age of a few orbital periods \citep{Guesten1987}.

\begin{figure}[ht]
\centering
\includegraphics[bb= -5 -5 250 290,clip=true,scale=0.8]{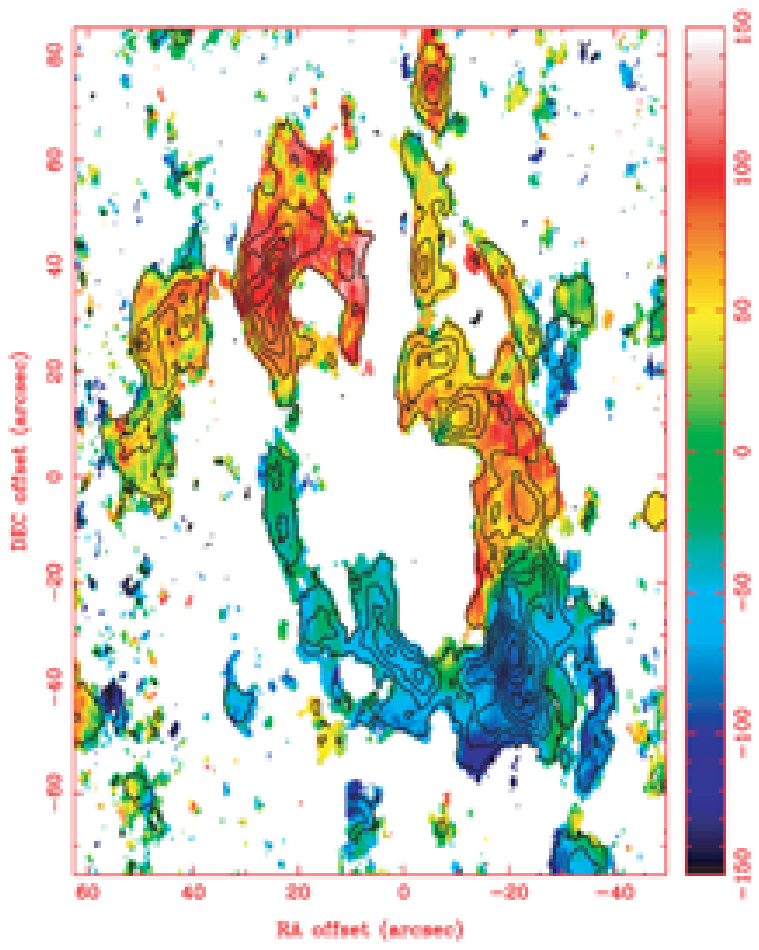} \caption[CND Integrated Velocity Map]{\label{Velmap}Flux weighted average velocity map of the HCN(1-0) emission in colour with integrated emission (black contours) \citet{Chris2005}. Blue colour indicates emission with negative velocities moving towards, red colour emission with positive velocities moving away from the observer and showing the disk's rotation. Velocities are relative to V$_{LSR}$ = 0 km s$^{-1}$ with magnitudes indicated by colour bar to the right of the map. Offsets are relative to SgrA$^{*}$ at (0,0).}
\end{figure}

\paragraph*{}
The ring's rotational velocity is $\sim$110 km s$^{-1}$ between 2 to 5pc from the galactic centre (see Fig.\ \ref{Velmap}). Lower velocities are indicated by CII and CO(7-6) at radii $\geq$ 4pc and higher velocities, 130--140 km s$^{-1}$, are indicated by HCN in the North Eastern part of the ring at 2pc \citep{Guesten1987}. \citet{Marshall1995} fitted a 3D rotating ring model to HCN (4-3) and (3-2) data and inferred a flat velocity profile, while noting that \citet{Harris1985} showed a velocity fall off between 2 and 6 pc from the centre with CO observations.

\begin{figure}[ht]
\centering
\includegraphics[scale=0.8]{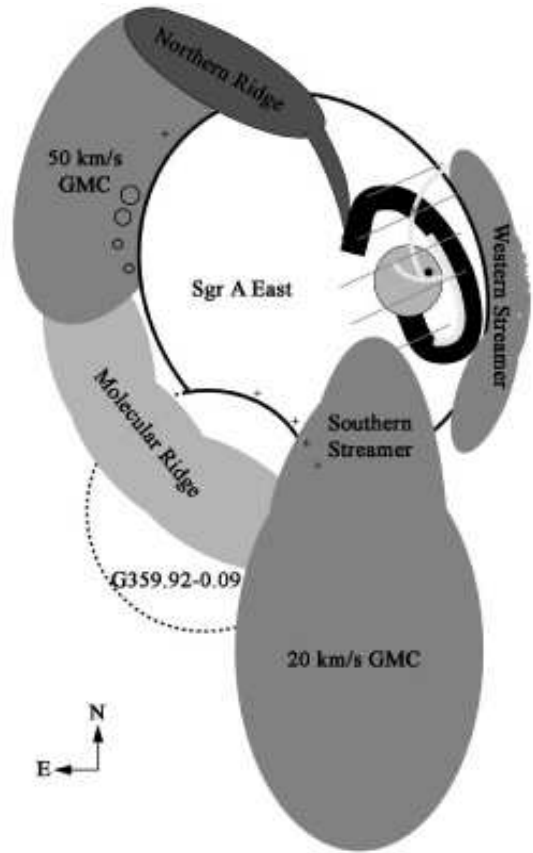} \caption[Central 10pc of Milky Way Galaxy]{\label{central 10pc} Schematic drawing of the central 10pc of the Milky Way Galaxy as seen in the plane of the sky \citep{HandH2002}. The CND is shown as a solid black torus, G359.92-0.09 is an adjacent Supernova Remnant to SgrA East. The central black hole is shown as a black dot.} 
\end{figure}

\begin{figure}[ht] 
\centering 
\includegraphics[bb= -5 -5 340 470,clip=true,scale=0.5] {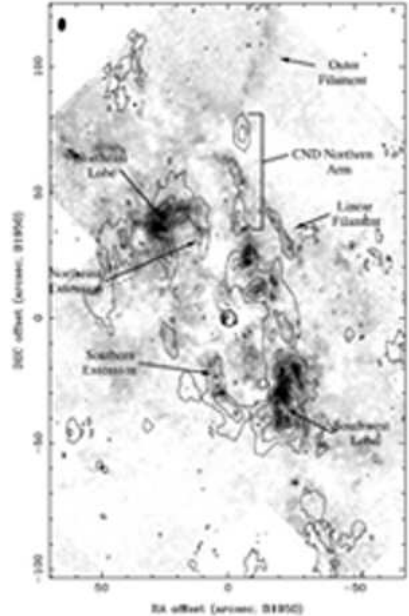} \fontsize{9} {9} \caption[CND Features]{\label{CND parts}HCN emission in contours superimposed on a reverse grey scale H$_2$(1-0)S($1$)line emission map \citep{Chris2005}. Significant features in both H$_2$ and HCN emission are labelled, and SgrA$^{*}$ is marked by a circle at(0,0).} 
\end{figure}

\paragraph*{}
The disk has been interpreted as an accretion disk transferring material from the giant molecular clouds located in the region 10--30pc from the galactic centre to the cavity located inside the the inner radius of the ring \citep{Guesten1987, Oka2011} (see Fig.\ \ref{central 10pc}). Parts of the disk's inner radius located in the East, North and West were found to be associated with the arms of the mini-spiral SgrA West that was suggested from the observations of line broadening of  HCN(4-3) core line spectra, possibly caused by a transient turbulent cascade of material falling towards the central cavity, in locations where they overlap, at least in projection, the mini-spiral arms \citep{MMC2009}. \citet{Chris2005} also noted a number of interactions between the ionised gas in the ring's inner radius and the western arc of the mini-spiral, which is both spatially and kinematically consistent with the CND's inner radius material. They also suggested a possible connection between the northern arm of the mini-spiral and the North Eastern extension of the CND (see Fig.\ref{CND parts}).

\paragraph*{}
The Northern and Southern lobes of the CND have different excitation levels and densities with the Southern warmer and denser than the Northern part. NH$_{3}$ (6-6) data confirm that molecular gas in the ring is denser and colder than the molecular gas in the cavity \citep{MMC2009}. The South Eastern portion of the ring is denser than the South Western part of the CND with the SE part becoming warmer and more diffuse as material heads NW approaching the central super massive black hole. The difference between the HCN(4-3) and NH$_{3}$(6-6) data indicates a probable infall of material from the CND toward the galactic centre through the Eastern part of the ring structure \citep{MMC2009}. 

\paragraph*{}
\citet{HandH2002} inferred that material inside the ring is hotter and denser than the mini spiral and appears unrelated, this seems at odds with the conclusions of \citet{Chris2005} and \citet{MMC2009} who infer a connection. NH$_{3}$(3-3) emission suggested an interaction between the southern arm of the mini-spiral and the southernmost part of the CND, with material in the ring undergoing compression as the arm approaches the CND before spiralling inward to the centre. The inner edge of the CND is more highly excited than the outer part of the ring most probably due to the  stellar cluster, inside, exciting the ring's inner edge \citep{MMC2009}.

\paragraph*{}
Core size estimates have varied from 0.05 to 0.12pc by \citep{Jacks1993} who used a telescope with a beam size incapable of resolving the core images, to 0.14 to 0.43pc \citep{Chris2005} and 0.14 to 0.5pc \citep{MMC2009}, with the larger core sizes based on resolved images.

\paragraph*{}
The dust continuum emission is well approximated by thermal emission at 20, 60 and 100\,K. Observations by \citet{Mezger1989} showed that warm dust $\geq$60\,K accounts for only $\sim$10\% of the total dust mass in the CND and is located near the ionisation front of SgrA West. The remaining dust in the disk appears to be rather cold at $\sim$20\,K. The south-western edge of the ring of dust surrounding SgrA East and the southern lobe of the CND overlap. Both have similar clumpy density structures but different radial velocities which allow them to be separated by observations of optically thin molecular lines \citep{Mezger1989}.

\paragraph*{}
\citet{Etxa2011} found the spectral energy distribution for the far IR emission from dust in the CND was best represented by a continuum summing dust temperatures at 90, 44.5 and 23\,K with the cold component accounting for $\sim$ 3.2$\times$10$^{4}$ M$_{\odot}$ out of the estimated total $\sim$5$\times$10$^{4}$ M$_{\odot}$ in the central 2pc of the CND and is similar to the findings of \citet{Mezger1989}.

\normalsize

%\clearpage

\section{Motivation for this Study}
\paragraph*{}
After twenty-five years of observing the CND there remains considerable uncertainty as to the molecular density of the CND's cores. \citet{Genzel2010} summarises the position by describing the two prevailing scenarios as
\begin{enumerate}
\item the original view of less dense (10$^{6}$cm$^{-3}$) warm gas ($>$ 100\,K) cores which are tidally unstable leading to a transient lifetime of $\sim$10$^{5}$ yr, or  
\item  the more recent idea of a denser (10$^{7}$-10$^{8}$ cm$^{-3}$) cool gas (50-100\,K) which provides stable cores with long lifetimes $\sim$10$^{7}$yr, that is long enough for the opportunity for star formation from core condensation. 
\end{enumerate}

\paragraph*{}
Evidence for the first scenario includes that there is no recorded evidence of precession of the orbits of the inner stellar cluster which would be expected if the CND's mass is 10$^{6}$ M$_{\odot}$. \citet{Subr2009} also noted that gravitational torque would have destroyed the central disk of young massive stars, inside the CND, within its age of 6\,Myr. Infra-red emission from the dust in the CND has the characteristics of an optically thin medium and is further evidence for scenario one.  For an optically thick medium (scenario two) cores would appear as dark spots in an infra-red image \citep{Mezger1996}. To date there have been no recorded observations of such dark spots and Scenario two relies on the size and density of the cores to provide stability against tidal forces and long lifetimes. 

\paragraph*{}
Large Velocity Gradient (LVG) modelling of CS(2-1), (3-2) and (5-4) transitions determined 10$^{6}$cm$^{-3}$ as the upper limit of molecular hydrogen density  \citep{SGE1989}; this is the same density proposed by \citet{Jacks1993}. \citet{Marr1993} proposed a density of 2$\times$10$^{6}$ cm$^{-3}$ for an optically thick ($\tau$=4) HCN(1-0) transition at a kinetic temperature of 250\,K. \citet{Chris2005} proposed a typical density of 3-4 $\times$ 10$^{7}$cm$^{-3}$ for an average core diameter of 0.25pc with a kinetic temperature of 50\,K and optically thick ($\tau$=4) HCN.  \citet{MMC2009} argued for virially stable cores based on HCN(4-3) data combined with HCN(1-0) \citet{Chris2005} data scaled to match. These papers show examples of both scenarios 1 \& 2 and indicate the need for further testing of the value for the hydrogen density in the CND. 

%\subsection{Current Analysis}
\paragraph*{}
The present thesis attempts to resolve the density question through LGV modelling of the HCN(1-0), (3-2) and (4-3) transitions together with the HCO$^{+}$(1-0) transition. Two sets of cores that are considered spatially and kinematically  matched are presented in the analysis. The first set of cores are taken from \citet{Marr1993} for the (1-0) transition of H$^{12}$CN, H$^{13}$CN and HCO$^{+}$  combined with the HCN(3-2) transition. The second set uses data for HCN(1-0) and HCO$^{+}$(1-0) \citep{Chris2005}, for HCN(3-2) \citep{Jacks1993} and for HCN(4-3) \citep{MMC2009}. The first set is  not resolved, however the transitions were scaled to a common resolution and hence is spatially and kinematically matched for all five cores. The second set consists of data from three sources with resolved data for the HCN(1-0) and (4-3) transitions and unresolved data for the HCN(3-2) transition which is corrected by a filling factor (based on the average core size of \citet{Chris2005}) to provide brightness temperature comparable with those for the resolved transitions. 

\paragraph*{}
LVG modelling for all transitions is carried out for a range of hydrogen densities and molecule column densities. The results favour an optically thin HCN and hence lower hydrogen density than advocated by \citet{Chris2005} and \citet{MMC2009}.

\section{Thesis Outline}
\paragraph*{}
Chapter \ref{Molex} outlines the radiative transfer equation and the equations describing emission and absorption, incorporating the Einstein A and B coefficients, optical depth, intensity, integrated intensity, excitation temperature, brightness temperature and collisions, which are used in the model for  tracing molecular rotational excitation from collisions with molecular hydrogen.  The molecular excitation model that is used to analyse the observed results is then described briefly. 

\paragraph*{}
Chapter \ref{diskgeom} uses the positions and velocities of twenty-six cores identified by \citet{Chris2005} to check the orientation of the disk's plane relative to the plane of the sky. The positions of each core in the plane of the sky  are converted to deprojected (true) offsets (from SgrA$^{*}$) in the disk's plane. Model core radial velocities were calculated for the adopted disk orientation of PA = 25$^{\circ}$ and inclination = 60$^{\circ}$ and are then compared with the observed core velocities.  The locations and radial velocities of  OH masers \citep{Sjman2008} and water and methanol masers \citep{FYZ2008} in and around the CND are also explored to assess their relationship to the disk. Two groups of cores are then identified as having consistent spatial and kinematic properties and selected for modelling. Five cores from \citet{Marr1993} were chosen as the first group and seven cores from \citet{Jacks1993} and \citet{MMC2009} that can be co-located with \citet{Chris2005} cores form the second. Four cores from \citet{Marr1993} are found to be co-located with members of the second group of seven cores 

\paragraph*{}
Chapter \ref{anal} describes the input parameters for the modelling, along with the results. The results indicate that the molecular hydrogen gas number density n(H$_{2}$) $\sim$ 10$^{6}$ cm$^{-3}$ agrees with scenario 1, with optically thin, inverted HCN(1-0) and HCO$^{+}$(1-0) transitions. Core densities are consistent with the \citet{Chris2005} optically thin scenario and as a consequence the cores are transient with little or no prospect of star formation from their condensation. The mass of the CND is also an order of magnitude less than the estimated 10$^{6}$ M$_{\odot}$. 

\paragraph*{}
Chapter \ref{concls} summarises the thesis results and conclusions, areas for future study are also identified.

\cleardoublepage

\chapter{\label{Molex} Molecular Excitation, Line Formation and Radiative Transfer} 

\section{Introduction}
\paragraph*{•}

This chapter outlines the radiative processes and their equations used to describe the radiative transfer process, particularly as it applies to emission arising from molecular rotation of tracers colliding with molecular hydrogen.

\paragraph*{•}
Section \ref{radcfs} describes the  instantaneous emission, absorption and collision processes which are associated with emission and absorption of photons from a material illustrated by Fig.\ \ref{translvls}. The Equation of Radiative Transfer which  describes how radiative intensity changes as radiation travels along a path through a medium with a known optical depth is then developed in Section \ref{trteqn}. Section \ref{backgrnd} follows explaining how the background radiation components are incorporated into radiative transfer equation.

\paragraph*{•}
Section \ref{bright} explains how brightness temperature, which is an alternative measure of intensity, is used in the LVG model. Integrated Intensity, which is the total radiation emitted by the medium and is the sum of intensities across the width of the spectrum, is then described in Section \ref{IntegI}.

\paragraph*{•}
The programme Molex, which was written by M. Wardle and modified by J. Chapman, that is used in this thesis to analyse the HCN observations reported in the published papers under review is described in the closing Section \ref{molexp}. The programme is based on the Large Velocity Gradient model (e.g. Section 14.10 of \citet{Rohlfs2006}). The model does not rely on the gas being in Local Thermal Equilibrium (LTE) and assumes that changes in local velocity predominate over thermal broadening of spectral lines. The probability of an emitted photon escaping the source is included in the radiative transfer equations for estimating the line intensities and optical depths of the radiating material. The model commonly assumes a spherical escape model and a Gaussian velocity profile but can cater for other escape models and velocity profiles.The adoption of an escape probability simplifies the analysis of the radiation transferred and promotes faster convergence to a solution of each calculation of the many energy transition levels of the trace molecule.

\paragraph*{•}
The HCN transitions of interest in this thesis are (1-0), (3-2) and (4-3). The (1-0) transition level observations of HCO$^{+}$ and H$^{13}$CN are also analysed. 
The calculations were checked using Radex, a similar programme to Molex.  
% which also uses the Cologne Data Base for Molecular Spectroscopy, CDMS, which
Radex is accessible on the web \footnote {URL http//:www.strw.leidenuniv.nl/\~{} moldata}, for calculations for static conditions of a species using  the underlying equations for this LVG model as outlined in \citet{vdeTak2004}. More recently a version of the programme has become  available in the public domain for download. Radex is also described in a more recent paper written by \citet{VderT2007}.
\paragraph*{•}
 
\section{Absorption and Emission Coefficients} \label{radcfs}
\begin{center}
\begin{figure}[ht] \centering \includegraphics[scale=1.0,bb= 0 20 370 190] 
%,clip=true,trim= 0 10 50 10] 
{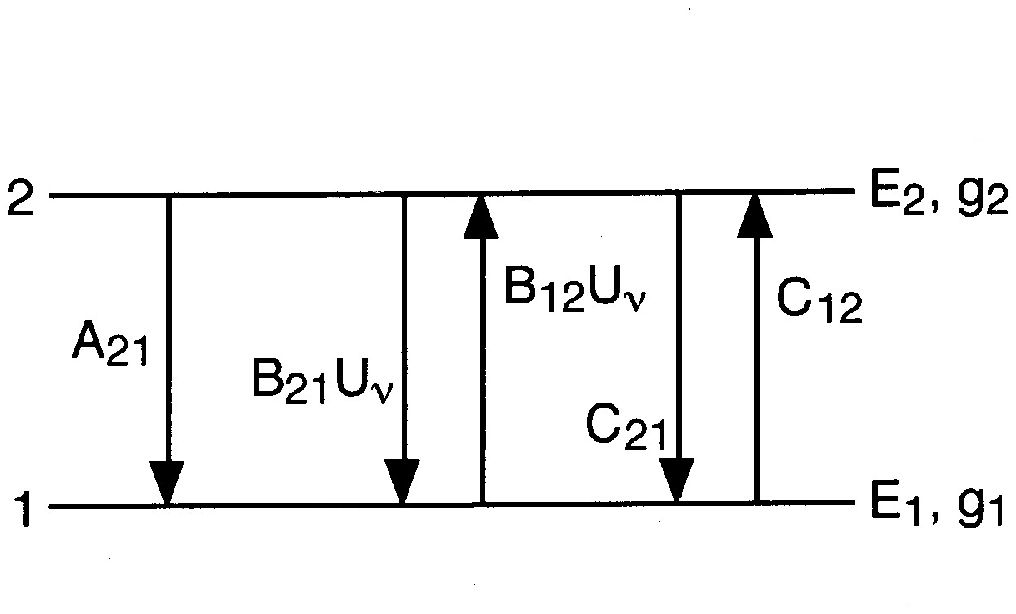} \fontsize{9} {9} \caption[Transitions between two energy levels]{\label{translvls} Transitions between two energy levels E$_{1}$ and E$_{2}$ above ground level with the statistical weights of the levels g$_{1}$ and g$_{2}$ and n$_{1}$ and n$_{2}$ molecules per unit volume. A$_{21}$ is the Einstein A probability coefficient for downward spontaneous transition from an upper to lower level; B$_{21}$ is the Einstein probability coefficient for stimulated emission; B$_{12}$ is the Einstein coefficient for absorption; U$_{\nu}$ is the energy density for photons per unit frequency interval; C$_{21}$ is the rate coefficient for de-excitation from level 2 to level 1 by collisions and C$_{12}$ is the rate coefficient for excitation from level 1 to level 2} 
\end{figure} 
\end{center}

The radiation intensity from a source for a simple case is a balance between emission and absorption of photons by the source. Collisions within the source also contribute indirectly to the intensity by excitation and de-excitation which change level populations. These processes are illustrated in Fig.\ \ref{translvls} and described in the following sub-sections.

\normalsize

\subsection{The Spontaneous Emission Coefficient j$_{\nu}$}
\paragraph*{•}
The spontaneous emission coefficient of isotropically emitted photons is given by the expression
\begin{equation} \label{jnu}
j_{\nu} = \frac{n_{2}A_{21}}{4\pi}h \nu \, \phi(\nu)\,,
\end{equation}
where n$_{2}$  is the population of the upper level 2, of a transition,		A$_{21}$ is the Einstein A probability coefficient of emission (see Fig.\ \ref{translvls}), h is Planck's constant, $\nu$ is the frequency of the emission of the nominated transition between an upper and lower level and $\phi(\nu)$ is the line function describing the shape of the spectral line centred on the frequency $\nu_{0}$, (e.g.\ a Gaussian) and can be defined as \\
\begin{equation} \label{lineprof}
\int_0^{\infty}\!\phi(\nu-\nu_{0})\,\mathrm{d}\nu = 1 \,.
\end{equation}

\paragraph*{•}
The intensity of emission at frequency $\nu$, I$_{\nu}$, can be expressed as the integral of the emission coefficient along a path, ds, through the medium (if it is optically thin) as
\begin{equation}\label{Inu}
I_{\nu} = \int \!j_{\nu}\,\mathrm{ds}
\end{equation}
so that the line integrated emission I is given by the expression 

\begin{equation}
I = \int I_{\nu}\,\mathrm{d}\nu
\end{equation}
and substituting for I$_{\nu}$ from Eqn.\ \ref{Inu} gives
\begin{eqnarray}\label{Itot}
 I &=& \int_{line}\left(\int\!\frac{\mathrm{n}_{2}A_{21}}{4\pi} h\nu_{21}\phi(\nu)\,\mathrm{d}s \right)\,\mathrm{d}\nu \\ 
 &=& \int_{line}\frac{A_{21}}{4\pi} h\nu_{21}\phi(\nu)\left( \int \mathrm{n}_{2}\,\mathrm{d}s\right)\,\mathrm{d}\nu \,, 
\end{eqnarray}
where n$_{2}$ is the population of the transition's upper level which when integrated along the emission's path through the medium produces the total column density, N$_{2}$ in the upper level. The line function, $\phi(\nu)$, is assumed independent of the distance,s, through the medium.

The column density, N$_{\mathrm{mol}}$, in all levels of the molecule is related to  N$_{2}$ by

\begin{equation}
N_{2} = N_{\mathrm{mol}} \times \mathrm{x}_{2}
\end{equation}
where x(2) is the fraction of the total population in the upper level Eqn.\ \ref{Itot} then becomes 
\begin{eqnarray} \label{Line}
 I &=& \frac{h\nu_{21}}{4\pi} A_{21}N_{2}\int\phi(\nu)\,\mathrm{d}\nu,  \\
 I &=& \frac{h\nu_{21}}{4\pi} A_{21}N_{\mathrm{mol}}x(u). 
\end{eqnarray}

\subsection{Absorption Coefficient $\alpha_{\nu}$}

\paragraph*{•}
Optical depth, \large{$\tau_{\nu}$} \normalsize , is defined as the integration of the absorption coefficient $\alpha_{\nu}$ over the depth, ds, of the radiation field as follows

\begin{equation}\label{taunt}
\tau_{\nu} = \int \alpha_{\nu}\,\mathrm{d}s \,,
\end{equation}
where $\alpha_{\nu}$, the coefficient of absorption is defined by 
\begin{equation}
\alpha_{\nu} = (n_{1}B_{12} - n_{2}B_{21})h\nu\phi(\nu) \,,
\end{equation}
and n$_{2}$ and n$_{1}$ are the upper and lower level populations, B$_{12}$ is the Einstein B probability coefficient for absorption and B$_{21}$ the Einstein probability coefficient for stimulated emission, (see Fig.\ \ref{translvls})\\% and $\phi(\nu)$ is the line profile (see Eqn.\ \ref{lineprof}).

Using the relationship
\begin{equation}
g_{1}B_{12} = g_{2}B_{21}
\end{equation}
where g$_{1}$ and g$_{2}$ are the statistical weights of the transition states and
\begin{equation}
A_{21} = \frac{2h\nu^{3}}{c^{2}}B_{21} \, ;
\end{equation}
substituting into the expression for  $\alpha_{\nu}$ produces
\begin{equation}
\alpha_{\nu} = \frac{A_{21}}{8\pi}\frac{c^{2}}{\nu^{2}}\left(\frac{n_{1}g_{2}}{g_{1}}-n_{2}\right) \phi(\nu) \,;
\end{equation}
and further substituting x$_{1}n_{\mathrm{tot}}$ for $n_{1}$ and  x$_{2}n_{\mathrm{tot}}$ for $n_{2}$, where x$_{1}$ and x$_{2}$ are the fractions of the total population n$_{\mathrm{tot}}$ produces,

\begin{equation}
\alpha_{\nu} = \frac{A_{21}}{8\pi}\frac{c^{2}}{\nu^{2}} n_{\mathrm{tot}}\left( \frac{x_{1}g_{2}}{g_{1}}-x_{2}\right) \phi_{\nu} \,. \\
\end{equation}
Integrating with respect to $\mathrm{ds}$ gives

\begin{eqnarray}
\tau_{\nu} &=& \int \alpha_{\nu}\,\mathrm{d}s \\
  &=& \frac{A_{21}}{8\pi}\frac{c^{2}}{\nu^{2}} \left( \frac{x_{1}g_{2}}{g_{1}}-x_{2}\right) \phi(\nu)\int n_{\mathrm{tot}}\,\mathrm{d}s \,.
\end{eqnarray}

Using the relationships \[ \frac{\mathrm{d}\nu}{\nu} = \frac{\mathrm{d}v}{c}  \] and

\[ \phi(\nu)\mathrm{d}\nu = \phi(\mathrm{v})\mathrm{dv} \]
where $\phi(\nu)$ is the frequency line profile, $\mathrm{d}\nu$ a small interval in line frequency, $\phi$(v) is the velocity line profile and $\mathrm{dv}$ a small interval in line velocity implies that,

\[ \phi(\nu) = \phi(\mathrm{v})\frac{\mathrm{dv}}{\mathrm{d}\nu} = \phi(\mathrm{v})\frac{\mathrm{c}}{\nu} \]

and so
\begin{equation}\label{taunu}
\tau_{\nu} = \frac{A_{21}c^{3}}{8\pi\nu^{3}} \left( \frac{x_{1}g_{2}}{g_{1}}-x_{2}\right) N_{\mathrm{mol}}\phi(v) \,.
\end{equation}
 The optical depth at the line centre of a Gaussian line profile is calculated from the expression

\begin{equation}
\phi(\mathrm{v}) = \frac{1}{\sqrt{2\pi}\sigma_{\mathrm{v}}}\exp\left( -\frac{\mathrm{v}^{2}}{2\sigma_{\mathrm{v}}^{2}}\right)  
\end{equation}
where $\sigma_{\mathrm{v}}$ is the standard deviation of the Gaussian. To relate the FWHM velocity $\Delta\mathrm{v}$ to $\sigma_{\mathrm{v}}$ the following expression is evaluated for the FWHM line width value, $\phi(\mathrm{v})$ = $\frac{1}{2}$ for a curve with central value = 1.

\begin{equation}
\exp\left( -\frac{(\Delta V/2)^{2}}{2\sigma_{V}^{2}}\right)  = \frac{1}{2}
\end{equation}
so that 

\begin{equation}
\sigma_{v} = \frac{\Delta V}{2\sqrt{2\mathrm{ln}2}} \,.
\end{equation}
At the line centre the value of the line profile $\phi$(0) is

\begin{equation} \label{profile}
\phi(0) = \frac{1}{\sqrt{2\pi}\sigma_{v}} = 2\sqrt{\frac{\mathrm{ln}2}{\pi}}\left( \frac{1}{\Delta V}\right) \,, 
\end{equation}
so that by substituting Eqn.\ \ref{profile} into Eqn.\ \ref{taunu}, the optical depth at line centre $\tau_{0}$ is
\begin{equation} \label{Tau0}
\tau_{0} = 2\sqrt{\frac{\mathrm{ln}2}{\pi}}\frac{A_{21}}{\Delta V}\frac{\lambda^{3}N_{mol}}{8\pi}\left( x_{1}\frac{g_{2}}{g_{1}}-x_{2}\right) \,.
\end{equation}

\subsection{Collisions} \label{colls}
\paragraph*{}
The number of effective collisions between two particles A and B per unit volume 
is given by 
\[ \mathrm{n_{A}}\mathrm{n_{B}}\sigma\mathrm{(v)V} \]
where 	
n$_{A}$ and n$_{B}$ are number densities of each in cm$^{-3}$, $\sigma$(v) is the collisional cross section (which depends on the energy level of the incoming particles) and V is the velocity of the incoming particles in cm s$^{-1}$ 
       
The collision rate \large{$\gamma$} \normalsize for a Maxwellian velocity distribution of moving particles at temperature T is given by the expression

\begin{equation}
\large{\gamma}\normalsize _{AB} = n_{A}n_{B} \int^{\infty}_0 V\sigma(\mathrm{v})4\pi V^{2}\Big(\frac{m}{2\pi kT}\Big)^{\frac{3}{2}}\exp\left( -\frac{mV^{2}}{2kT}\right) \mathrm{d}V \,;
\end{equation}
for a large number of collisions and applying the principle of detailed balance the following equation applies

\begin{equation}
 \frac{\gamma_{12}}{\gamma_{21}} = \frac{g_{2}}{g_{1}}\exp-\left( \frac{\mathrm{E}_{21}}{\mathrm{kT}}\right) \,.
\end{equation}

Collision rates are used in Molex to calculate the populations for the range of transition levels of the rotating trace molecules colliding with molecular hydrogen in the CND. The collision rates for all transitions at a range of specific temperatures are included in the molecular properties data file input to Molex. Molex uses an extrapolation routine to calculate collision rates for the kinetic temperature of the molecule to be modelled.

\paragraph*{•}
The population of a series of levels $j = 1,2,3,4 \cdots $ is determined by the rates of collisional and radiative excitation and de-excitation between the levels  (see Fig.\ \ref{translvls}).  

\paragraph*{•}
Changes in all levels of the molecule contribute to changes in the population of level j according to

\begin{equation}\label{levpop}
\begin{split}
\dfrac{\mathrm{d}n_{j}}{\mathrm{d}t} = \sum_{k>j}A_{kj}n_{k} + \sum_{k<j}(B_{kj}U_{kj}n_{k}-B_{jk}U_{kj}n_{j}) + \sum_{k>j}(C_{kj}n_{k} - C_{jk}n_{j}) \\
  - \sum_{i<j}A_{ji}n_{j} + \sum_{i<j}(B_{ij}U_{ji}n_{i} - B_{ji}U_{ji}n_{j}) + \sum_{i<j}(C_{ij}n_{i} - C_{ji}n_{j}) 
\end{split} 
\end{equation}
\begin{equation}
\begin{split}
 = \sum_{k>j}\left[A_{kj}n_{j}+B_{kj}U_{kj}\left(n_{k}-\dfrac{g_{k}}{g_{J}}n_{j}\right)\right] - \sum_{i<j}\left[A_{ji}n_{j} + B_{ji}U_{ji}\left(n_{j}-\dfrac{g_{j}}{g_{i}}n_{i}\right)\right] \\
 + \sum_{k>j}(C_{kj}n_{k} - C_{jk}n_{j}) +\sum_{i<j}(C_{ij}n_{i} - C_{ji}n_{j}) \,. \\
\end{split} 
\end{equation}
U$_{kj}$ and U$_{ji}$ are the radiation densities integrated over their respective levels kj and ji and defined as,

\begin{equation}
U_{ji} = \int \phi_{ji}(\nu)U_{\nu}\mathrm{d}\nu \,,
\end{equation}
where U$_{\nu}$, the energy density for photons per unit frequency interval is,

\begin{equation}
U_{\nu} = \dfrac{1}{c}\int I_{\nu}\mathrm{d}\Omega \,.
\end{equation}
 
Assuming full redistribution, so that the one dimensional velocity distribution function is the same Maxwellian for all transitions, then \\
\begin{equation}
\phi_{ji}(\nu) = \dfrac{\lambda_{ji}}{(2\pi)^{1/2}\sigma} \mathrm{exp}(-v^{2}/2\sigma^{2})
\end{equation}
where 
\begin{equation}
\sigma = (kT/m)^{1/2}   
\end{equation}
and
\begin{equation}
v = -(\nu - \nu_{ij})c/\nu_{ij} \,.
\end{equation}
The flux from the transition j to i can then be defined as \\
\begin{equation}
F_{ji} = \dfrac{B_{ji}U_{ji}}{A_{ji}} = \dfrac{c^{2}}{2h\nu^{3}}\int\phi_{ji}I_{\nu}\mathrm{d}\nu\dfrac{\mathrm{d}\Omega}{4\pi}
\end{equation}
\\
where the integral is the angle-averaged intensity, J$_{\nu}$, integrated over the line profile. The radiation field is determined by the radiative transfer Eqn.\ \ref{RTE1}. The energy density U$_{\nu}$ enters Eqn.\ \ref{levpop} as the level populations determine the emission and absorption coefficients j$_{\nu}$ and $\alpha_{\nu}$. 

\paragraph*{}
While it is very difficult to simultaneously solve the equations for statistical equilibrium and radiative transfer in the general case it can be done for simple geometries as follows.
Any transition 2 $\rightarrow$ 1 contributes\\

\begin{equation}
A_{21}\left[n_{2} +F_{21}\left(n_{2}-\dfrac{g_{2}}{g_{1}}n_{1}\right)\right]
\end{equation} \\
positively to $\mathrm{d}n_{1}/\mathrm{d}t$ and negatively to $\mathrm{d}n_{2}/\mathrm{d}t$. For a spatially uniform source function $S_{\nu}= j_{\nu}/\alpha_{\nu}$, \\

\begin{equation}
I_{\nu} = I_{0}e^{-\tau_{\nu}} + S_{\nu}(1-e^{-\tau_{\nu}}) \,.
\end{equation}

For full redistribution in the absence of overlapping lines $S_{\nu} = B_{\nu}(T_{ex})$ and does not change significantly over the line, this usually true for the background radiation I$_{0}$ so that\\

\begin{eqnarray}
F_{21} &=&  \frac{c^{2}}{2h\nu^{3}}\left[S_{\nu} + (I_{0} - S_{\nu})\int\phi_{21}e^{-\tau_{\nu}}\mathrm{d}\nu\dfrac{\mathrm{d\Omega}}{4\pi}\right]  \\
&= &\frac{c^{2}}{2h\nu^{3}}\left[ \beta_{21}I_{0} + (1 - \beta_{21})S_{\nu}\right] \,. \\
\end{eqnarray}

\[ \beta_{21} =\int\phi_{21}e^{-\tau_{\nu}}\mathrm{d\nu}\dfrac{\mathrm{d\Omega}}{4\pi} \] \\
is the escape probability and in this case the contribution to $\mathrm{d}n_{1}/\mathrm{d}t$ is \\

\begin{equation}
 \beta_{21}A_{21}n_{2} + \dfrac{c^{2}}{2h\nu^{3}}I_{0}\left(n_{2} - \dfrac{g_{2}}{g_{1}}n_{1}\right)\beta_{21}A_{21}. \\
\end{equation}

The escape probability for a sphere has been adopted for the analysis performed in this thesis and in Molex to calculate $\beta$ where the absolute value of $\tau > 0.001$ in Molex  and for a radius of $\tau/2$  the escape probability is

\begin{equation}
\beta = \frac{1.5}{\tau}\left( 1-\frac{2}{\tau^{2}}+ \left( \frac{2}{\tau}+\frac{2}{\tau^{2}}\right)e^{-\tau}\right) 
\end{equation}
 \\
which reduces to $\frac{1.5}{\large{\tau}}$ \normalsize under optically thick conditions.

\section{The Radiative Transfer Equation} \label{trteqn}
\paragraph*{•}
The radiative transfer equation relates the change in intensity along a path to the net effect of spontaneous emission and absorption. It is given by
\begin{equation} \label{RTE1} 
\frac{dI_{\nu}}{\mathrm{ds}} = -\alpha_{\nu}I_{\nu} + j_{\nu}
\end{equation} 
%\begin{center}
where $\alpha_{\nu}$ is the absorption coefficient, j$_{\nu}$ is the spontaneous emission coefficient, I$_{\nu}$ is the intensity and ds is the distance along the radiation's path through the medium. \\
%\end{center}

Optical depth in the differential form of Eqn.\ \ref{taunt} is 
%\begin{center}
\begin{equation}
\mathrm{d}\tau_{\nu} = \alpha_{\nu}\mathrm{ds}
\end{equation}
and dividing Eqn.\ \ref{RTE1} by $\alpha_{\nu}$ leads to
\begin{equation} \label{RTE2}
\frac{\mathrm{d}I_{\nu}}{\mathrm{d}\tau_{\nu}} = -I_{\nu} + S_{\nu}
\end{equation} 
where the source function is 
%\begin{center} 
\begin{equation}
S_{\nu} = \frac{j_{\nu}}{\alpha_{\nu}} \,.
\end{equation}

Transposing the terms of Eqn.\ \ref{RTE2} and multiplying by e$^{\tau_{\nu}}$ gives
\begin{equation}
\left( I_{\nu} + \frac{\mathrm{d}I_{\nu}}{\mathrm{d}\tau_{\nu}}\right) e^{\tau_{\nu}} = S_{\nu}e^{\tau_{\nu}}; 
\end{equation}
the expression on the left hand side (LHS) is  $\dfrac{\mathrm{d}}{\mathrm{d}\tau}\left( I_{\nu}e^{\tau_{\nu}}\right) $ so that integrating with respect to $\tau_{\nu}$ produces

\begin{equation} 
\left[ I_{\nu}e^{\tau_{\nu}^{'}} \right]_0^{\tau_{\nu}} = \int_0^{\tau_{\nu}}\!S_{\nu}e^{\tau_{\nu}^{'}}\ \mathrm{d}\tau_{\nu}^{'}
\end{equation}
and assuming that $S_{\nu}$ is constant,
\begin{equation} \label{RTE3}
I_{\nu} = I_{0}e^{-\tau_{\nu}} + S_{\nu}(1-e^{-\tau_{\nu}})
\end{equation} 
where $I_{0} = I_{\nu}$ at $\tau_{\nu} = 0$ \,. \\

The first and second terms on the right hand side (RHS) of Eqn.\ \ref{RTE3} are the contributions of the background radiation and emission from within the medium , correspondingly corrected for optical depth.

\paragraph*{•}
The radiative transfer equation is a central part of the calculation of intensities in Molex, and S$_{\nu}$ depends on the molecular excitation discussed in Section \ref{colls}.

\section{Background Radiation} \label{backgrnd}
\paragraph*{•}
Background radiation arises from two sources :

\begin{enumerate}
\item the cosmic microwave background, CMB, at the relevant frequency calculated from the Planck function  
\begin{equation} \label{cmb}
I_{\mathrm{CMB}} = B_{\nu}(T_{\mathrm{CMB}}) = \dfrac{2h\nu^{3}/c^{2}}{(\exp(h\nu/kT_{\mathrm{CMB}})-1)}
\end{equation}
where $T_{\mathrm{CMB}} = 2.728\,\mathrm{K}$,  and\\
\item IR radiation from the irradiation of the CND's dust by sources located in the central cavity which is surrounded by the CND. The contribution of external dust radiation is calculated from  
\begin{equation} \label{dust}
I_{\mathrm{dust}} = B_{\nu}(T_{\mathrm{dust}})(1-\exp(-\tau_{\mathrm{dust}})) \,,
\end{equation}
where T$_{\mathrm{dust}}$ is the dust temperature and $\tau_{\mathrm{dust}}$ is the dust's optical depth.
\end{enumerate}

\paragraph*{•}
The total background radiation, I$_{0}$, is the sum of the contribution from these background sources. The LVG model subtracts the background radiation (I$_{0}$) from total radiation to provide a radiation intensity for comparison with the gas core intensities. This given by \\
\begin{equation}
I_{\nu} - I_{0} = S_{\nu}(1-e^{-\tau}) + I_{0}e^{-\tau} - I_{0},
\end{equation}
which reduces to
\begin{equation}\label{Intlinc}
I_{\nu} - I_{0} = \left(S_{\nu}-I_{0}\right)\left(1-e^{-\tau}\right)\,, \\
\end{equation}
where the LHS of Eqn.\ \ref {Intlinc} represents the net intensity at line centre. 

\paragraph*{•}
The effect of dust in the LVG model is taken into account by specifying total extinction in the V filter band (Av) and temperature (T$_{dust}$). Av depends on the dust's optical depth and temperature that \citet{Marr1993} assumed to be 75\,K and together with the observed intensity of $2.0\times 10^{5}$ M Jys$^{-1}$ at 100$\mu$m (both estimates inferred from \citet{Becklin1982}) allowed calculation of the dust's optical depth using, 
\begin{equation} \label{Inutau}
I_{\mathrm{dust}} = \large{\tau}\normalsize_{\mathrm{dust}} \times B_{\nu} (T_{\mathrm{dust}}) \,,
\end{equation}
where B$_{\nu}$ is calculated using the Planck equation for T$_{\mathrm{dust}}$ = 75\,K and $\lambda$ = 100$\mu$m. The result, $\tau_{\mathrm{dust}} = 3.43 \times 10^{-2}$ that is consistent with the use of the optically thin expression in Eqn.\ \ref{Inutau} and also with the estimates of \citet{Becklin1982} who found that the mean optical depth over their map(Fig.\ 1(c)) was 0.05 and the value at the position of the galactic centre was $\backsim 0.03$.   

\paragraph*{•}
Av is related to the dust's optical depth ($\tau_{\mathrm{dust}}$) and the column density for neutral hydrogen (N$_{H}$) substituting typical values of N$_{H}$ = 5.8$\times$10$^{21}$ E(B-V) cm$^{-2}$ mag$^{-1}$ \citep{Bohlin1978}, $\tau_{\mathrm{v}}$ = 0.4$\ln$ 10 Av \citep{Lockett1989} and Av = 3.1 E(B-V) \citep{Rieke1985} into the relationship between the dust optical depth and extinction was established from an adopted fit of the total graphite and silicate extinction curve in .\ 9 of \citet{Draine1984} produces,
\begin{equation} \label{Av1}
\large{\tau}\normalsize_{\mathrm{dust}} = 12.769 f_{x} \lambda^{\!-2} \mathrm{Av} \,,
\end{equation}
where 
\begin{equation}
f_{x} = -11.478 +20.1769x -12.2355x^{2} +3.2809x^{3} -0.33010654x^{4} \,,
\end{equation}
is a polynomial fit to the extinction curve (M. Wardle, private communication) and 
\begin{equation}
\mathrm{x} = \mathrm{log}_{10} \lambda (\mu\mathrm{m}) \,.
\end{equation}

Eqn.\ \ref{Av1} is valid for $30 \mu \mathrm{m} <\lambda < 1000 \mu \mathrm{m}$, for longer wavelengths $f_{x} = 1$ and entering $3.43 \times 10^{-2}$  into Eqn.\ \ref{Av1} for $\tau_{dust}$ gives a value of Av = 26.9 which is used in Molex. 

\section{Brightness Temperature} \label{bright}
\paragraph*{•}
The brightness temperature B$_{\nu}$(T$_{b}$) (the equivalent black body radiation from a source at temperature T) is usually derived from the Planck function
\begin{equation}\label{Planck}
B_{\nu}(T_{b}) = I_{\nu}
\end{equation}
however in the millimetre and sub millimetre wavelength ranges, the brightness temperature is frequently defined as
\begin{equation}\label{TRJ}
T_{b}(RJ) = \frac{\lambda^{2}}{2\mathrm{k}} I_{\nu},
\end{equation}
where T$_{b}$(RJ) is the brightness temperature which is the intensity based on the Rayleigh-Jeans approximation to the Planck function where h$\nu$ $\ll$ kT. In this equation $\lambda$ is the radiation's wavelength, k is the Boltzmann constant and I$_{\nu}$ is radiation's intensity. This relationship is used at sub mm wavelengths where it does not strictly apply but corresponds to observer conventions. The peak brightness temperature at line centre is T$_{b}$(0) where the shorthand version T$_{b}$ will be used and is the value of the temperature used in the model to match the observed brightness temperatures in Chapter \ref{anal}. 

\section{Integrated Intensity} \label{IntegI}
\paragraph*{}
Molex calculates integrated intensity by using the line's wavelength ($\lambda$), Einstein A coefficient (A$_{21}$), HCN column density (N$_{\mathrm{HCN}}$), upper level population (x$_{u}$) and escape probability ($\beta$). 

The equation is derived as follows from the definition: 
\begin{equation} \label{intInu}
 \int I_{\nu}\,\mathrm{d}\nu = 2k\frac{\nu^{2}}{c^{2}}\int T_{b}\,\mathrm{d}\nu \,, 
\end{equation}
and using the relationship 
%\begin{equation}
\[\frac {\mathrm{d}\nu}{\nu} = \frac{\mathrm{d}V}{c} \,,\]

the RHS of Eqn.\ \ref{intInu} can be integrated with respect to $\mathrm{d}\nu$ :
\begin{equation}
\int I_{\nu}\,\mathrm{d}\nu = 2k\frac{\nu^{2}}{c^{2}}\frac{\nu}{c}\int T_{b}\,\mathrm{d}V
\end{equation}
so that
\begin{equation} \label{IntTb}
\int T_{b}\,\mathrm{d}V = \frac{c^{3}}{2k\nu^{3}}I_{\nu} \,.
\end{equation}

Substituting the intensity from Eqn.\ \ref{Line} and adjusting for the likelihood of re-absorption using the escape probability term $\beta$ Eqn.\ \ref{IntTb} becomes:
\begin{equation}
\int T_{b}\,\mathrm{d}V = \frac{c^{3}}{2k\nu^{3}}\frac{h\nu_{21}}{4\pi} A_{21}N_{\mathrm{mol}}x_{u}\beta \,.
\end{equation}
This reduces to

\begin{equation}\label{INTI}
\int T_{b}\,\mathrm{d}V = \frac{hc\lambda^{2}}{8\pi}A_{21}N_{\mathrm{mol}}x_{u}\beta \,,
\end{equation}
which is used in Molex expressed in units of K km s$^{-1}$.

\section{The Molex Programme} \label{molexp}
\paragraph*{}
Molex uses one of a number of molecular data files from the Cologne Data Base for Molecular Spectroscopy that contains details of transition frequencies, Einstein A  coefficients and collision rates for a range of gas temperatures for all the molecule's transitions. Data from a second input file prepared by the user specifies the molecule and the transitions to be analysed, the kinetic temperature of the gas, the dust temperature, dust Av extinction, the fraction of molecular hydrogen in ortho form (0.75), the initial and final hydrogen densities along with the step interval, and the initial, final and step intervals for the column density of the rotating trace molecule. 

\paragraph*{}
The programme starts at a point where the HCN molecule is in a state of LTE and operates in specified steps of decreasing hydrogen number density and increasing HCN column density to cover the range of interest. Level populations are calculated along with escape probabilities for all levels to produce intensities, opacities and brightness for the gas temperatures. This author used $10^{12}$  as the starting value for both the HCN column density and atomic hydrogen number density to ensure initial LTE conditions. Log column densities were then incrementally increased from 12 to 18 for each increment of log hydrogen densities which is decreased from 12 to 3, increments of 0.1 are specified for both densities. 

\paragraph*{}
Selected output parameters, notably the brightness temperature T$_{b}$ and optical depth \large{$\tau$} \normalsize for multiple transitions were plotted as contours on graphs with an ordinate of log\,n$_{H}$ cm$^{-3}$ and abscissa log (N$_{\mathrm{mol}}$/$\delta$V) cm$^{-2}$ per (km s$^{-1}$) using an IDL routine which was written to accept the output from Molex. The graphs shown in Chapter \ref{anal} make the thousands of lines of Molex output more easily interpreted so that the trends in the data can be analysed.

\paragraph*{}
The next chapter will outline the geometry of the CND and its orientation in relation to the plane of the sky before proceeding to the choice of cores for analysis with Molex in Chapter \ref{anal}.
\cleardoublepage  
\chapter{Disk Geometry, HCN Cores and Masers} \label{diskgeom}
\section{Introduction}
\paragraph*{}
This chapter explores the geometry and kinematics of the Circumnuclear Disk using the co-ordinates and deprojected distances from SgrA$^{*}$ of the HCN cores listed in Table 2 of \citet{Chris2005}. The aim of this work is to establish which of the observed cores are located in the CND by comparing the cores' radial velocities with a set of modelled radial velocity curves for the CND and then  identify suitable cores for LVG modelling.
\paragraph*{}
The velocity curves for points in an inclined rotating disk are based on a flat rotation curve, where the rotational velocity v$_{\phi}$ is constant and its thickness is zero. The orientation of the disk is determined by the position angle (PA) of its major axis East of North in the plane of the sky, and the inclination angle of its axis of rotation to an observer's line of sight.  
\paragraph*{}
Only projected co-ordinates and the deprojected distances of the cores from SgrA$^{*}$ were listed in Table 2 of \citet{Chris2005}. The deprojected co-ordinates have to be calculated from their projected co-ordinates by assuming values for the above two angles and comparing the deprojected distances with the listed values. In Section \ref{cordtran} generic transformation equations are established for conversion of the plane of sky, projected, co-ordinates to the plane of disk (deprojected) co-ordinates, which lead to a trial and error determination of the disk's inclination to the plane of the sky for the HCN (1-0) cores (see Section \ref{2005cores}). 
\paragraph*{}
The reverse transformation equations are then formulated and the expression for the radial velocity is derived in Section \ref{radv} by differentiating the equation that describes the depth of field in the sky's plane in terms of the deprojected core's position in the disk with respect to SgrA$^{*}$. This is done to compare  core radial velocities from the model with observed values from \citet{Chris2005}.

%disk's co-ordinates ie. $\mathrm{d}z/\mathrm{d}t$ where z is the distance along the observer's line of sight and the radial velocity is the rate of change in this distance generated by the disk's rotation causing small changes in core positions in the disk's plane.
\paragraph*{}
The observed core radial velocities are then superimposed on the the plot of model radial velocity curves to show the anomaly between the observed and theoretical value for their position in the disk. The disk's angular parameters are varied to  assess their effects on the theoretical velocity curve and to produce an envelope of possible model values based on previously published values of the angles \citep{Marshall1995}.

\paragraph*{}
In Section \ref{cordconv}  J2000 co-ordinates are selected as the reference grid and B1950 co-ordinates are converted to J2000 offsets from SgrA$^{*}$. Table \ref{Core co-ords} lists the co-ordinates and SgrA$^{*}$ offsets for all cores and masers of interest for this thesis. 

\paragraph*{}
The locations of water, OH and methanol masers in the vicinity of the CND to and their relationship with the disk are discussed in Section \ref{secmasers}.

 \paragraph*{}
Finally two sets of CND cores that have observations of HCN in multiple transitions and that have the same spatial and kinematical properties are identified in Section \ref{coreselect}, so that their physical properties can be deduced using the Large Velocity Gradient (LVG) model in Chapter \ref{anal}.  

\section {Co-ordinate Transformation } %from the Sky's  to the Disk's Plane} 
\begin{figure}[ht] 
\centering
\vspace*{35pt}
\begin{pspicture}(-1,-1)(1,2)
  \psset{Alpha=5,Beta=-10,xMax=2, yMax=14, zMax=3}
\pstThreeDCoor[linecolor=blue,nameX=x,nameY=z,nameZ=y]
\ThreeDput[normal=1 -1 1](-2,0,2){Plane of Sky}
\ThreeDput[normal=1 0.55 0](0.2,0.8,2.5){\LARGE Line of Sight}
\pstThreeDPut(-0.2,-0.2,-0.2){o}
\end{pspicture}  
\caption[Astronomical 3D axes system]{Astronomical 3D axes system. ox points eastward. oy points northward and oz is positive away from the observer.} 
\label{3daxes}
%\end{center}
\end{figure}
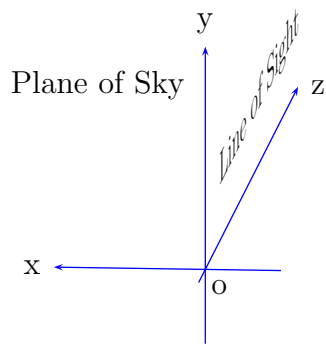

\label{cordtran}
\paragraph*{}
Astronomical figures use a 3D axes convention where the x axis runs East, the y axis runs North and the line of sight (z) axis, runs into the page (see Fig.\ \ref{3daxes}).

\normalsize

\paragraph*{}
Rotation matrices can be used to transform plane of sky to plane of disk co-ordinates where co-ordinates are expressed as offsets from SgrA*. The process is performed in two rotations, the first about the the line of sight (oz) so that the new axes, x$^{'}$ and  y$^{'}$, align with the major and minor axes of the projected disk and the second about the ox$^{'}$ axis to incline the disk's axis of rotation out of the plane of the sky to produce a disk with true or deprojected dimensions and  core offsets, x$_{D}$ and y$_{D}$ from SgrA$^{*}$.

\begin{figure}[!h] 
\begin{center}
\vspace*{35pt}
\begin{pspicture}(-2,-2)(2,2)
 \psset{Alpha=60,Beta=-90}
%\pstThreeDCoor[linecolor=blue,nameX=x,nameY=z,nameZ=y]
\pstThreeDCoor[linecolor=blue,xMin=0,xMax=3,yMin=0,yMax=3,zMax=5,nameX=x$^{'}$,nameY=y$^{'}$,
nameZ= ]
\pstThreeDCoor[linecolor=red,RotSequence=quaternion,RotSet=concat,
RotAngle=-60,xRotVec=0,yRotVec=0,zRotVec=1,xMin=0,xMax=3,yMin=0,yMax=3,zMin=0, zMax=3,nameZ=]
%,nameZ=]
%,nameZ= oz ]
\ThreeDput[normal=1 -1 1](-0.65,0,0.15){$\phi$}
\psarcn[arcsep=1.0pt,linecolor=black]{->}{0.8}{180}{120}
\psset{SphericalCoor=true}
\pstThreeDPut(0.0,180,0){$\oplus$}
\pstThreeDPut(0.5,190,0.2){oz = oz$^{'}$}
\end{pspicture}    
\caption[Rotation of x and y axes through an angle $\phi$ about the Line of Sight, oz] {Rotation of x and y axes through an angle $\phi$ about the line of sight, oz. ox$^{'}$ is the disk's projected semi-major axis and oy$^{'}$ is its projected semi-minor axis.}
\label{xyrot}
\end{center}
\end{figure}
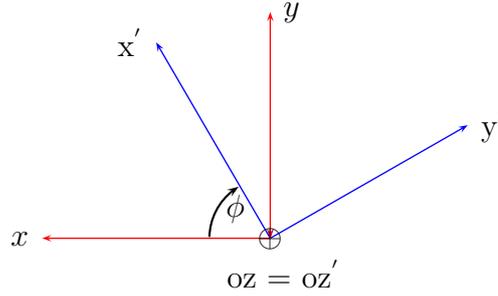

%\subsection{1 Rotation of sky co-ordinates to align with the disk's x$^{'}$, y$^{'}$ axes}
\paragraph*{}
Rotation of the x and y axes in the plane of the sky about the line of sight,(z axis), through an angle $\phi$ from the x axis towards the y axis with the transformed axes labelled x$^{'}$ and y$^{'}$  to align with the CND's major and minor axes, respectively, projected on the sky and z$^{'}$ axis remains the line of sight (see Fig.\ \ref{xyrot}).

\normalsize
The first transformation is given by
\begin{equation} \label{xytran}
\left( \begin{array}{c}x'\\y'\\z' \end{array}\right) = \left( \begin{array}{ccc}\cos\phi &\sin\phi &0\\-\sin\phi &\cos\phi &0\\0&0&1
\end{array} \right)\left( \begin{array}{c}x\\y\\z \end{array}\right) \,.
\end{equation}
%\[\  \]

%\subsection{2 Rotation of the line of sight to align with the disk's rotational axis z$_{D}$}
Rotation of y$^{'}$ and z$^{'}$ axes to align the z axis with the disk's axis of rotation z$_{D}$. The y$^{'}$z$^{'}$ axes are rotated through an angle $\alpha$ degrees away from the y$^{'}$ axis bring the x$^{'}$ and y$^{'}$ axes into the plane of the disk. The disk's axes are labelled x$_{D}$, y$_{D}$ and z$_{D}$.  
\begin{figure}[!h] 
\centering
\vspace*{40pt}
\begin{pspicture}(-2,-2)(2,2)
 \psset{Alpha=90,Beta=-90}
%\pstThreeDCoor[linecolor=blue,nameX=x,nameY=z,nameZ=y]
\pstThreeDCoor[linecolor=blue,xMin=0,xMax=3,yMin=0,yMax=3,zMax=5,nameX=y$^{'}$,nameY=z$^{'}$,
nameZ=]
%\ThreeDput(0,0,-0.3){x$^{'}$}
\pstThreeDCoor[linecolor=red,RotSequence=quaternion,RotSet=concat,
RotAngle=-60,xRotVec=0,yRotVec=0,zRotVec=1,xMin=0,xMax=3,yMin=0,yMax=3,zMin=0, zMax=3,nameX=y$_{D}$,nameY=z$_{D}$,nameZ= ]
\ThreeDput[normal=1 -1 1](-0.6,0,0.4){$\alpha$}
\psarc[arcsep=1.0pt,linecolor=black]{->}{1.0}{90}{150}
\psset{SphericalCoor=true}
\pstThreeDPut(0.0,180,0){$\oplus$}
\pstThreeDPut(0.5,180,-0.5){ox$^{'}$ = ox$_{D}$}
\end{pspicture}    
\caption[Rotation of y$^{'}$ and z$^{'}$ axes about the disk's major axis ox$^{'}$]
{Rotation of y$^{'}$z$^{'}$ axes through an angle $\alpha$ about the ox$^{'}$ axis oz$_{D}$ is the disk's axis of rotation, oy$_{D}$ is the disk's deprojected vertical axis and ox$_{'}$ is also ox$_{D}$ the disk's deprojected horizontal axis.}
\label{yzrot}
\end{figure}
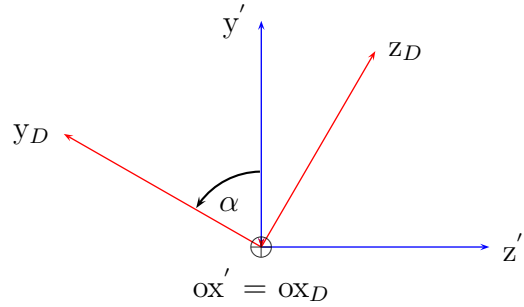
\normalsize
This second transformation is given by

\begin{equation} \label{yztran}
\left( \begin{array}{c}x_{D}\\y_{D}\\z_{D} \end{array}\right) = \left( \begin{array}{ccc}1&0&0\\0&\cos\alpha&-\sin\alpha\\0&\sin \alpha&\cos\alpha \end{array} \right) \left( \begin{array}{c}x'\\y'\\z' \end{array}\right) \,.
\end{equation}
%\[ \]

\paragraph*{}
The transformation from sky to disk co-ordinates is then derived by substituting the matrices from Eqn.\ \ref{xytran} into Eqn.\ \ref{yztran} to produce

%\[
\begin{equation}\label{TM}
 \left( \begin{array}{c}x_{D}\\y_{D}\\z_{D} \end{array}\right) = \left(
\begin{array}{ccc} \cos\phi &\sin\phi&0\\-\sin\phi\cos\alpha&\cos\phi\cos\alpha&-\sin\alpha
\\-\sin\phi\sin\alpha& \cos\phi\sin\alpha&\cos\alpha \end{array}\right)
\left( \begin{array}{c}x\\y\\z \end{array}\right) \,. \\
\end{equation}
%\]
The inverse transformation of Eqn \ref{TM} is

\begin{equation} \label{ITM}
\left( \begin{array}{c}x\\y\\z \end{array}\right) = \left(
\begin{array}{ccc} \cos\phi &-\sin\phi\cos\alpha&-\sin\phi\sin\alpha\\\sin\phi&\cos\phi\cos\alpha&\cos\phi\sin\alpha
\\0& -\sin\alpha&\cos\alpha \end{array}\right)
\left( \begin{array}{c}x_{D}\\y_{D}\\z_{D} \end{array}\right) \,, 
\end{equation}
\\

where x$_{D}$ and y$_{D}$, x and y offset  are offsets from SgrA$^{*}$ and z$_{D}$ is the disk's z co-ordinate which is aligned with its axis of rotation.

\section{Radial Velocity} \label{radv}
\paragraph*{}

The radial velocity calculation for the cores assumes that the rotational velocity, v$_{\phi}$ is independent of the radius within the disk. This flat rotation curve has been adopted by a number of authors including \citet{Marshall1995} and \citet{Guesten1987}. In \citet{Harris1985} a decline in rotational velocity by a factor of 1.4 to 2 is assumed between a disk radius of 2 to 6pc which is consistent with a ``Keplerian'' (R$^{-\frac{1}{2}}$) decline. This thesis adopts the flat velocity model on the basis that the majority of cores are within 2pc of SgrA$^{*}$ where the rotational velocity is considered constant and that the differences in radial velocity produced by a declining rotational velocity would be negligible due to the small changes in distances from  SgrA$^{*}$. Velocities along the disk's radii and velocities normal to the disk's plane are assumed to be zero.

%only four of the twenty-six cores in the study have deprojected distances more than 2pc from the galactic centre which justified the adoption of the flat rotational velocity model.    

\paragraph*{}
  A core's velocity along the line of sight or radial velocity is caused by the change in position of the core in the disk's plane generating a change in the distance along the line of sight in the sky. The position of a core in the disk is specified by its x$_{D}$ and y$_{D}$ co-ordinates. Evaluation of x$_{D}$ is straightforward as it is only dependent on the projected x and y co-ordinates. The y$_{D}$ co-ordinate is dependent on x, y and z co-ordinate values (see Eqn.\ \ref{ITM}. The projected z co-ordinate along the line of sight is not directly observable, but can be expressed as a function of the projected x and y co-ordinate values and y$_{D}$ calculated by assuming z$_{D}$ is zero as there is no independent information available for this quantity and it seems reasonable to adopt this assumption. 
  
  %to simplify the expression.

\paragraph*{•}
From the third component of Eqn.\ \ref{TM} 

\[ z_{D} = -x\sin\phi\sin\alpha + y\cos\phi\sin\alpha + z\cos\alpha = 0 \] \\  
so that 
\begin{equation} \label{zed}
z =  \frac{(x\sin\phi\sin\alpha - y\cos\phi\sin\alpha)}{\cos\alpha} \,;
\end{equation}
\\
and substituting Eqn.\ \ref{zed} for z into the second row of Eqn \ref{TM} produces
\[ y_{D} = -x\sin\phi\cos\alpha + y\cos\phi\cos\alpha - \sin\alpha((x\sin\phi\sin\alpha - y\cos\phi\sin\alpha)/\cos\alpha) \,, \] %\\ 
which reduces to
\begin{equation} \label{ydisk}
 y_{D} = \frac{(-x\sin\phi + y\cos\phi)}{\cos\alpha} \,. \\
\end{equation}

\paragraph*{}
The radial velocity is generated by the disk's rotation, which has a currently accepted rotational velocity v$_\phi$ of $110\pm5$ km s$^{-1}$ \citep{Marshall1995}. The disk's rotation produces movement (velocity) along the line of sight that is calculated from the inverse relationship (see Eqn \ref{ITM}).

%translates to a line of sight or radial velocity in the plane of the sky which is the following equations derived from 

\paragraph*{}
The position of a point in the disk can be expressed in cylindrical co-ordinates (r, $\phi_{D}$, z$_{D}$ as 
\[ x_{D} = r\cos\phi_{D} \qquad\mbox{ }\qquad y_{D} = r\sin\phi_{D} \qquad\mbox{and}\qquad z_{D}=0 \] \\  
where \[ r = \sqrt{x_{D}^2 + y_{D}^2} \]
and
\begin{equation} \label{phiD}
\phi_D= \arctan (y_{D}/x_{D}) \,.
\end{equation}
  
The third component of Eqn.\ \ref{ITM} produces
\begin{eqnarray}\label{RV1}
z = -y_{D}\sin\alpha \,,
\end{eqnarray}
 and substituting  r$\sin\phi_D$ for y$_{D}$ into Eqn \ref{RV1} 
gives 
\begin{eqnarray}
  z = -r\sin \phi_D\sin\alpha \,.
\label{dof}
\end{eqnarray}
Differentiating with respect to time gives the disk's rate of rotation, $ \mathrm{d}\phi/\mathrm{d}t $ \\
so that,
%\begin{align}
\begin{equation}
\frac{\mathrm{d}z}{\mathrm{d}t} = -r\cos \phi_D\sin \alpha \frac{\mathrm{d}\phi_D}{\mathrm{d}t} \,,
\end{equation}
%\nonumber \\
where for a constant rotational velocity v$_{\phi}$,
%\nonumber \\
\begin{equation}
\quad \frac{\mathrm{d}\phi_D}{\mathrm{d}t} = -\mathrm{v}_\phi/r \,.
\end{equation}
%\nonumber
%\end{align}
Since the disk is rotating anti-clockwise, $\mathrm{d}\phi_D/\mathrm{d}t$ is negative as a consequence of the three dimensional axes orientation defined at the start of Section \ref{cordtran} and shown in Fig.\ \ref{3daxes}. The model radial velocity is then obtained from the following expression:
\begin{eqnarray}\label{RV2}
 v_z = \frac{\mathrm{d}z}{\mathrm{d}t} = \mathrm{v}_\phi\cos \phi_D\sin \alpha \,.
\end{eqnarray}
%\vspace*{5pt}

\normalsize
%\afterpage{\clearpage}
%\newpage 
\section{Disk Parameters}
\begin{center}

\begin{figure}[ht] \includegraphics[bb= -125 0 600 220,scale=0.5]{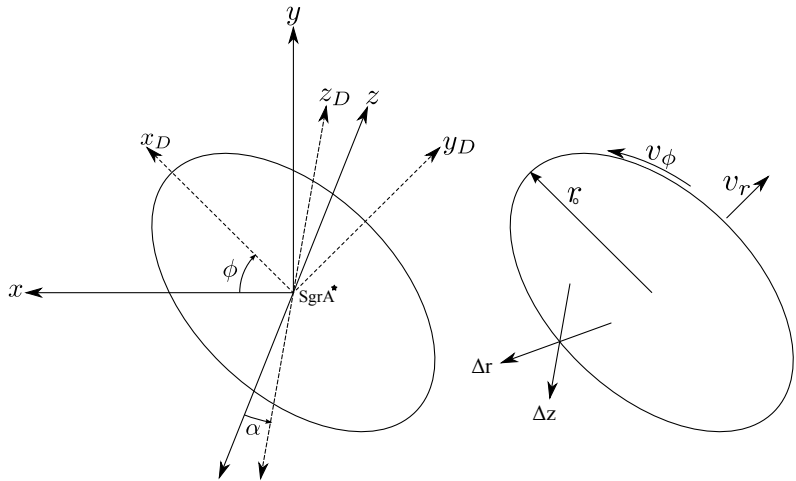} \fontsize{9} {9} \caption[Illustration of CND model parameters]{\label{CND params}Illustrates model parameters, the left diagram shows the plane of the sky xy with the z axis as the line of sight. x$_{D}$, y$_{D}$ define the disk's plane with z$_{D}$ the axis of rotation, $\alpha$ is the angle between z and z$ _{D} $; $\phi$ is the angle of the disk's major axis measured clockwise from East in the plane of the sky xy and is the complement of the PA (which is measured East of North). 
The right diagram shows the disk's physical dimensions: r$_{0} $ is the mean radius of the ring; v$ _{\phi} $ is the rotational velocity; v$ _{r} $ is radial velocity, $\Delta$r is radial thickness with FWHM thickness $ = r_{0}\sqrt{\mathrm{log4}} = 1.18r_{0} $; $\Delta$z is thickness perpendicular to the plane of the ring and is assumed to be zero. Typical values for these parameters are listed in Table \ref{Ring pars}. This figure and Table \ref{Ring pars} are based on Fig.\ 9 and Tables 1 \& 2 of \citet{Marshall1995}.} 
\end{figure}
\end{center}

\paragraph*{}
The direction of the disk's major axis is usually specified by its PA East of North, which in this study equates to the complementary angle of $\phi$, and its inclination to the plane of the sky, $\alpha$, as illustrated in Fig.\ \ref{CND params}. Common values of the parameters are given in Table \ref{Ring pars}. 
\paragraph*{}

\vspace*{50pt}
\begin{landscape}
\begin{center}
%\small
\begin{threeparttable}[!h]

%\begin{table}[!h] 
 
\caption{\label{Ring pars} CND's Defining Parameters.}
\begin{tabular}{c|c|c|c|c|c|c|c|c} 
\hline
\hline
\multicolumn{1}{c|}{ }&\multicolumn{1}{c|}{$\alpha$}&\multicolumn{1}{c|}{$\phi$ }&\multicolumn{1}{c|}{v$_{\phi}$}&\multicolumn{1}{c|}{v$_{r}$}&\multicolumn{1}{c|}{r$_{0}$}&\multicolumn{1}{c|}{$\Delta$r}&\multicolumn{1}{c|}{$\Delta$z}&\multicolumn{1}{c}{v$_{disp}$} \\
{}&\multicolumn{1}{c|}{(deg})&\multicolumn{1}{c|}{(deg)}&\multicolumn{1}{c|}{(km s$^{-1}$)}&\multicolumn{1}{c|}{(km s$^{-1}$)} &\multicolumn{1}{c|}{(arcsec)}&\multicolumn{1}{c|}{(arcsec)}&\multicolumn{1}{c|}{(arcsec)}&\multicolumn{1}{c}{(km s$^{-1}$)} \\
 & & & & & & & & \\
\hline
\hline
\multicolumn{1}{c|}{Previous Estimates\tnote{a}}&60-70&60-65&100-110&{$<20$}&40-50&\multicolumn{1}{c|}{extends from 30 arcs}&\multicolumn{1}{c|}{10-15 at inner edge}&10-70 \\
{Larger Ring\tnote{b}}&70& &110&52&96&18&27&10 \\
\multicolumn{1}{c|}{HCN$_{3-2}$\tnote{c}}&59&64&76&-12&38&36&15&47 \\ 
\multicolumn{1}{c|}{HCN$_{4-3}$\tnote{c}}&60&65&72&-13&44&13&16&40 \\ 
%\hline
%\hline
%{\citet{Jacks1993}}& & & & & & & & \\
\multicolumn{1}{c|}{HCN$_{3-2}$\tnote{d}}&70$ \pm 5 $&$ \sim65 $&110$ \pm 5 $&$ < 19$&39-52& & & \\
\hline
\hline
\end{tabular}
\begin{tablenotes}
\item[a] based on values from \citet{Harris1985}, \citet{Guesten1987}, \citet{Sutton1990} and \citet{Jacks1993} \\
\item[b] HI absorption data from \citet{Liszt1985} \\
\item[c] data from Table 2 of \citet{Marshall1995} \\
\item[d] data from Table 1 of \citet{Jacks1993}
\end{tablenotes}
  
\end{threeparttable}
\end{center}
%\end{table} 
\end{landscape}
 
%\normalsize
\normalsize 
%\afterpage{\clearpage}
%\newpage

%\newpage 
\section{HCN(1-0) Cores} \label{2005cores}
\paragraph*{}
This section describes how the angle of inclination of the disk's axis of rotation to the line of sight is calculated from data listed in Table 2 of \citet{Chris2005}.    

\paragraph*{}
\citet{Chris2005} identified twenty-six HCN (1-0) cores in their Table 2 with the size, spectral central velocity, width and integrated flux measured for each core using the twenty-six letters of the English alphabet to label the cores in their tabulation. Their observations were made between November 1999 and April 2005 using the Owens Valley Radio Observatory (OVRO). Their criteria for a core's inclusion was that it was a bright emission source which was isolated in position and velocity space. Their list of cores is not exhaustive but rather a good representative sample containing the majority of bright sources and a few lower emission sources. The sample was also restricted to cores in the CND, except for cores X and Y which are located in the linear filament (see Fig.\  \ref{CND parts}). \citet{Chris2005} list the core positions by their $\alpha$ and $\delta$ offsets from SgrA* in arcsecs in B1950 epoch co-ordinates. The  plane of the sky (projected) distances, of the cores from SgrA* are given in parsecs (pc), assuming a distance to SgrA* of 8kpc.

\paragraph*{}
The inclination angle, $\alpha$, together with the slope of the disk's major axis to the x axis, $\phi$, is needed to generate: 
\begin{itemize}
\item  the deprojected core offsets from SgrA$^{*}$, and then
\item  a radial velocity curve for comparison with observed core radial velocities 			 using Eqn \ref{RV2}. Such a comparison gives an indication of whether or 			 not a particular core lies in the disk.  
\end{itemize}
Deprojected x$_{D}$ and y$_{D}$ offsets from SgrA$^{*}$ were calculated using Eqns.\ \ref{TM} and \ref{ydisk} with $\phi$ = 65$^{\circ}$, i.e. the complementary angle to the PA, and a range of values for $\alpha$, the inclination angle. These offsets were used to calculate core deprojected distances from SgrA$^{*}$ that were compared with those values published in Table 2 of \citet{Chris2005}. A value of 60$^{0}$ together with a major axis inclination angle, $\phi$ = 65$^{0}$ produced deprojected distances that agreed with the  values published in \citet{Chris2005} to two decimal places. Table \ref{Core props} lists  ID, projected offsets and projected distances from SgrA$^{*}$, deprojected offsets and distances from SgrA$^{*}$, angular position  in the disk $\phi_{D}$, angular position in the sky $\phi_{sky}$, observed and modelled radial velocities and their differences for all twenty-six HCN (1-0) cores and is based on B1950 co-ordinates. The B1950 co-ordinates are converted to J2000 co-ordinates to allow comparison of core positions with those published in other references (see Section \ref{cordconv}).    

\paragraph*{}
\citet{Chris2005} specify the position angle value of the disk's major axis as $\sim$ 25$^{\circ}$, which equates to $\phi$ = 65$^{\circ}$ and is the same as that used by \citet{Jacks1993}. \citet{Chris2005} give a value ranging from 50 to 75 degrees, consistent with the value of 60$^{\circ}$ obtained here. 

\begin{landscape}
\begin{center}
%\tiny
%\scriptsize
\footnotesize
%\small
\begin{threeparttable}[!h]
 \caption[Table of Core Properties]{\label{Core props}B1950 HCN Core Positions Relative to  SgrA$ ^{*} $ and Radial Velocities } 
%\end{center}
%\fontsize{9} {9} 
%\vspace{5mm} 
%\begin{center}
%\begin{tabular}{|d{-1}|d{2}|d{2}|d{2}|d{2}|d{2}|d{2}|d{1}|d{1}|d{-1}|d{-1}|d{-1}|} 
\begin{tabular}{c|d{2}|d{2}|d{2}|d{2}|d{2}|d{2}|d{1}|d{1}|d{-1}|d{-1}|d{-1}} 
%\begin{tabular}[10cm]{|c|c|c|c|c|c|c|c|c|c|c|c|} 
%\hline
\hline
\multicolumn{1}{c|}{Core}&\multicolumn{1}{c|}{$\alpha$\tnote{a}}& \multicolumn{1}{c|}{$\delta$\tnote{a}}&\multicolumn{1}{c|}{Proj Dist\tnote{a}}&\multicolumn{1}{c|}{x$_{D}$\tnote{b}}&\multicolumn{1}{c|}{y$_{D}$\tnote{b}}&\multicolumn{1}{c|}{Deproj\tnote{a}}&\multicolumn{1}{c|}{$\phi_{D}$\tnote{c}}&\multicolumn{1}{c|}{$\phi_{sky}$\tnote{d}}&\multicolumn{1}{c|}{Obs Radial\tnote{a}} &\multicolumn{1}{c|}{Model v$_{z}$\tnote{e}}&\multicolumn{1}{c}{Difference\tnote{f}}  \\ 
%\hline
\multicolumn{1}{c|}{ }&\multicolumn{1}{c|}{pc}&\multicolumn{1}{c|}{pc}&\multicolumn{1}{c|}{pc}&\multicolumn{1}{c|}{pc}&\multicolumn{1}{c|}{pc}&\multicolumn{1}{c|}{Dist pc}&\multicolumn{1}{c|}{Degrees}&\multicolumn{1}{c|}{Degrees}&\multicolumn{1}{c|}{Vel km s$^{-1}$}&\multicolumn{1}{c|}{km s$^{-1}$}&\multicolumn{1}{c}{v$_{z}$ km s$^{-1}$} \\ 
\hline
%\hline 
A&0.36&1.24&1.29&1.28&0.40&1.34&17.4&74.4&107&91&16 \\ %\hline
B&0.42&1.55&1.61&1.58&0.55&1.68&19.2&75.5&139&90&49 \\ %\hline
C&0.93&1.58&1.84&1.83&-0.35&1.86&349.2&60.0&105&94&11 \\ %\hline
D&1.07&1.35&1.72&1.68&-0.80&1.86&334.5&52.0&101&86&15\\ %\hline						
E&0.98&1.04&1.43&1.36&-0.89&1.62&326.8&47.3&78&80&-2 \\ %\hline						
F&1.89&1.01&2.14&1.71&-2.58&3.10&303.5&28.4&72&53&19 \\ %\hline
G&1.94&0.34&1.97&1.13&-3.23&3.42&289.3&10.3&65&31&34 \\ %\hline
H&0.85&-0.03&0.85&0.33&-1.57&1.61&281.9&358.2&-17&20&-37 \\ %\hline
I&0.85&-0.04&0.94&0.01&-1.87&1.81&270.3&335.7&-18&1&-19 \\ %\hline 
J&0.19&-0.98&0.99&-0.81&-1.16&1.42&235.1&281.0&-37&-55&18\\ %\hline 
K&0.09&-1.27&1.28&-1.11&-1.24&1.67&228.2&274.2&-71&-64&-7 \\ %\hline 
L&0.06&-1.29&1.29&-1.14&-1.20&1.66&226.5&272.8&-38&-66&28 \\ %\hline 
M&-0.12&-1.46&1.46&-1.37&-1.01&1.70&216.4&265.4&-64&-77&13 \\ %\hline 
N&-0.64&-1.69&1.81&-1.80&-0.28&1.82&188.8&249.6&-64&-94&30 \\ %\hline 
O&-0.82&-1.33&1.57&-1.56&0.36&1.60&167.0&238.4&-108&-93&-15\\ %\hline 
P&-0.82&-0.96&1.27&-1.22&0.68&1.39&150.9&229.5&-73&-83&10 \\ %\hline
Q&-1.04&-0.78&1.30&-1.14&1.23&1.68&132.8&216.9&-38&-65&27 \\ %\hline 
R&-0.70&-0.30&0.76&-0.563&1.02&1.16&118.9&203.2&-37&-46&9 \\ %\hline 
S&-0.90&-0.23&0.93&-0.59&1.43&1.55&112.4&194.7&89&-36&125 \\ %\hline 
T&-0.93&-0.22&0.96&-0.59&1.50&1.62&111.5&193.3&44&-35&79 \\ %\hline 
U&-0.65&-0.22&0.69&-0.47&1.00&1.10&115.2&198.6&79&-41&120\\ %\hline 
V&-0.42&0.42&0.59&0.20&1.11&1.13&79.8&135.4&58&17&41 \\ %\hline 
W&-0.30&0.90&0.95&0.69&1.29&1.47&61.9&108.7&56&45&11 \\ %\hline 
X&-0.95&1.06&1.42&0.56&2.61&2.67&77.9&132.3&64&20&44\\ %\hline 
Y&-0.73&1.52&1.69&1.07&2.61&2.82&67.7&115.9&78&36&42 \\ %\hline 
Z&-0.14&1.57&1.57&1.36&1.58&2.08&49.3&95.5&58&62&-4 \\ %\hline 
\hline 
\end{tabular} 
\begin{tablenotes}
\item[a] sourced from Table 2 \citet{Chris2005}  
\item[b] values calculated using Eqn \ref{TM} 
\item[c] values calculated using Eqn \ref{phiD} 
\item[d] values based on projected x,y offsets 
\item[e] values based on Eqn \ref{RV2} 
\item[f] Observed minus Model Radial Velocities
\end{tablenotes} 
\end{threeparttable}
\end{center}
\end{landscape}
%\paragraph*{}

\normalsize 
%\afterpage{\clearpage}
%\newpage

\section{Conversion from B1950 to J2000 co-ordinates} \label{cordconv}
\paragraph*{}
The reference point for celestial co-ordinates is adjusted every fifty years to account for the precession of the earth's axis over this period B1950 and J2000 refer to the reference years 1950 and 2000. The J2000 co-ordinate system was chosen as the reference for this system to facilitate comparisons of data from different papers including some that used B1950 co-ordinates.
\paragraph*{}
The conversion from B1950 to J2000 epoch co-ordinates involved three steps 
\begin{enumerate}
\item conversion of $\Delta\alpha$ and $\Delta\delta$ offsets from SgrA$^{*}$ in arcsecs to RA and Dec in the B1950 epoch used 
Eqn.\ \ref{dxconv} for the conversion of the $\alpha$ offset, the $\delta$ offset was simply added to or subtracted from the declination of SgrA$^{*}$ as appropriate.
\item conversion of B1950 to J2000 co-ordinates using the conversion tool at HEASARC ( www.astronomy.csdb.cn/heasarc/docs/tools.html).
\item conversion of J2000 co-ordinates to RA and Dec offsets from SgrA$^{*}$, differences in RA expressed as seconds need to be multiplied by 15 to convert to $\alpha$ offsets in arcsecs.
\end{enumerate} 
These processes are described below in more detail.
%\subsection{Conversion B offsets from SgrA$^*$ to RA and Dec offsets}

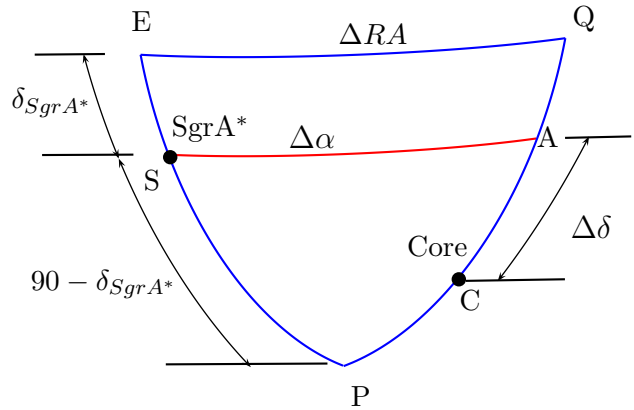
\begin{figure}[ht] 
\begin{center}

\begin{pspicture}(-6,-3) (6,0)
\psset{unit=3.4cm,drawCoor=true,Alpha=58,Beta=175}
\psset{dotstyle=*,dotscale=1.5,drawCoor=false}
\pstThreeDCoor[drawing=false,
    linecolor=black,linestyle=dashed,xMin=0,xMax=1.5,yMin=0,
    yMax=1.0,zMin=0,zMax=1.5,nameX=E,nameY=Q,nameZ= ]
\newcommand\oA{\pstThreeDLine[linecolor=black,
    SphericalCoor=true]
    (1.0,100,24)(1.2,110,20)}
\newcommand\oC{\pstThreeDLine[linecolor=black,
    SphericalCoor=true]
    (1.8,-10,-0.2)(1.55,-5,0)}
\newcommand\oB{\pstThreeDLine[linecolor=black,
    SphericalCoor=true]
    (1.1,100,63)(1.3,115,49)}
\newcommand\oD{\pstThreeDLine[linecolor=black,
    SphericalCoor=true]
    (1.6,0,14)(1.8,-10,12.5)}
\newcommand\oE{\pstThreeDLine[linecolor=black,
    SphericalCoor=true]
    (1.40,25,65)(2.2,10,32.2)}
\pstThreeDEllipse[beginAngle=-1,endAngle=12,linewidth=0.5pt,linecolor=black,arrows=<->]
    (0,0,0)(1.9,0.0,0.0)(0.0,0.55,1.70)
\pstThreeDEllipse[beginAngle=12,endAngle=44,linewidth=0.5pt,linecolor=black,arrows=<->]
    (0,0,0)(1.9,0.0,0.0)(0.0,0.55,1.70)
\pstThreeDEllipse[beginAngle=18,endAngle=43.5,linewidth=0.5pt,linecolor=black,arrows=<->]
   (0,0,0)(0,1.3,0)(0.0,-0.8,1.85)
\pstThreeDEllipse[beginAngle=0,endAngle=90,linecolor=blue]
    (0,0,0)(1.5,0,0)(0,1,0)
\pstThreeDEllipse[beginAngle=1.5,endAngle=90.5,linecolor=red]
    (0,0,0.4)(1.36,0,0)(0,0.86,0)
\pstThreeDEllipse[beginAngle=0,endAngle=70,linecolor=blue]
    (0,0,0)(1.5,0,0)(0,0.3,1.35)
\pstThreeDEllipse[beginAngle=0,endAngle=78.5,linecolor=blue]
    (0,0,0)(0,1,0)(0,-0.3,1.385)
\psset{SphericalCoor=true}
\pstThreeDPut(1.2,37,-3.5){\small $\Delta RA$}
\pstThreeDPut(1.2,5,15){\small SgrA$^{*}$}
\pstThreeDPut(0.6,15,41.5){$\Delta \alpha$}
\pstThreeDPut(1.3,95,35){$\Delta \delta$}
\pstThreeDPut(1.0,70,55){\small Core}
\pstThreeDPut(1.2,85,60){\small C}
\pstThreeDDot(1.08,105,65) 
\pstThreeDDot(1.35,0,18) 
\pstThreeDPut(1.5,0,-5){\small E}
\pstThreeDPut(1.5,70,-5){\small Q}
\pstThreeDPut(1.45,45,80){\small P}
\pstThreeDPut(1.1,-20,30){\small S}
\pstThreeDPut(1.0,90,23){\small A}
\pstThreeDPut(1.9,-5,5){\small $\delta_{SgrA^*}$}
\pstThreeDPut(1.8,-5,30){\small $90-\delta_{SgrA^*}$}
\oA \oB \oC \oD \oE
\end{pspicture}
\vspace*{50pt}
\caption [A celestial spherical triangle for the Southern Sky]{ EQP is a Celestial Spherical Triangle for the Southern Sky. Side EQ represents the celestial equator; the apex of the triangle, P represents the south celestial pole. Side EP is the line of constant RA for SgrA$^{*}$ and QP the line of constant RA for a HCN core. EQ is the difference in RA between SgrA$^{*}$ and the core, $\Delta RA$; SA is the $\Delta\alpha$ offset between SgrA$^{*}$ and the core.  ES is the declination of SgrA$^{*}$, $\delta_{SgrA^{*}}$; QC is the core's declination and AC is the difference between the declination of the core and declination of SgrA$^{*}$, which is the $\Delta\delta$ offset. The relationship between $\Delta\alpha$ and $\Delta RA$ is given by Eqn.\ \ref{dxconv}.}   
\label{xtoRA} 
\end{center}
\end{figure}

Transformation of $\Delta\alpha$ and $\Delta\delta$ offsets to spherical coordinate offsets requires the following relationship which is based on spherical trigonometry's sine rule. This is illustrated in Fig.\ \ref{xtoRA}. 
%\vspace*{25pt}
\begin{equation}
%\begin{align}
\left [\frac{\sin SA}{\sin \angle APS}\right ] = \left [\frac{\sin SP}{\sin \angle SAP }\right ]
\nonumber
%\end{align}
\end{equation}
%\vspace*{25pt}
Substituting for the above terms with values from the spherical triangle gives  
\begin{equation}
%\begin{align}
\left [\frac{\sin \Delta\alpha}{\sin \Delta RA}\right ] = \left [\frac{\sin \left (90-\delta_{SgrA^*} \right )}{\sin 90}\right ] = \cos \delta_{SgrA^*} \, ,
\nonumber
%\end{align}
\end{equation}
and transposing terms results in

%\vspace*{25pt}
\begin{eqnarray}\label{dxconv}
\Delta RA= \frac{1}{15}\arcsin \left[\frac{\sin \Delta \alpha}{\cos \delta_{ SgrA^{*}}} \right] \,. 
\end{eqnarray}

\normalsize
\paragraph*{}
The $\Delta$ RA in seconds calculated using Equation \ref{dxconv} is then added to RA$_{SgrA^{*}}$ when East of SgrA$^{*}$ and subtracted if West, to determine the core's RA . The core's declination is derived by adding the y offset to $\delta_{SgrA^{*}}$ when South of SgrA$^{*}$ and subtracted when North.
\paragraph*{}
%\subsection{Conversion of Core Co-ordinates from B1950 to J2000 epoch}
The core RA and declinations in the B1950 epoch,  were entered into a text file for batch processing by HEARSARC's co-ordinate conversion tool to produce J2000 epoch co-ordinates and listed in columns 6 \& 7 of Table \ref{Core co-ords}. 
\paragraph*{}
Core $\Delta\alpha$ and $\Delta\delta$ offsets in J2000 co-ordinates were calculated by using the inverse of Eqn.\ \ref{dxconv} and listed in columns 8 \& 9 of Table \ref{Core co-ords}.   
\paragraph*{}
Table \ref{Core co-ords} summarises the above co-ordinate conversions for cores detected in HCN(1-0) by \citet{Chris2005}, H$^{12}$CN(1-0), H$^{13}$CN(1-0) and HCO$^{+}$(1-0) from \citet{Marr1993},  HCN(3-2) from \citet{Jacks1993} and HCN(4-3) from \citet{MMC2009},listing core positions in both B1950 and J2000 epochs.
The object labels follow the quoted papers with the exception of \citet{Jacks1993} where the features are identified by their offsets in arcseconds from SgrA$^{*}$. In this case the present author assigned alphabetic names to each core.  

\begin{landscape}
%\small
\scriptsize
%\footnotesize
%\tiny
%\begin{center}
\tablefirsthead{%
\hline
Object & \multicolumn{2}{c|}{B1950 Offsets from SgrA$^{*}$} & \multicolumn{2}{c|}{B1950 Co-ordinates} &\multicolumn{2}{c|}{J2000 Co-ordinates} &\multicolumn{2}{c}{ J2000 Offsets from SgrA$^{*}$} \\ %\hline
 & $\alpha$ arcsecs& $\delta$ arcsecs&RA h m s &Dec $^{0}$ ${'}$ ${''}$  & RA h m s& Dec $^{0}$ ${'}$ ${''}$& $\alpha$ arcsecs& $\delta$ arcsecs \\ \hline}
\tablehead{%
\hline
\multicolumn{9}{c}{\small\slshape continued from last page}\\ \hline
Object & \multicolumn{2}{c|}{B1950 Offsets from SgrA$^{*}$} & \multicolumn{2}{c|}{B1950 Co-ordinates} &\multicolumn{2}{c|}{J2000 Co-ordinates} &\multicolumn{2}{c}{ J2000 Offsets from SgrA$^{*}$} \\ %\hline
& $\alpha$ arcsecs& $\delta$ arcsecs&RA h m s &Dec $^{0}$ ${'}$ ${''}$  & RA h m s& Dec $^{0}$ ${'}$ ${''}$& $\alpha$ arcsecs& $\delta$ arcsecs\\ \hline}
\tabletail{%
\hline
\multicolumn{9}{c}{\small\slshape continued on next page}\\ \hline}
\tablelasttail{\hline}
\topcaption{\label{Core co-ords} Core and Maser Positions in the Circumnuclear  Ring} 
%\topcaption{\label{Core co-ords} } 
\begin{supertabular}{c|c|c|c|c|c|c|c|c}
%\begin{supertabular}{c|.|.|c|c|c|c|.|.}
% & & & & & & & & \\
SgrA$^{*}$&0&0&17 42 29.30&-28 59 46.7&17 45 40.03&-29 28.30&0&0 \\ \hline
% & & & & & & & & \\
& & & & & & & & \\ \hline
\citet{Chris2005} Core A &9.2&32.0&17 42 30.00&$-$28 58 46.7&17 45 40.72&$-$28 59 56.2&8.9&32.1 \\ %\hline
Core B&10.8&40.0&17 42 30.12&$-$28 58 38.7&17 45 40.83&$-$28 59 48.2&10.4&40.1 \\ %\hline
Core C&24.0&40.8&17 42 31.13&-28 58 37.9&17 45 41.84&-28 59 47.3&23.6&41.0 \\ %\hline
Core D&27.6&34.8&17 42 31.40&-28 58 43.9&17 45 42.12&-28 59 53.3&27.3&35.0 \\ %\hline
Core E&25.2&26.8&17 42 31.22&-28 58 51.9&17 45 41.94&-29 00 01.3&25.0&27.0 \\ %\hline
Core F&48.0&26.0&17 42 33.02&-28 58 52.7&17 45 43.74&-29 00 02.0&48.6&26.3 \\ %\hline
Core G&50.0&8.8&17 42 33.11&-28 59 09.9&17 45 43.84&-29 0 19.2&49.9&9.1 \\ 
%\hline
Core H&22.0&-0.8&17 42 30.98&-28 59 19.5&17 45 41.71&-29 00 29.0&21.9&-0.7 \\ %\hline
Core I&22.0&-10.0&17 42 30.98&-28 59 28.7&17 45 41.71&-29 00 38.2&21.9&-9.9 \\ %\hline
Core J&4.8&-25.2&17 42 29.67&-28 59 43.9&17 45 40.41&-29 00 53.5&4.9&-25.2 \\ %\hline
Core K&2.4&-32.8&17 42 29.48&-28 59 51.5&17 45 40.22&-29 01 01.1&2.4&-32.8 \\ %\hline
Core L&1.6&-33.2&17 42 29.42&-28 59 51.9&17 45 40.16&-29 01 01.5&1.6&-33.2 \\ %\hline
Core M&-3.2&-37.6&17 42 29.06&-28 59 56.3&17 45 39.81&-29 01 05.9&-3.0&-37.6 \\ %\hline
Core N&-16.4&-43.6&17 42 28.05&-29 00 02.3&17 45 38.80&-29 01 12.0&-16.2&-43.7 \\ %\hline
Core O&-21.2&-34.4&17 42 27.68&-28 59 53.1&17 45 38.42&-29 01 02.8&-21.2&-34.5 \\ %\hline
Core P&-21.2&-24.8&17 42 27.68&-28 59 43.5&17 45 38.42&-29 00 53.2&-21.2&-24.9 \\ %\hline
Core Q&-26.8&-20.0&17 42 27.26&-28 59 38.7&17 45 38.00&-29 00 48.4&-26.7&-20.1 \\ %\hline
Core R&-18.0&-7.6&17 42 27.93&-28 59 26.3&17 45 38.67&-29 00 36.0&-18.0&-7.7 \\ %\hline
Core S&-23.2&-6.0&17 42 27.53&-28 59 24.7&17 45 38.26&-29 00 34.4&-23.3&-6.1 \\ %\hline
Core T& -24.0&-5.60&17 42 27.47&-28 59 24.3&17 45 38.20&-29 00 34.0&-24.1&-5.7 \\ %\hline
Core U&-16.8&-5.6&17 42 28.02&-28 59 24.3&17 45 38.75&-29 00 34.0&-16.9&-5.7 \\ %\hline
Core V&-10.8&10.8&17 42 28.48&-28 59 10.8&17 45 39.21&-29 00 17.6&-10.9&10.7 \\ %\hline
Core W&-7.6&23.2&17 42 28.72&-28 58 55.5&17 45 39.44&-29 00 05.1&-7.9&23.2 \\ %\hline
Core X&-24.4&27.2&17 42 27.44&-28 58 51.5&17 45 38.16&-28 59 01.2 &-24.6&27.1 \\ %\hline
Core Y&-18.8&39.2&17 42 27.87&-28 58 39.5&17 45 38.59&-28 59 49.2&-19.0&39.1 \\ %\hline
Core Z&-3.6&40.4&17 42 29.03&-28 58 38.3&17 45 39.74&-28 59 47.9&-3.9&40.4 \\ \hline
 & & & & & & & & \\
& & & & & & & & \\ \hline
\citet{Marr1993} Core A&25.3&36.5&17 42 31.23&-28 58 42.2&17 45 41.95&-28 59 51.7&25.1&36.6\\ %\hline
Core B&-4.6&18.9&17 42 28.95&-28 58 59.8&17 45 39.67&-29 00 09.4&-4.8&18.9\\ %\hline
Core C&-5.5&41.7&17 42 28.88&-28 58 37.0&17 45 39.59&-28 59 46.6&-5.9&41.7\\ %\hline
Core D&-21.6&-9.4&17 42 27.65&-28 59 28.1&17 45 38.39&-29 00 37.8&-21.6&-9.5\\ %\hline
Core E&-20.8&-36.5&17 42 27.71&-28 59 55.2&17 45 38.46&-29 01 04.9&-20.7&-36.6\\ \hline
& & & & & & & & \\ \hline
\citet{Jacks1993} Core A&20.0&10.0&17 42 30.82&-28 59 08.7&17 45 41.55&-29 00 18.2&19.8&10.1 \\ %\hline
Core B&30.0&40.0&17 42 31.59&-28 58 38.7&17 45 42.30&-28 59 48.1&29.7&40.2 \\ %\hline
Core C&0.0&40.0&17 42 29.30&-28 58 38.7&17 45 40.01&-28 59 48.3&-0.4&40.0 \\ %\hline
Core D&-10.0&20.0&17 42 28.54&-28 58 58.7&17 45 39.26&-29 00 8.3&-10.2&20.0 \\ %\hline
Core E&-20.0&0.0&17 42 27.78&-28 59 18.7&17 45 38.51&-29 00 28.4&-20.1&-0.1 \\ %\hline
Core F&-20.0&-20.0&17 42 27.78&-28 58 38.7&17 45 38.52&-29 00 48.4&-19.9&-20.1 \\ %\hline
Core G&-20.0&-40.0&17 42 27.78&-28 59 58.7&17 45 38.53&-29 01 08.4&-19.8&-40.1 \\ %\hline
Core H&-10.0&-40.0&17 42 28.54&-28 59 58.7&17 45 39.29&-29 01 08.3&-9.8&-40.0 \\ %\hline
Core I&-30.0&-60.0&17 42 27.01&-29 00 18.7&17 45 37.77&-29 01 28.4&-29.8&-60.1 \\ %\hline
Core J&10.0&-60.0&17 42 30.06&-29 00 18.7&17 45 40.82&-29 01 28.2&10.3&-59.9 \\ %\hline
Core K&10.0&-40.0&17 42 30.06&-28 59 58.7&17 45 40.81&-29 01 08.2&10.1&-39.9 \\ %\hline
Core L&20.0&-10.0&17 42 30.82&-28 59 28.7&17 45 41.56&-29 00 38.2&20.0&-9.9 \\ %\hline
Core M&20.0&0.0&17 42 30.82&-28 59 18.7&17 45 41.55&-29 00 28.2&19.8&0.1 \\ %\hline
Core N&60.0&-20.0&17 42 33.87&-28 59 38.7&17 45 44.61&-29 00 47.9&60.0&-19.6 \\ %\hline
Core O&40.0&-10.0&17 42 32.35&-28 59 28.7&17 45 43.09&-29 00 38.1&40.0&-9.8 \\ %\hline
Core P&50.0&10.0&17 42 33.31&-28 59 08.7&17 45 43.84&-29 00 18.0&49.9&10.3\\ \hline
% & & & & & & & & \\
& & & & & & & & \\ \hline
\citet{MMC2009} Clump A& & & & &17 45 42.13&-28 59 53.4&27.4&26.9 \\ %\hline
Clump C& & & & &17 45 41.09&-29 00 05.8&13.8&22.5 \\ %\hline
Clump D& & & & &17 45 40.91&-29 00 14.2&11.4&14.1 \\ %\hline
Clump E& & & & &17 45 39.81&-28 59 49.8&-3.0&38.5 \\ %\hline
Clump F& & & & &17 45 39.48&-29 00 03.8&-7.4&24.5 \\ %\hline
Clump G& & & & &17 45 38.56&-28 59 50.6&-19.4&37.7 \\ %\hline
Clump H& & & & &17 45 39.17&-29 00 16.6&-11.4&11.7 \\ %\hline
Clump I& & & & &17 45 40.06&-29 00 25.0&0.2&3.3 \\ %\hline
Clump K& & & & &17 45 38.17&-29 00 33.4&-24.7&-5.1 \\ %\hline
Clump N& & & & &17 45 38.41&-29 00 49.8&-21.4&-21.5 \\ %\hline
Clump Q& & & & &17 45 38.41&-29 01 03.0&-21.4&-34.7 \\ %\hline
Clump R& & & & &17 45 38.41&-29 01 11.4&-21.4&-43.1 \\ %\hline
Clump U& & & & &17 45 39.78&-29 01 05.8&-3.4&-37.5 \\ %\hline
Clump W& & & & &17 45 40.15&-29 01 01.8&1.4&-33.5 \\ %\hline
Clump X& & & & &17 45 40.18&-29 00 54.6&15.0&-26.3 \\ %\hline
Clump Z& & & & &17 45 40.73&-29 00 44.6&9.0&-16.3 \\ %\hline
Clump AA& & & & &17 45 41.76&-29 00 37.8&22.6&-9.5 \\ %\hline
Clump BB& & & & &17 45 43.78&-29 00 19.0&49.0&9.3 \\ %\hline
Clump CC& & & & &17 45 43.53&-29 00 06.6&45.8&21.7 \\ %\hline
Clump DD& & & & &17 45 41.61&-29 00 18.2&20.6& 10.1 \\ %\hline
Clump EE& & & & &17 45 42.04&-29 00 01.4&26.2&26.9 \\ \hline
& & & & & & & & \\ \hline
\citet{Sjman2008}& & & & & & & & \\
OH Masers& & & & & & & &  \\
359.925-0.044& & &17 42 26.21&-29 00 11.2&17 45 36.96&-29 01 20.9&-40.3&-52.6 \\
359.926-0.045& & &17 42 26.53&-29 00 12.7&17 45 37.28&-29 01 22.4&-36.1&-54.1 \\
359.929-0.048& & &17 42 27.70&-29 00 07.7&17 45 38.45&-29 01 17.4&-20.7&-49.1 \\ 
359.930-0.048& & &17 42 28.02&-29 00 05.8&17 45 38.76&-29 01 15.5&-16.7&-47.2 \\
359.952-0.035& & &17 42 28.70&-28 58 33.4&17 45 38.76&-28 59 43.0&-16.1& 45.3  \\ 
359.955-0.040& & &17 42 29.67&-28 58 32.2&17 45 40.37&-28 59 41.7&  4.5& 46.6  \\
359.960-0.037& & &17 42 29.71&-28 58 11.6&17 45 40.40&-28 59 21.1&  4.9& 67.2  \\
359.955-0.041& & &17 42 29.91&-28 58 34.4&17 45 40.62&-28 59 44.0&  7.7& 44.3  \\
& & & & & & & & \\ \hline
\citet{FYZ2008}& & & & & & & & \\
Water Masers& & & & & & & & \\
SgrA-CND-NE& & & & &17 45 42.0&-28 59 48.0& 25.8& 40.3 \\
SgrA-CND-SW2& & & & &17 45 38.4&-29 00 48.0&-21.4&-19.8 \\
SgrA-CND-NN& & & & &17 45 39.7&-28 59 15&-4.33& 73.3 \\
SgrA-CND-EE& & & & &17 45 43.9&-29 00 03.0& 50.8& 25.3 \\
SgrA-CND-N& & & & &17 45 40.0&-29 00 00.0&-0.4& 28.3 \\
SgrA-CND-NE2& & & & &17 45 41.8&-29 00 08.0& 23.2& 20.3 \\
& & & & & & & & \\ \hline
Methanol Masers& & & & & & & \\
SgrA-CND-NW& & & & &17 45 39.3&-29 00 16.0&-9.6& 12.3 \\
SgrA-CND-EE& & & & &17 45 43.9&-29 00 03.0& 50.8& 25.3 \\
SgrA-CND-NW& & & & &17 45 43.6&-29 00 28.0& 46.8& 0.3 \\ 
SgrA-CND-NW C& & & & &17 45 44.0&-29 00 30.0& 52.1&-1.7 \\
SgrA-CND-NW N& & & & &17 45 44.0&-29 00 10.0& 52.1& 18.3 \\
SgrA-CND-NW S& & & & &17 45 44.0&-29 00 20.0& 52.1& 8.3 \\
SgrA-CND-NW E& & & & &17 45 45.1&-29 00 30.0& 66.5&-1.7 \\
SgrA-CND-NW W& & & & &17 45 43.0&-29 00 30.0& 39.0&-1.7 \\
SgrA-CND-NW NW& & & & &17 45 44.0&-29 00 00.0& 52.1& 28.3 \\
SgrA-CND-NW SS& & & & &17 45 44.0&-29 00 40.0& 52.1&-11.7 \\ \hline
\end{supertabular}
%\end{center} 
\end{landscape}
\clearpage
\normalsize

%\clearpage
\section{Water, Methanol and OH Masers} \label{secmasers}
\paragraph*{}
This section collates the positions of water (22 GHz) and methanol (44 GHz) masers in and around the CND observed using the Green Bank telescope \citep{FYZ2008} and 
OH (1720 MHz) masers from papers by \citet{Karlsson2003,FYZ1999} based on VLA observations from 1986 to 2005 and collated with new observations by \citep{Sjman2008}.

\begin{figure}[ht] \includegraphics[scale=0.8,bb=  5 125 595 730] 
{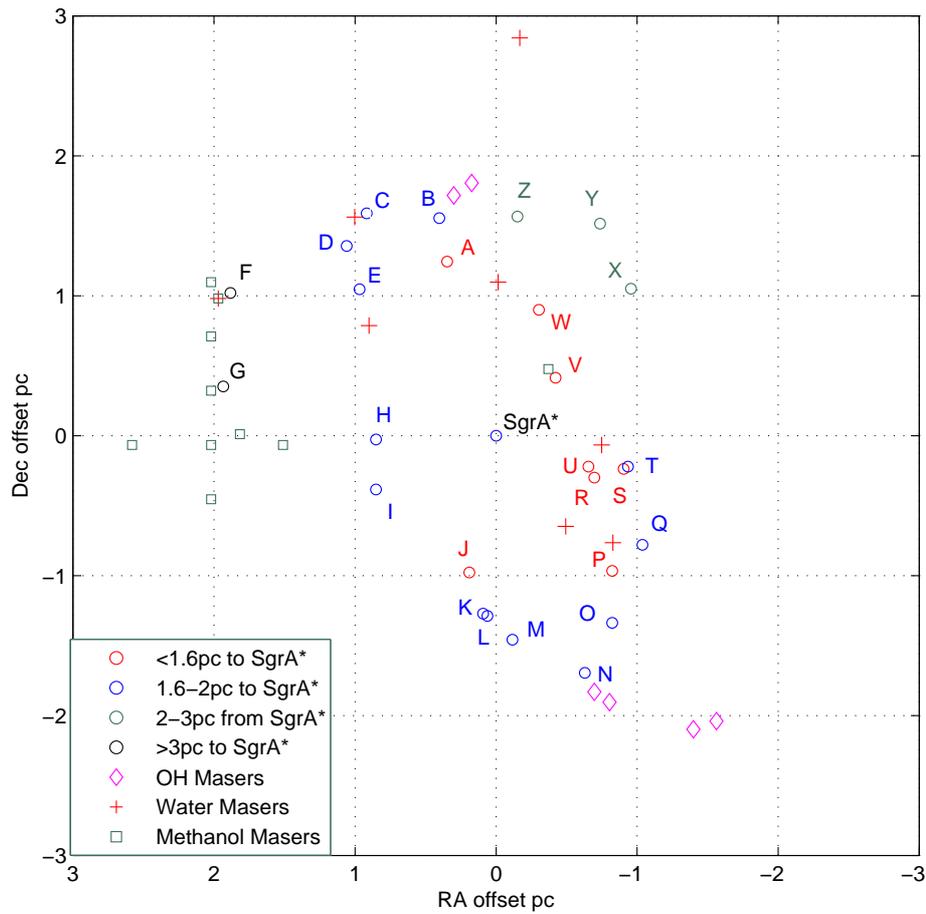} \fontsize{9} {9} \caption[Plot of HCN Cores and Masers]{\label{masers}  CND HCN(1-0) cores, Water, Methanol and OH masers are shown as offsets from SgrA$^{*}$ in parsecs J2000 epoch. Positive RA offsets are East of SgrA$^{*}$ and positive Dec offsets are North of SgrA$^{*}$.} 
\end{figure}

\paragraph*{}
The maser positions based on co-ordinates from \citet{Sjman2008} and Yusef-Zadeh (private communication) were converted, where needed, to J2000 offsets from SgrA$^{*}$ and are listed in Table \ref{Core co-ords}. The maser and HCN (1-0) core positions, based on J2000 co-ordinates, are shown in Fig.\ \ref{masers}.

\begin{figure}[!ht] \includegraphics[scale=0.85,bb= 5 125 595 730]
{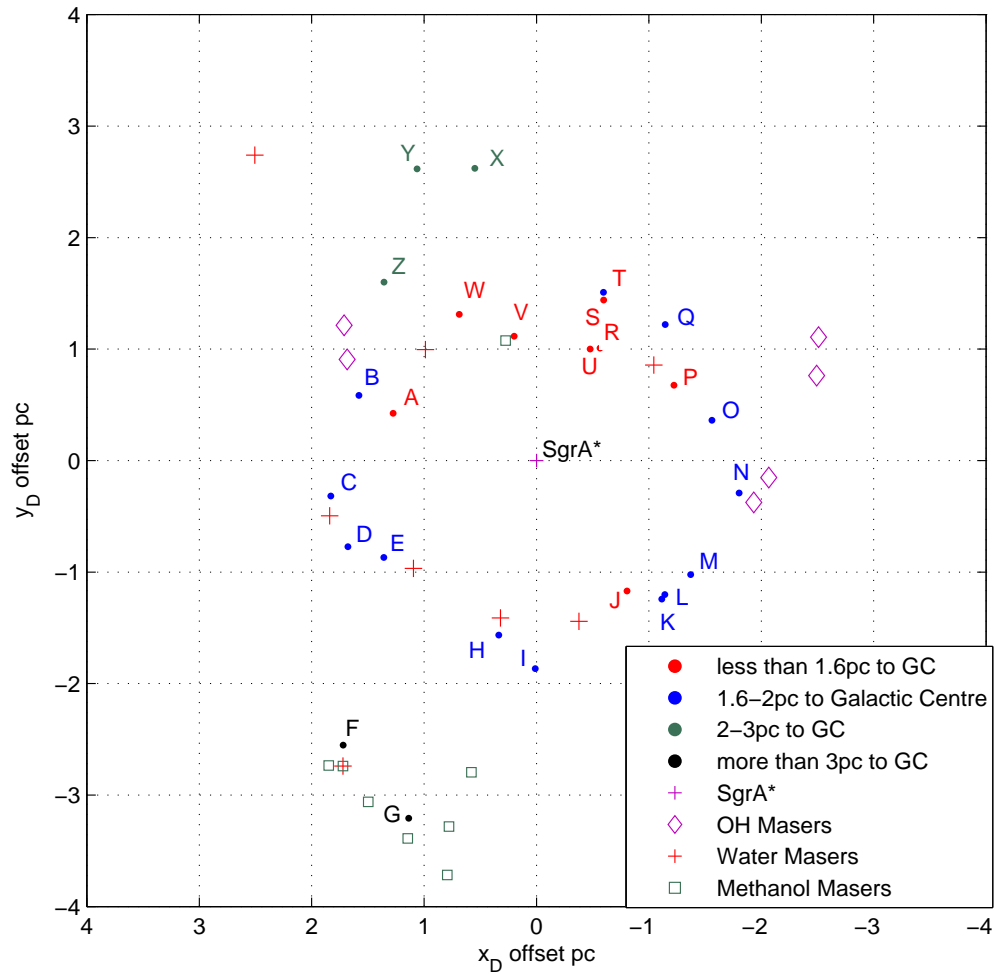}  
 \fontsize{9} {9} \caption[Plot of deprojected CND cores]{\label{Deproj cores}Deprojected HCN(1-0) core locations in the plane of the circumnuclear disk (see Section \ref{cordtran}). The co-ordinates are in the J2000 epoch. Masers were only plotted  where the deprojected distances from SgrA$^{*}$ were less than 4pc and their observed radial velocities were compatible with CND radial velocities.} 
\end{figure}

\begin{landscape}
\begin{figure}[ht] \includegraphics[scale=0.75,bb= -86 179 682 665,clip=true,trim= 30 0 30 0] 
{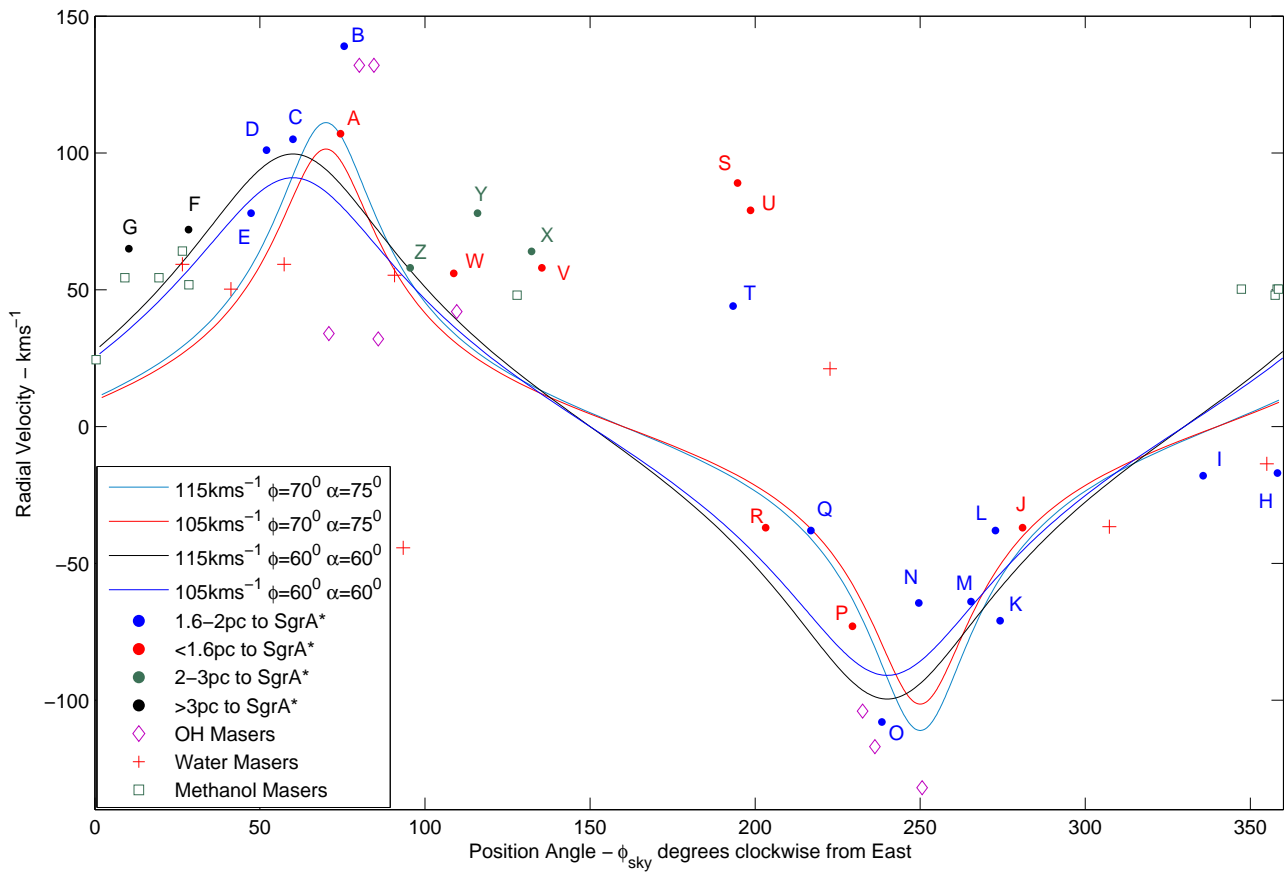} \fontsize{9} {9} \caption[Plot of HCN Core and Maser Radial Velocities]{\label{radvmaser} 
Plots of model radial velocities generated by varying the disk's major axis angle to the sky's horizon, $\phi$, and its inclination $\alpha$. As listed in the figure's legend, rotational velocities of 105 and 115 km s$^{-1}$ were adopted as the upper and lower limits bracketing 110 km s$^{-1}$ see Table \ref{Ring pars} . The effect of increasing the major axis orientation is to shift the curve horizontally to the left. Increasing the tilt angle increases the peak value of the radial velocity. The observed radial velocities of HCN(1-0)cores, Water, Methanol and OH masers are also plotted.} 
\end{figure} 
\end{landscape}

\paragraph*{•}
The methanol masers identified by \citet{FYZ2008} near Cores F, G and V are marked by green squares on Figs.\ \ref{masers}, \ref{Deproj cores} and \ref{radvmaser}. Both the projected and deprojected plots confirm the masers are in the vicinity of their respective cores and their observed radial velocities are within $\leq$ 10km s$^{-1}$ from their respective cores. The masers near Cores F and G are part of a group of eight methanol masers that are on the eastern side of the CND, while Core V is close to the inner western edge of the CND and is a site of shocked H$_{2}$ emission. A redshifted wing of HCN emission from Core V in the direction of the methanol maser could be the signature of a classic one sided outflow as occurs in star forming regions \citep{Mehringer1997}. \citet{FYZ2008} propose that the Class I Methanol masers close to the three HCN cores are evidence of a protostar about 10$^{4}$ years after gravitational collapse.  
Water masers (red crosses) were found close to cores B-Z, D, E, F, O and W. 

\normalsize

\paragraph*{•}
\citet{Sjman2008} identified OH masers near Cores B and N (cyan diamonds in Figs.\ \ref{masers}, \ref{Deproj cores} and \ref{radvmaser}). The two masers in the NW lobe have highly positive radial velocities of +132km s$^{-1}$ which closely match the observed radial velocity of +139kms$^{-1}$ for Core B. The masers near Core N have highly negative radial velocities, --141 and --132km s$^{-1}$, compared with --64km s$^{-1}$ for Core N so appear unrelated to this core while associated with the two masers in the NW lobe. Two other OH masers with radial velocities of --104 and --117km s$^{-1}$ lie about a parsec outside Core O which has a radial velocity of +108 km s$^{-1}$ and could be in the same rotating streamer. \citet{Sjman2008} argue that the high velocities together with their symmetry of positive and negative values indicate that these masers are rotating in the CND structure. They contended that the source of excitation is collisions of CND cores and is not related to the supernova shell of SgrA East. Clumps of OH masers SE  of the CND are indications of interaction between the expanding supernova shell and the +50km s$^{-1}$ molecular gas cloud. %Masers outside Cores F \& G make up a group of NE SNR masers while masers outside Cores B, Z \& Y form the NW group of SNR masers.

%\afterpage{\clearpage}
%\clearpage
%\normalsize

\section{HCN Core and Maser Radial Velocities} \label{coreselect}
\paragraph*{}
Fig.\ \ref{radvmaser} shows model radial velocity curves based on combinations of  $\phi$ = 60$^{\circ}$ and 70$^{\circ}$ clockwise from East and inclination angles, $\alpha$ = 60$^{\circ}$ and 75$^{\circ}$ to produce an envelope of curves that constrain the range of model radial velocity values. The core positions and observed radial velocities are superimposed for comparison with the model and an assessment made as to the likelihood of particular cores being located in the disk. 

\paragraph*{}
Both Table \ref{Core props} and Fig.\ \ref{Deproj cores} indicate only five cores (F, G, X, Y and Z) lie outside a distance of 2pc from SgrA$^{*}$ and eight cores (A, J, P, K, S, U, V and W) lie within 1.6pc from SgrA$^{*}$. Six of the cores, (A, C, D, E, F and Z), are located in the NE section of the ring and nine cores, (J, K, L, M, N, O, P, Q and R) are in the SW section. This indicates that most of the detected HCN cores are located in the inner section of the CND(i.e.\ within 2pc of SgrA$^{*}$). 
 
\paragraph*{}
Fig.\ \ref{radvmaser} shows that the radial velocities of eight cores (B, H, S, T, U, V, X and Y) have large discrepancies, $>$35km s$^{-1}$, between observed and modelled radial velocities and do not appear to fit the model of a group of cores rotating about the galactic centre in circular orbits as part of the CND. All these cores, except U and V, lie between $\sim$ 1.6 and 3.5pc from SgrA$^{*}$. Five cores (H, S, T, U and V) are located within a deprojected distance of $\sim$ 1.6pc of SgrA$^{*}$ and may be influenced by the movement of ionised gas in the western arm of the mini-spiral which has positive radial velocities at these positions see \citet{Zhao2009} in contrast to the observed mainly  negative velocities of these cores by \citet{Chris2005}. Cores X and Y are located in the linear filament that is located adjacent to the NW section of the CND (see Fig\ \ref{CND parts}). Cores F and G are two outlier cores at deprojected distances more than 3pc east of the galactic centre and some 2pc inside the NE group of methanol  masers reported by \citet{Sjman2008} as marking the shock front of the SgrA East supernova remnant (SNR) shell. Core B is located in the Northern Lobe close to the ring's northern gap. The Northern Arm of the mini spiral is in the vicinity about 0.2pc to the west and 0.3pc to the South. Elements N1 and N2 of this feature have observed radial velocities of +78 and +100 km s$^{-1}$ respectively (see Table 3 of \citet{Zhao2009}) compared to the observed +139 km s$^{-1}$ for Core B. Assuming Core B and the two methanol masers are part of the CND requires that they be located in a CND streamer circulating at a much higher rotational velocity ($\sim$ 150 km s$^{-1}$) than the average 110 km s$^{-1}$. 

%\paragraph*{•}

%\afterpage{\clearpage}

\section{Selection of HCN Cores for Analysis}
\paragraph*{}
The analysis using the LVG model relies on finding cores that have been observed in multiple transitions of HCN so that intensities for the three transitions can be matched to infer HCN column densities and hydrogen number densities. A literature review led to a choice of two groups of cores that had been observed in multiple transitions of HCN and that were physically and kinematically related. The first group of five cores were collated by \citet{Marr1993} who observed cores in H$^{13}$CN(1-0) and HCO$^{+}$(1-0) and convolved H$^{12}$CN data \citep{Guesten1987} and HCN(3-2) \citep{Jacks1993} data with their observations. All the data was produced from unresolved images but had a consistent set of related intensities, spatial and kinematic properties which were modelled by \citet{Marr1993}. This writer's modelling is in effect a re-evaluation of the earlier analysis. The second group of seven cores were identified by the writer who selected data from three separate papers that described observations in three HCN transitions and HCO$^{+}$(1-0). Core positions were established as described in Section \ref{cordconv} and kinematic properties established from data in the papers. Although four cores (A, B, C and E) from the first group are common to cores (D, W, Z and O) of the second  group it has been decided to analyse these cores in their separate groups to provide a comparison between the results from unresolved data with results largely derived from resolved data.

\paragraph*{}
The second group of cores were labelled independently by each author and located using a mixture of co-ordinates for the different HCN transitions (viz. B1950 for the (1-0) and (3-2) transitions and J2000 for the (4-3) transition). The cores observed in multiple transitions were identified by comparing locations using J2000 offsets from SgrA$^{*}$ initially given in arcseconds (see Table \ref{cordconv}) and subsequently converted to parsecs by dividing by 25.8 (based on a distance of 8 kpc to SgrA$^{*}$).

\paragraph*{•}
The variety of telescopes used to observe the three transitions of HCN cores in the CND are listed in Table \ref{scopes}. The (1-0) and (4-3) observations are displayed with resolved, while the (3-2) observations are shown with unresolved maps.

\begin{center}
\footnotesize
%\small
\begin{threeparttable}[ht]
\caption{\label{scopes} HCN Observations}
\begin{tabular}{c|c|c|c|c|c} 
\hline
\hline
\multicolumn{1}{c|}{HCN}&\multicolumn{1}{c|}{Paper}&\multicolumn{1}{c|}{Telescope}&\multicolumn{1}{c|}{Type}&\multicolumn{2}{c}{Beam Size}\tnote{a} \\
 & & & &\multicolumn{1}{c|}{$\Delta$x}&\multicolumn{1}{c}{$\Delta$y} \\
\hline
\multicolumn{1}{c|}{(1-0)}&\multicolumn{1}{c|}{\citet{Chris2005}}&\multicolumn{1}{c|}{Ovens Valley Radio Obs}&\multicolumn{1}{c|}{Interferometer}&\multicolumn{1}{c|}{$5.1''$}&\multicolumn{1}{c}{$2.7''$} \\
\multicolumn{1}{c|}{(3-2)}&\multicolumn{1}{c|}{\citet{Jacks1993}}&\multicolumn{1}{c|}{IRAM 30m}&\multicolumn{1}{c|}{Single Dish}&\multicolumn{1}{c|}{$12''$}&\multicolumn{1}{c}{$12''$} \\
\multicolumn{1}{c|}{(4-3)}&\multicolumn{1}{c|}{\citet{MMC2009}}&\multicolumn{1}{c|}{ Sub mm Array}&\multicolumn{1}{c|}{Interferometer}&\multicolumn{1}{c|}{$4.6''$}&\multicolumn{1}{c}{$3.0''$} \\
\hline
\hline
\end{tabular}
\begin{tablenotes}
\item[a]  Beam area is calculated as $\pi\Delta$x$\Delta$y/(4Ln2)

\end{tablenotes}
\end{threeparttable}
\end{center}
\normalsize
\paragraph*{•}
Identifying correspoding cores in the second group relies on the proximity of their spatial co-ordinates and their central spectral or radial velocities. Four of the seven objects, i.e.\ Cores D, M, W and Z have the strongest velocity space correlation and are undoubtedly cores observed in multiple HCN transitions. Cores I and O have anomalous central velocities for the (3-2) transition. These observations have been included on the basis that the spectra are from unresolved data, which can leave greater room for discrepancies given that they are visual estimates from the \citet{Jacks1993} figures. Core P has a lower central velocity in the HCN(4-3) transition, but has been included on the basis that it was one of the cores matched by \citet{MMC2009} with the HCN(1-0) observations by \citet{Chris2005}. Fig.\ 7 in \citet{MMC2009} shows the (1-0) and (4-3) spectra with double peaks and absorption occurs between the peaks in the (1-0) spectrum which combining the effects can explain the discrepancies (see Table\ref{Vels}). 

\paragraph*{•}
It should be noted that central velocities were only quantified by \citet{Chris2005} in Table 2 of their paper for the (1-0) transition. Central velocities for the (3-2) and (4-3) transitions required visual estimates from the spectra provided in the relevant papers. Line widths were specified for the (1-0) and (4-3) transitions and had to be estimated for the (3-2) transition. The (3-2) transition spectral widths are large, due in some measure to the larger beam size.
 
\begin{center}
\small
\begin{threeparttable}[ht]
\caption{\label{Vels} Selected 2nd Core Group Spectral Properties} 
\begin{tabular}{c|c|c|c|c|c|c|c|c}
\hline
%\hline
\multicolumn{3}{c|}{HCN Core}&\multicolumn{3}{c|}{Central Velocity}&\multicolumn{3}{c}{Spectral Width} \\
\multicolumn{3}{c|}{ }&\multicolumn{3}{c|}{km s$^{-1}$}&\multicolumn{3}{c}{km s$^{-1}$} \\
\multicolumn{1}{c|}{(1-0)\tnote{a}}&\multicolumn{1}{c|}{(3-2)\tnote{b}}&\multicolumn{1}{c|}{(4-3)\tnote{c}} &\multicolumn{1}{c|}{(1-0)}&\multicolumn{1}{c|}{(3-2)\tnote{a}}&\multicolumn{1}{c|}{(4-3)\tnote{d}}&\multicolumn{1}{c|}{(1-0)}&\multicolumn{1}{c|}{(3-2)\tnote{d}}&\multicolumn{1}{c}{(4-3)} \\
\hline
{D}&{B}&{A}&101&100&110&45.5&80.0&38.5 \\
{I}&{L}&{AA}&-18&-50&-25&15.0&80.0&49.5 \\
{M}&{H}&{U}&-64&-50&-50&19.0&75.0&47.0 \\
{O}&{G}&{Q}&-108&-70&-90&36.5&90.0&51.0 \\
{P}&{F}&{N}&-73&-75&-40&28.2&75.0&97.0 \\
{W}&{D}&{F}&56&45&50&27.9&45.0&40.0 \\ 
{Z}&{C}&{E}&58&50&40&39.6&50.0&55.0 \\
\hline
%\hline
\end{tabular}
\begin{tablenotes}
\item[a] core IDs from \citet{Chris2005}
\item[b] core IDs assigned by this author by labelling spectra alphabetically from the top left in Fig.\ 1 of \citet{Jacks1993} 
\item[c] core IDs from \citet{MMC2009}
\item[d] Visual estimates 
\end{tablenotes}
\end{threeparttable}
\end{center}
\normalsize
%\afterpage{\clearpage}
%\newpage 

\begin{figure}[ht] \includegraphics[scale=0.80,bb=  20 170 500 670] %,clip=true,trim= 104 0 104 0]
{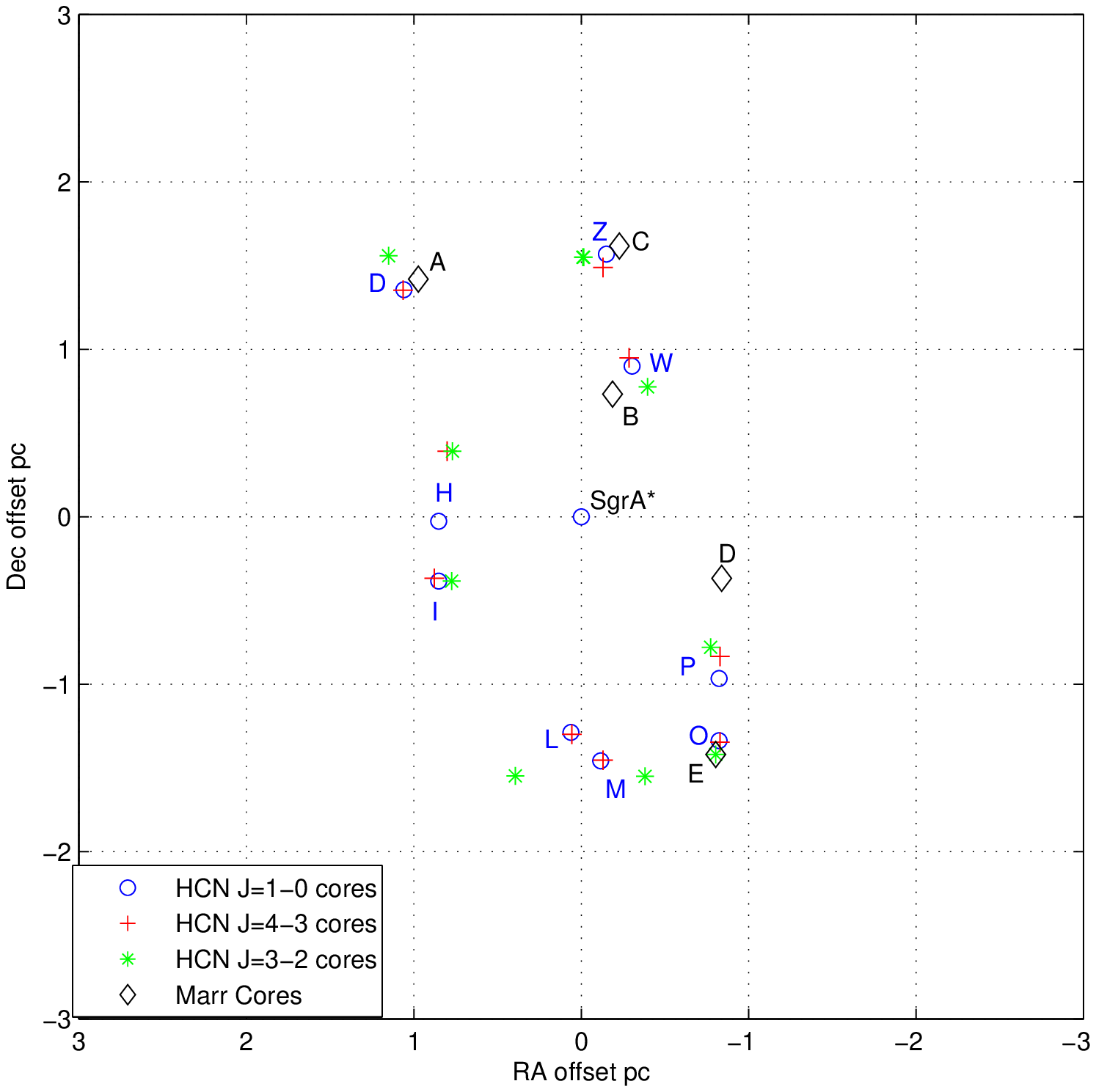} \fontsize{9} {9} \caption[Plot of HCN Cores Selected for LVG Modelling]{\label{Cocores}  The figure shows positions of the two HCN core groups selected for LVG modelling. Locations of the first group are marked with black diamonds and have black labels useed by \citet{Marr1993}. The location of the second core group transitions are marked with their respective symbols as shown in the legend and have blue labels used by \citet{Chris2005}. Offsets from SgrA$^*$ are in parsecs based on the J2000 epoch.} 
\end{figure}

\paragraph*{} 
Table \ref{Coincores} and Fig.\ \ref{Cocores} show nine cores with the observations in three HCN transitions located in reasonable proximity of one another together with the offsets from SgrA$^{*}$ in parsecs for the transition observations. Seven of these nine cores (D, I, M, O, P, W and Z) have their different transition observations within $\lesssim$ 7 arcseconds and generally within 2.6 arcsecs of their mean location and can be regarded as the same core especially given that the uncertainty of the positions of the HCN(3-2) cores are $\pm 0.25$pc or 6.5 arcsecs due to the beam size of the telescope. Cores H and L have the (1-0) and (4-3) transitions in close proximity but the (3-2) transition is too distant to be considered from the same core. %The remaining seven HCN core groups will be studied as the second group of cores in Chapter \ref{anal} where hydrogen densities and opacities from modelling HCN molecular rotational collisions with hydrogen will be compared with observations from the four papers cited in this chapter. 
For convenience the cores chosen for analysis shall be referred to by their \citet{Chris2005} labels. 
\clearpage

%\begin{landscape} 
\begin{center} 
\begin{threeparttable} [!h]
%\small
\scriptsize

\caption{\label{Coincores} Positions of Cores Identified in Multiple Transitions of HCN } 
\begin{tabular}{c|c|c|c|c|c|c|d{2}|d{2}}
\hline
\hline
\multicolumn{1}{c|}{Core group}&\multicolumn{1}{c|}{Core ID for}&\multicolumn{1}{c|}{$\Delta$ RA}&\multicolumn{1}{c|}{$\Delta$ Dec}&\multicolumn{1}{c|}{Central Vel\tnote{a}}&\multicolumn{2}{c|}{Mean Offset Position}&\multicolumn{2}{c}{Core Posn rel to} \\ \hline
\multicolumn{1}{c|}{ID}&\multicolumn{1}{c|}{Transition}&\multicolumn{1}{c|}{Offset}&\multicolumn{1}{c|}{Offset}&\multicolumn{1}{c|}{VLSR}&\multicolumn{1}{c|}{$\Delta$RA}&\multicolumn{1}{c|}{$\Delta$Dec}&\multicolumn{2}{c}{Mean Position} \\ \hline
 & &\multicolumn{1}{c|}{pc}&\multicolumn{1}{c|}{pc}&\multicolumn{1}{c|}{km s$^{-1}$}&\multicolumn{1}{c|}{pc}&\multicolumn{1}{c|}{pc}&\multicolumn{1}{c|}{pc}&\multicolumn{1}{c}{pc}  \\ \hline
 &{D(1-0)}&1.06&1.36&101& & &-0.03&0.04 \\
{D\tnote{b}}&{B(3-2)}&1.15&1.56&100&1.09&1.32&0.06&0.24 \\
 &{A(4-3)}&1.06&1.04&100& & &0.08&-0.07 \\ \hline
 &{H(1-0)}&0.85&-0.03&-17& & &0.05&-0.28 \\
{H\tnote{f}}&{A(3-2)}&0.75&0.39&0&0.80&0.25&-0.05&0.14 \\
 &{DD(4-3)}&0.80&0.39&-5& & &0.00&0.14 \\ \hline
 &{I(1-0)}&0.85&-0.38&-18& & &0.02&0.0 \\
{I}&{L(3-2}&0.77&-0.38&-50&0.83&-0.38&-0.06&0.0 \\
 &{AA(4-3)}&0.88&-0.37&-25& & &0.03&0.01 \\ \hline
 &{L(1-0)}&0.06&-1.29&-38& & &-0.11&0.09 \\
{L\tnote{f}}&{K(3-2)}&0.39&-1.55&-50&0.17&-1.38&0.22&-0.17 \\
 &{W(4-3)}&0.06&0.40&-50& & &-0.11&0.08 \\ \hline
 &{M(1-0)}&-0.12&-1.48&-64& & &0.09&0.01 \\
{M}&{H(3-2)}&-0.38&1.55&-50&-0.21&-1.49&-0.17&-0.06 \\
 &{U(4-3)}&-0.13&-1.45&-50& & &0.08&0.04 \\ \hline
 &{O(1-0)}&-0.82&-1.24&-108& & &0.01&0.07 \\
{O\tnote{c}}&{G(3-2)}&-0.77&-1.55&-70&-0.81&-1.41&0.04&-0.14 \\
 &{Q(4-3)}&-0.83&-1.34&-90& & &-0.02&0.06 \\ \hline
 &{P(1-0)}&-0.82&-0.97&-73& & &-0.01&0.07 \\
{P}&{F(3-2)}&-0.77&0.78&--75&-0.81&-0.86&0.04&0.08 \\
 &{N(4-3)}&-0.83&-0.83&-40& & &-0.11&0.06 \\ \hline
 &{W(1-0)}&-0.30&0.90&56& & &0.03&0.02 \\
{W\tnote{d}}&{D(3-2)}&-0.40&0.78&45&-0.33&0.88&-0.07&-0.10 \\
 &{F(4-3)}&-0.29&0.95&50& & &0.04&0.07 \\ \hline
 &{Z(1-0)}&-0.15&1.57&58& & &-0.05&0.03 \\
{Z\tnote{e}}&{C(3-2)}&-0.01&1.55&50&-0.10&1.54&0.09&0.01 \\
 &{E(4-3)}&-0.11&1.49&40& & &-0.01&-0.05 \\ %\hline
\hline
\hline
\end{tabular}
\begin{tablenotes}
%\\
\item[a] Central velocity for (1-0) cores from Table 2 \citet{Chris2005} velocities for (3-2) and (4-3) estimated visually by the thesis writer from spectra 
\item[b]  also associated with Core A \citet{Marr1993} %see Fig\ref{Cocores}
\item[c]  also associated with Core E \citet{Marr1993} %see Fig\ref{Cocores}
\item[d]  also associated with Core B \citet{Marr1993} %see Fig\ref{Cocores}
\item[e]  also associated with Core C \citet{Marr1993} %see Fig\ref{Cocores}
\item[f]  position of (3-2) transition too distant from (1-0) and (4-3) transitions not selected for group two
\end{tablenotes} 

\end{threeparttable}
\end{center}
 
%\end{landscape}
\normalsize    

\section{Summary}
\paragraph*{•}
A comparison of core radial velocities with model velocities corresponding to their deprojected positions in the CND showed that eighteen of the twenty-six cores have radial velocities consistent with being part of a disk rotating at $\sim$ 110 km s$^{-1}$. The spread of core radial velocities, when compared to the envelope of model velocity curves (see Fig.\ \ref{radvmaser}) is consistent with a disk composed of a series of rotating warped rings or streamers of gas  (see Fig.\ \ref {CND parts} and \citet{Genzel1989}). The methanol masers in the vicinity of Cores F, G and V have radial velocities consistent with the velocities of their neighbouring cores. Core B together with the OH masers in its vicinity form part of  a higher rotational velocity streamer ($\sim$ 150 km s$^{-1}$) than the mean of 110 km s$^{-1}$ 

\paragraph*{•}
Fig.\ \ref{Deproj cores} showed the cores distributed in a circular pattern about SgrA$^{*}$ with an inner cavity of about 1.6pc. 

\paragraph*{•}
Two groups of cores have been selected for analysis based on having published data in three HCN transitions, the first group of five cores (A, B, C, D and E) have been taken from \citet{Marr1993}, the second group of seven cores are Cores D, I, M, O, P, W and Z with their positions and central velocities listed in Table \ref{Coincores}.

\paragraph*{•}
The next chapter covers the analysis of the core groups and the conclusions that can be drawn from the results.

\cleardoublepage

\chapter{\label{anal} Analysis of HCN Cores}

\section{Introduction}
In this chapter the Large Velocity Gradient model, described in Chapter \ref{Molex}, is used to simultaneously fit HCN and HCO$^{+}$ line strengths in order to infer hydrogen number densities, optical depths and column densities of HCN and HCO$^{+}$ in the CND cores.
\paragraph*{•}
Data for two separate HCN core groups selected from Chapter \ref{diskgeom} are analysed:
\begin{enumerate}
\item Unresolved observations of the five cores (A to E) reported in \citet{Marr1993} that were observed in the (1-0) transitions of H$^{12}$CN, H$^{13}$CN, HCO$^{+}$ and the (3-2) transition of HCN from \citet{Jacks1993}. These results are given in Section \ref{Marr}
\item  Seven cores (D, I, M, O, P, W and Z) identified  as having observations in the (1-0), (3-2) and (4-3) transitions of HCN and the (1-0) transition of HCO$^{+}$ reported in \citet{Chris2005}, \citet{Jacks1993} and \citet{MMC2009}. These results are given in Section \ref{Chris}.
\end{enumerate}
Modelling was performed using the parameters reported in the papers and peak brightness temperature contours were plotted for values derived from the integrated intensity maps contained in the papers.
\paragraph*{•}
The modelling implies that the HCN(1-0), H$^{13}$CN(1-0) and HCO$^{+}$(1-0) lines are optically thin and weakly inverted. This is contrary to the findings of both \citet{Marr1993} and \citet{Chris2005} who argued that the HCN (1-0) was optically thick. Reasons for the different outcomes are discussed in Section \ref{discuss}. 
\paragraph*{•}
Results are presnted and comments made separately for each group before the Chapter closes with general discussion of both groups.  
\section{Group One}
\subsection{Input Data from \citet{Marr1993}}
\paragraph*{•}
Peak brightness temperatures and FWHM velocities for Marr cores (A to E) identified by \citet{Marr1993} are reproduced here in Table \ref{Input4}. Core F was excluded from analysis by \citet{Marr1993} as its spectrum was considered to be affected by foreground absorption. The table summarises their observations of H$^{13}$CN and HCO$^{+}$ as well as data for H$^{12}$CN from \citet{Guesten1987} and data for HCN (3-2) from \citet{Jacks1993} both remapped to scale with the \citet{Marr1993} H$^{13}$CN and HCO$^{+}$ data. 

%\begin{table}[!h] 
\begin{center} 
\begin{threeparttable}[!h]

%\end{Huge}

\caption{\label{Input4} Input Data from \citet{Marr1993}}
\vspace{0.5cm}
\scriptsize

\begin{tabular}{c|d{-1}|c|c|c|c|c} \hline
\hline
\multicolumn{1}{c|}{ }&\multicolumn{1}{c|}{Central Vel}&\multicolumn{1}{c|}{FWHM $\Delta$ V}&\multicolumn{4}{c}{Peak Brightness Temperature K} \\
%\hline
\multicolumn{1}{c|}{ }&\multicolumn{1}{c|}{km s$^{-1}$}&\multicolumn{1}{c|}{km s$^{-1}$}&\multicolumn{1}{c|}{H$^{12}$CN\tnote{a}}&\multicolumn{1}{c|}{HCO$^{+}$}&\multicolumn{1}{c|}{H$^{13}$CN}&\multicolumn{1}{c}{HCN (3-2)\tnote{b}} \\
\hline
\multicolumn{1}{c|}{Core A}&102&50&3.6&3.8&0.8&4.4 \\
%\hline
\multicolumn{1}{c|}{Core B}&65&60&4.8&3.4&1.3& \\ 
%\hline
\multicolumn{1}{c|}{Core C}&55&60&2.9&3.0&1.3&2.5 \\
%\hline
\multicolumn{1}{c|}{Core D}&85&75&3.0&{$\leq 1.0$}&0.7&2.1 \\
%\hline
\multicolumn{1}{c|}{Core E}&-96&88&4.4&2.2&0.7&5.0 \\
\hline
\hline
\end{tabular}
\begin{tablenotes}
\item[a]original data from \citet{Guesten1987} convolved for \citet{Marr1993} beam 
\item[b]original data from \citet{Jacks1993} convolved for \citet{Marr1993} beam 
\end{tablenotes}

\end{threeparttable}
\end{center}
\normalsize

\paragraph*{•}
Molecular data for HCN and HCO$^{+}$ were sourced from the Cologne Data Base for Molecular Spectroscopy. This data included excitation level information, Einstein A coefficients and collision rates for a range of kinetic temperatures and formed part of the input to the molecular rotation excitation model described in Chapter \ref{Molex} that is used to analyse the observations.

\paragraph*{•}
150\,K was chosen to be the fiducial value of the kinetic temperature for modelling as this was midway between 250\,K, chosen by \citet{Marr1993} after considering a temperature range of (150 to 450\,K) and 50\,K, used by \citet{Chris2005}. The sensitivity to  kinetic temperature was tested and the differences between 150 and 250\,K can be seen by comparing Core A parameter values in Table \ref{Marrcprops}. The excitation temperatures for all tracers increased at 250\,K by between 7 and 18$\%$, optical depths decreased by 17$\%$ for H$^{12}$CN and $7\%$ for H$^{13}$CN, while they increased 0.3$\%$ for HCO$^{+}$ and 3$\%$ for HCN (3-2). Peak brightness temperatures increased  by 8$\%$ for H$^{12}$CN but decreased by values ranging from 1 to 8$\%$ for the other three tracers. 
 
\subsection{Results} \label{Marr}
\paragraph*{•}
Each molecule was modelled by starting with local thermal equilibrium (LTE) by choosing a high atomic hydrogen density (10$^{12}$ cm$^{-3}$) and decreasing the density in steps of (logn$_{H}$ = 0.1 to 10$^{3}$ cm$^{-3}$) for each increasing step in molecular column density of (logN$_{col}$ = 0.1) from 10$^{12}$ to 10$^{18}$ cm$^{-2}$. Peak brightness values for the transitions were collected and plotted as contours to show how the specified brightness line (see Table \ref{Input1}) shifted with changing values of the two density parameters. The co-ordinates of the average of the intersection points of the brightness curves were taken as indicative values of a core's hydrogen density and molecular column density per unit line width. The co-rdinate values from the  core's brightness plot were then applied to the plot of optical depths for the transitions and values taken where each optical depth curve intersected the co-ordinate point. 

\begin{figure}[ht] \includegraphics[scale=0.7,bb=  0 0 580 760] 
{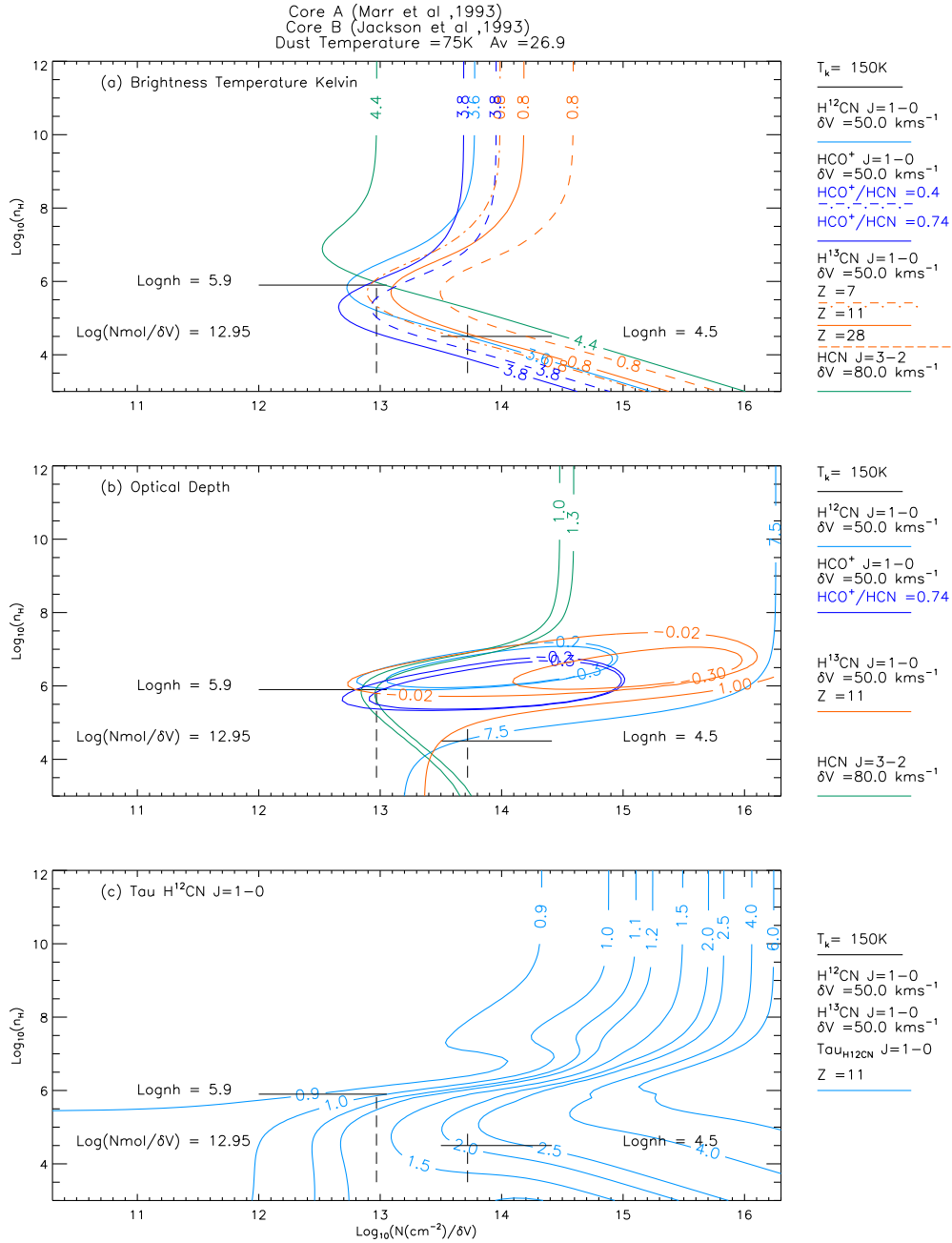} \fontsize{9} {9} \vspace{-1cm} \caption[Marr Core A Tb and Optical Depth Plots]{\label{MarrA}   Panel (a) shows brightness temperature contours for H$^{12}$CN, H$^{13}$CN, HCO$^{+}$ all (1-0) and HCN(3-2). H$^{13}$CN with Z = 7 (chain dotted) and Z = 28 (dashed) and HCO$^{+}$ with Z = 0.4  (dashed) contours are also plotted. The average intersection point of the species is logn$_{\mathrm{h}}$ = 5.9, log(N$_{\mathrm{mol}}$/dV) = 12.95 with the closest point of approach of the contours at logn$_{\mathrm{h}}$ = 4.5, logN$_{\mathrm{Mol}}$/dV) = 13.7.
Panel (b) shows optical depth contours for the tracers. One abundance for both H$^{13}$CN and HCO$^{+}$ are plotted as indicated for clarity. H$^{12}$CN, H$^{13}$CN and HCO$^{+}$ are optically thin with inverted transition populations, and HCN(3-2) is optically thin with \large{$\tau$} \normalsize $\sim$ 1.3.
H$^{12}$CN is optically thick with \large{$\tau$} \normalsize $\sim$ 7.5 at the closest point of approach of brightness contours. 
Panel (c) shows the optical depth contours of H$^{12}$CN calculated using brightness temperatures for H$^{12}$CN and H$^{13}$CN from Molex and a [$^{12}$C]/[$^{13}$C] abundance ratio, Z = 11. \large{$\tau^{12}$} \normalsize = $\sim$ 1 at the average intersection point and $\sim$ 2.0 for the closest point of approach of the H$^{12}$CN \& H$^{13}$CN brightness contours.} 
\end{figure}

\begin{figure}[ht] \includegraphics[scale=0.8,bb=  0 0 580 760] 
{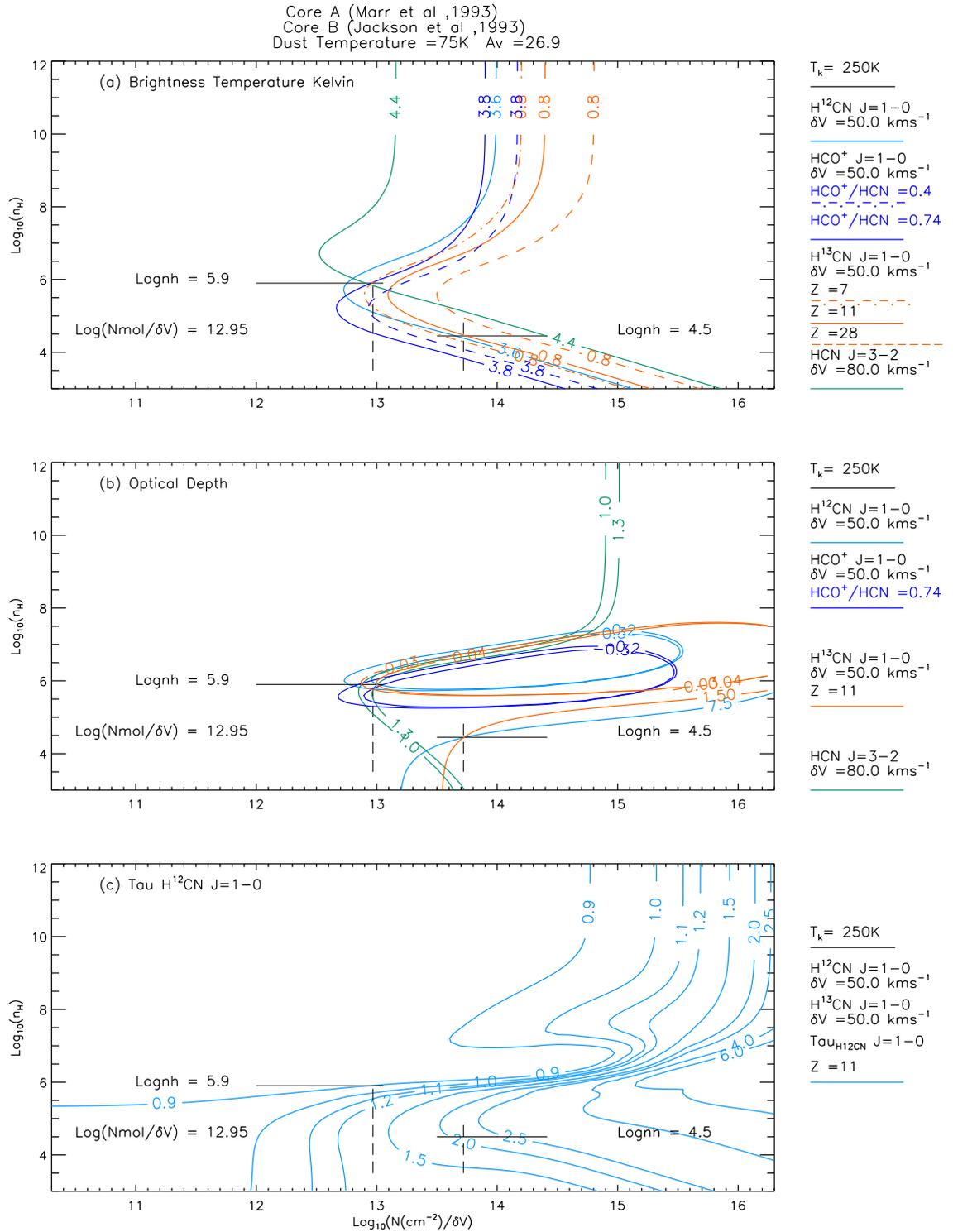} \fontsize{9} {9} \vspace{-1cm} \caption[Marr Core A Tb and Optical Depth Plots for Tk 250\,K]{\label{MarrF}   
As for Fig.\ \ref{MarrA} with a raised kinetic temperature of 250\,K.
Panel (a) shows that the average intersection and closest approach points of the brightness temperature contours are unchanged from Fig.\ \ref{MarrA}.  
Panel (b) shows optical depth contours the same as for the tracers in Fig.\ \ref{MarrA} Panel(c) shows the optical depth contours of H$^{12}$CN from brightness temperatures for H$^{12}$CN and H$^{13}$CN calculated by Molex and a [$^{12}$C]/[$^{13}$C] abundance ratio Z = 11. Opacities are similar to Fig.\ \ref{MarrA}} 
\end{figure}

\begin{figure}[ht] \includegraphics[scale=0.8,bb=  0 0 580 760] 
{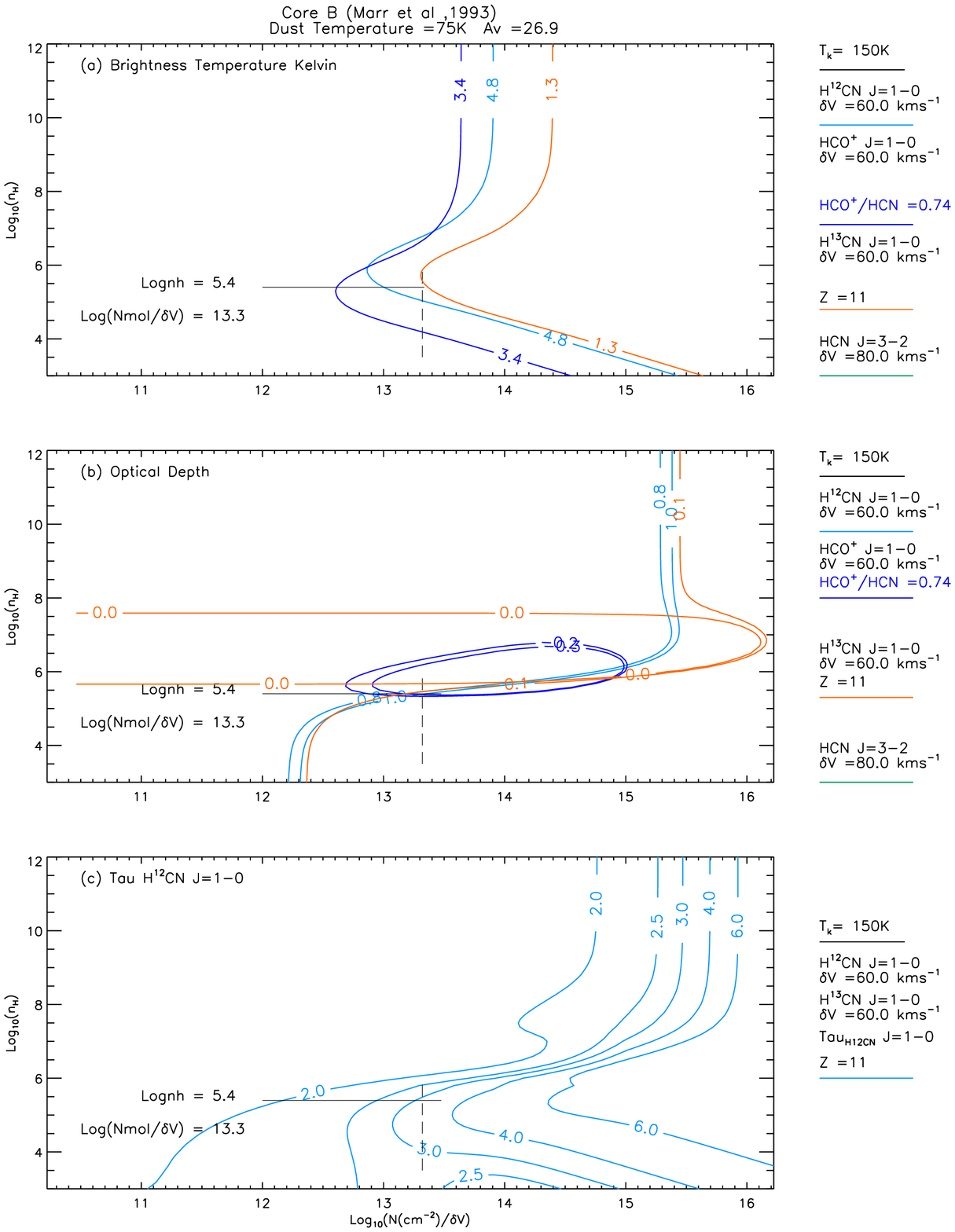} \fontsize{9} {9} \vspace{-1cm} \caption[Marr Core B Tb and Optical Depth Plots]{\label{MarrB}  As in Fig.\ \ref{MarrA} but for Core B.} 
\end{figure}

\begin{figure}[ht] \includegraphics[scale=0.8,bb=  0 0 580 760] 
{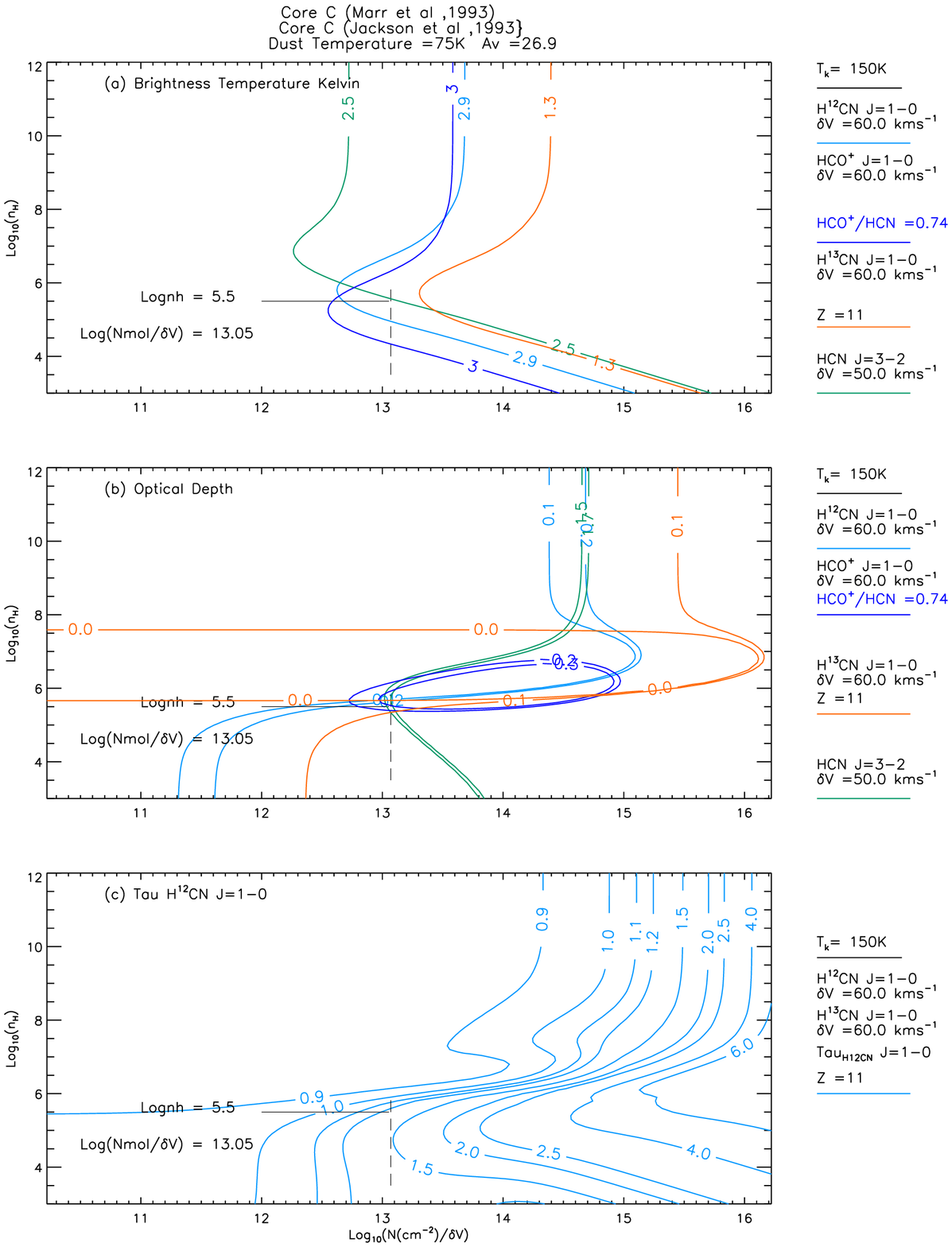} \fontsize{9} {9} \vspace{-1cm} \caption[Marr Core C Tb and Optical Depth Plots]{\label{MarrC}  As in Fig.\ \ref{MarrA} but for Core C.} 
\end{figure}

\begin{figure}[ht] \includegraphics[scale=0.8,bb=  0 0 580 760] 
{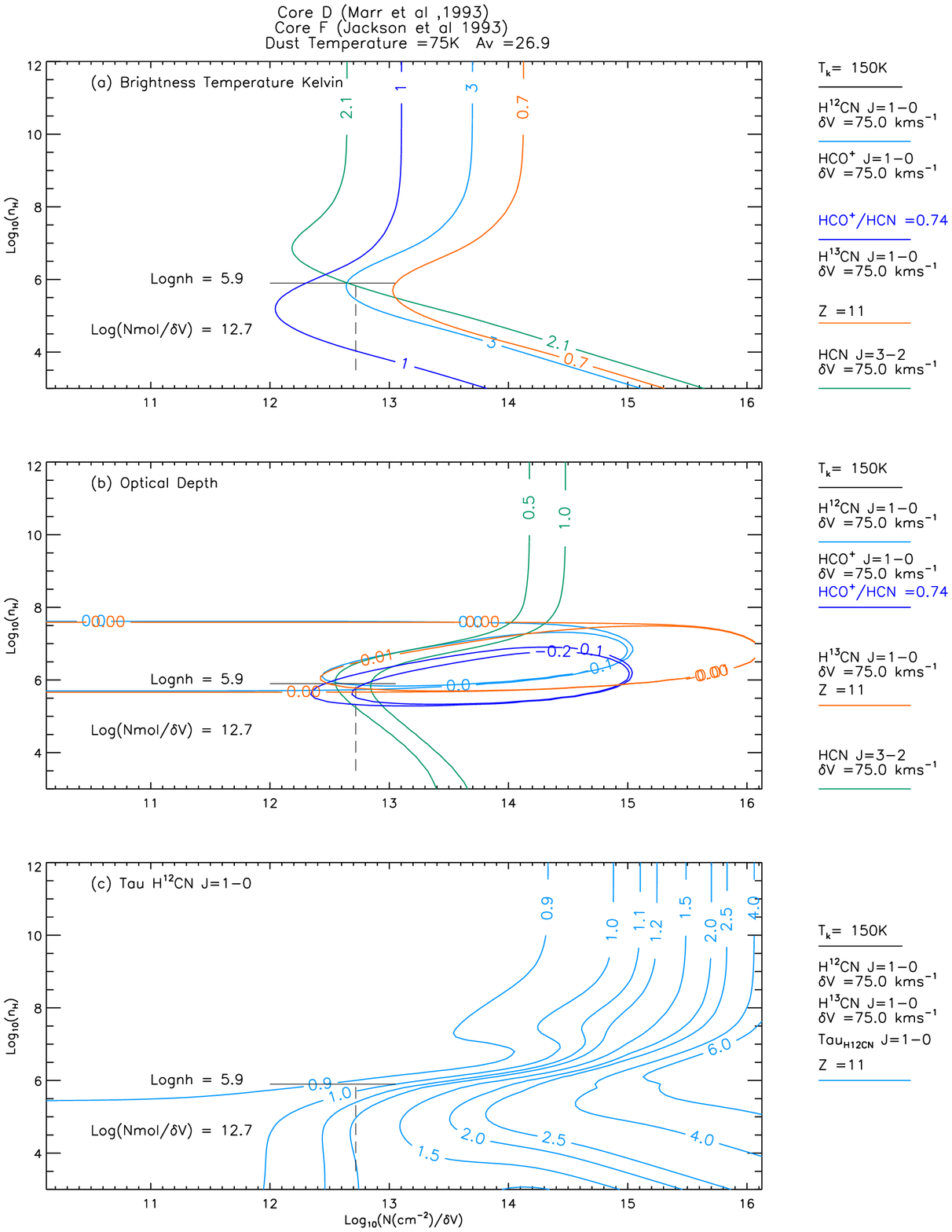} \fontsize{9} {9} \vspace{-1cm} \caption[Marr Core D Tb and Optical Depth Plots]{\label{MarrD}  As in Fig.\ \ref{MarrA} but for Core D.} 
\end{figure}

\begin{figure}[ht] \includegraphics[scale=0.8,bb=  0 0 580 760] 
{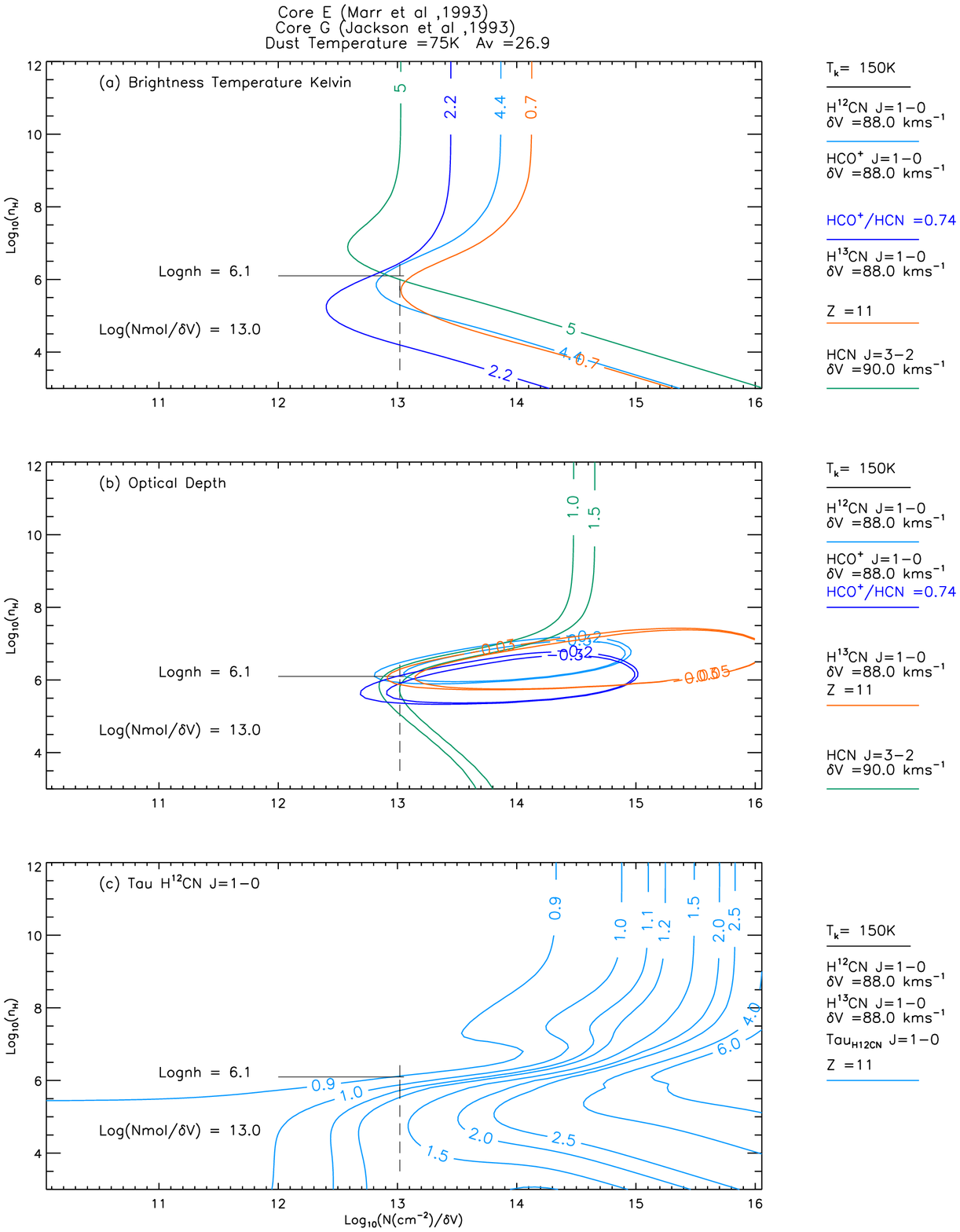} \fontsize{9} {9} \vspace{-1cm} \caption[Marr Core E Tb and Optical Depth Plots]{\label{MarrE}  As in Fig.\ \ref{MarrA} but for Core E.} 
\end{figure}

\paragraph*{•}
The modelling results for HCN column densities, hydrogen densities, excitation temperatures, optical depths and  peak brightness temperatures for each transition 
are summarised in Table \ref{Marrcprops} and are shown in Figs.\ \ref{MarrA} to \ref{MarrE} which are contour plots of:
\paragraph*{•}
(a) The peak brightness temperature values for each of the molecules reported in Table 1 of \citet{Marr1993}. The brightness values have an estimated accuracy of $\pm0.2$\,K and are derived from unresolved observations of the cores. 

\paragraph*{•}
(b) Optical depths calculated by the model for the tracers, based on Eqn.\ \ref{Tau0}.
 
\paragraph*{•}
The outputs from the model were plotted as contours on a log-log plot of HCN column density per unit line width  (abscissa cm$^{-2}$/(km s$^{-1}$)) and atomic hydrogen number density (ordinate cm$^{-3}$).

\paragraph*{•}
 
The intersection of three contours can be regarded as a reliable indicator of the column densities of the molecular transitions and the hydrogen number density. Although the observations produced unresolved data the ratios of the brightness temperatures are assumed to be valid given that \citet{Marr1993} scaled data from other sources to match their H$^{13}$CN(1-0) and HCO$^{+}$(1-0) to allow comparison.

\paragraph*{•}
This approach is endorsed in section 4.1 of the Radex notes where the authors suggest the easiest way of solving the problem of unresolved images is to model intensity ratios and hope/argue that the beam dilution factor for the lines are comparable \citep{vdeTak2004}. This approach has support from \citet{Marr1993} who stated in the results section of their paper that 
\begin{itemize}
\item An overlay of HCO$^{+}$ and H$^{12}$CN data shows that HCO$^{+}$ is distributed similarly to the HCN emission. Both are concentrated in the CND and are clumped at the same locations.   
\item Signals from H$^{13}$CN are much weaker and closer to the noise level, with their brightest peaks located in the ring and with spectral centroids at the same velocities as the HCO$^{+}$ and H$^{12}$CN emission at the core positions. This means that the H$^{13}$CN brightness data can be included in the analysis and provide reliable results when used with the brightness data from the other transitions.
\end{itemize}

\paragraph*{•}
The lesser abundant molecules were plotted over a lower range of column densities (see Table \ref{Abundances}) and plotting consistency maintained by multiplying  by the relevant abundance ratio to convert them to their H${12}$CN equivalent column densities. All tracers were plotted on the HCN column density scale divided by the relevant tracer's line width.

\begin{center} 
\begin{threeparttable}[ht]
\caption{\label{Abundances} Abundance Ratios of HCN to Other Modelled Molecules}  %\vspace{0.5cm}
%\small
%\scriptsize
\begin{tabular}{c|c|c|c|c}
\hline
\hline
\multicolumn{1}{c|}{Molecule}&\multicolumn{1}{c|}{Normal Ratio\tnote{a}}&\multicolumn{1}{c|}{Model Ratio}&\multicolumn{2}{c}{Log Column Density} \\
\multicolumn{1}{c|}{}&\multicolumn{1}{c|}{}&\multicolumn{1}{c|}{}&\multicolumn{1}{c|}{Start Value}&\multicolumn{1}{c}{Finish Value} \\
\multicolumn{1}{c|}{}&\multicolumn{1}{c|}{}&\multicolumn{1}{c|}{}&\multicolumn{1}{c|}{(cm$^{-2}$)}&\multicolumn{1}{c}{(cm$^{-2}$)} \\ 
\hline 
{H$^{12}$CN\tnote{b}}&1&1&12.0&18.0 \\
H$^{13}$CN&30&7.0&11.2&17.2 \\
 & &{11.0\tnote{c}}&10.9&16.9 \\
 & &{28\tnote{d}}&10.6&16.6 \\
\hline
HCO$^{+}$&$\sim$ 1&{1.35\tnote{e}}&11.9&17.9 \\
  & &{2.5\tnote{e}}&11.6&17.6 \\
\hline
\hline 
\end{tabular}
\begin{tablenotes}
\item[a] as occurs in galactic clouds %\\
\item[b] commonly referred to as HCN %\\
\item[c] lower value in range \citet{Marr1993} %\\
\item[d] higher value in range \citet{Marr1993} %\\
\item[e] inverse of 0.74 quoted as average for CND value by \citet{Chris2005}
\item[f] inverse of 0.4 quoted as the ratio corresponding to locations in the CND with peak HCN(1-0) emission \citet{Chris2005}
\end{tablenotes}
\end{threeparttable}
\end{center}   
%\end{landscape}

\normalsize

\paragraph*{•}
Uncertainties arise with the [$^{12}$C]/[$^{13}$C] (Z ratios) as determined in \citet{Marr1993} from the rms noise in their data and standard error propagation through their equations relating $\tau$(H$^{13}$CN), $\tau$(H$^{12}$CN) and $\tau$(HCO$^{+}$). \citet{Marr1993} noted that the lower boundary value of [HCN]/[H$_{2}$] = 6 $\times$ 10$^{-9}$ for their modelling occurred with values of Z = 4 to 7 which was subsequently superseded for a more reasonable value for [HCN]/[H$_{2}$] = 8 $\times$ 10$^{-8}$ and higher values for Z = 20. The effects of varying the value for Z are shown in Fig.\ \ref{MarrA} where brightness contours shift to the right for rising values of Z for both H$^{13}$CN and HCO${+}$. The [$^{12}$C]/[$^{13}$C] abundance ratio is addressed more fully in Section \ref{discuss}.

\paragraph*{•}
The effect of varying the kinetic temperature from 150 to 250\,K and Z values in the current model is shown for Core A in Figs.\ \ref{MarrA} and \ref{MarrF} and Table \ref{Marrcprops} with increases for H$^{12}$CN of 12\% in excitation temperature and 8\% in brightness temperature and a decrease of 17\% in optical depth occur.  At Z = 7 the H$^{13}$CN brightness temperature contour passes through the average  of the intersection points and intersects the H$^{12}$CN contour indicating that this is a more representative Z value for the prevailing conditions. 
  
\paragraph*{•}
Table \ref{Marrcparam} lists the model parameters for Core A, where the H$^{12}$CN and H$^{13}$CN brightness contours are at their closest point of approach (logn$_{H}$ =0.1). Ideally these curves should intersect and the fact that they do not can be attributed to the uncertainty in the value for Z:

\begin{itemize}
\item Fig.\ \ref{MarrA} Panel(a): for a kinetic temperature of T = 150\,K this point occurs at about logn$_{\mathrm{h}}$ = 4.5 and logN$_{\mathrm{mol}}$/dV = 13.7 where the excitation temperature for  H$^{12}$CN, T$_{\mathrm{ex}}$ = 6.5\,K and optical depth, \large{$\tau$} \normalsize = 7.8. The corresponding values for H$^{13}$CN are 3.8\,K and 1.6 (see Table \ref{Marrcparam}).
\item Fig.\ \ref{MarrF}  Panel(a): for 250\,K this point occurs at about the same values of logn$_{\mathrm{h}}$ = 4.5 and logN$_{\mathrm{mol}}$/dV = 13.7 where  the excitation temperature for H$^{12}$CN, T$_{\mathrm{ex}}$ is 7.2\,K and optical depth, \large{$\tau$} \normalsize is 6.75. Values for H$^{13}$CN are 4\,K and 1.48 (see Table \ref{Marrcparam}).
\end{itemize}

\paragraph*{•}
The modelling results for the points of closest approach show  high H$^{12}$CN optical depths ranging from about $\simeq$3 to 9. At these points the molecular hydrogen number density of 1.58$\times$ 10$^{4}$cm$^{-3}$ is much lower than the density of $\simeq$ 10$^{6}$cm$^{-3}$ inferred by \citet{Marr1993}.
\paragraph*{•}

Table \ref{Marrcparam} includes Radex model results for the above points of closest approach for comparison with the Molex model. The agreement between the models is close for H$^{12}$CN and acceptable for H$^{13}$CN. For H$^{12}$CN at 150\,K the excitation temperature and optical depth from Radex are 0.5$\%$ higher while the brightness temperature is 0.6$\%$ lower than Molex. At 250\,K the excitation temperature from Radex is 0.14$\%$ higher, the optical is depth 0.3$\%$ higher and the brightness temperature is 0.24$\%$ lower than Molex. The variations for H$^{13}$CN at both kinetic temperatures are greater, with values up to 8$\%$ lower for Radex than Molex. The close agreement confirms that the Molex model is providing reliable results, and as Radex does not include dust extinction, it confirms that these dust parameters have a very small influence on the Molex results. This will be discussed further in Section \ref{Chris} for the second core group where the dust parameters are varied in the Molex model to show minimal effect. 

\paragraph*{•}
Panel (c) of Figs.\ \ref{MarrA} to \ref{MarrE} for the Marr cores shows contours
 of H$^{12}$CN opacity calculated using  brightness temperature values for
 H$^{12}$CN and H$^{13}$CN from the Molex model. The optical depths ranged from 0.9 to 1.3 when using a [C$^{12}$]/[C$^{13}$] ratio Z = 11, which is the minimum considered by \citet{Marr1993}. \citet{Marr1993} cite an optical depth of 4, a kinetic temperature of 250\,K and a molecular hydrogen density of $\approx$
 2.6$\times10^{6}$ cm$^{-3}$ as more reasonable values from their modelling. For
 Core A, Eqn \ref{H12CN} produces \large{$\tau$}\normalsize(H$^{12}$CN) = 2.5  (see Table \ref{Marrcparam}). An optical thickness of\large{$\tau$}\normalsize(H$^{12}$CN) = 4 would require a column density at least an order of magnitude greater. The reasons for the different results are again covered in Section \ref{discuss}.   

\begin{landscape}
\begin{center} 
\begin{threeparttable}[ht]

\caption{\label{Marrcprops} \citet{Marr1993} Core Properties.}
\vspace{0.5cm}
%\small
\scriptsize
\begin{tabular}{c|c|c|c|c|c|c|c|c|c|c|c|c|c|c|c}
%|c|d{2}|d{2}|d{2}|d{2.4}|d{2.4}|d{2.4}|} \hline
%\begin{tabular}{|c|c|c|c|c|c|c|} \hline
\hline
\hline
\multicolumn{1}{c|}{Marr Core}&\multicolumn{1}{c|}{Gas Temp}&\multicolumn{1}{c|}{Average Log}&\multicolumn{1}{c|}{n(H$_{2}$)}&\multicolumn{4}{c|}{Excitation Temperature Tex}&\multicolumn{4}{c|}{Optical Depth $\tau$}&\multicolumn{4}{c}{Peak Brightness Temp T$_{\mathrm{b}}$} \\
%\hline
\multicolumn{1}{c|}{ }&\multicolumn{1}{c|}{Tk}&\multicolumn{1}{c|}{Col Density}&\multicolumn{1}{c|}{ }&\multicolumn{1}{c|}{H$^{12}$CN}&\multicolumn{1}{c|}{ H$^{13}$CN}&\multicolumn{1}{c|}{HCO$^{+}$}&\multicolumn{1}{c|}{HCN}&\multicolumn{1}{c|}{H$^{12}$CN}&\multicolumn{1}{c|}{ H$^{13}$CN}&\multicolumn{1}{c|}{HCO$^{+}$}&\multicolumn{1}{c|}{HCN}&\multicolumn{1}{c|}{H$^{12}$CN}&\multicolumn{1}{c|}{ H$^{13}$CN}&\multicolumn{1}{c|}{HCO$^{+}$}&\multicolumn{1}{c}{HCN} \\
%\multicolumn{1}{c|}{H$^{13}$CN}&\multicolumn{1}{c|}{HCO$^{+}$}&\multicolumn{1}{c|}(HCN}&\multicolumn{1}{c|}{H$^{12}$CN}&\multicolumn{1}{c|}{H$^{13}$CN}&\multicolumn{1}{c|}{HCO$^{+}$}&\multicolumn{1}{c|}(HCN}&\multicolumn{1}{c|}{H$^{12}$CN}&\multicolumn{1}{c|}{H$^{13}$CN}&\multicolumn{1}{c|}{HCO$^{+}$} &\multicolumn{1}{c|}(HCN} \\
%\hline
\multicolumn{1}{c|}{ }&\multicolumn{1}{c|}{ }&\multicolumn{1}{c|}{per line}&\multicolumn{1}{c|}{ }&\multicolumn{1}{c|}{1-0}&\multicolumn{1}{c|}{1-0}&\multicolumn{1}{c|}{1-0}&\multicolumn{1}{c|}{3-2}&\multicolumn{1}{c|}{1-0}&\multicolumn{1}{c|}{1-0}&\multicolumn{1}{c|}{1-0}&\multicolumn{1}{c|}{3-2}&\multicolumn{1}{c|}{1-0}&\multicolumn{1}{c|}{1-0}&\multicolumn{1}{c|}{1-0}&\multicolumn{1}{c}{3-2} \\
%\hline
\multicolumn{1}{c|}{ }&\multicolumn{1}{c|}{K}&\multicolumn{1}{c|}{cm$^{-2}$kms$^{-1}$}&\multicolumn{1}{c|}{ 10$^{6}\times$ cm$^{-3}$}&\multicolumn{1}{c|}{K}&\multicolumn{1}{c|}{K}&\multicolumn{1}{c|}{K}&\multicolumn{1}{c|}{K}&\multicolumn{1}{c|}{ }&\multicolumn{1}{c|}{ }&\multicolumn{1}{c|}{ }&\multicolumn{1}{c|}{ }&\multicolumn{1}{c|}{K}&\multicolumn{1}{c|}{K}&\multicolumn{1}{c|}{K}&\multicolumn{1}{c}{K} \\
\hline
\hline
{A}&150&12.95&0.397&-12.5&-16.3&-11.2&13.9&-0.195&-0.043&-0.214&0.777&3.42&0.85&3.49&4.51 \\
{A}&250&12.95&0.397&-11.0&-13.4&-9.95&13.0&-0.228&-0.046&-0.208&0.803&3.69&0.84&3.09&4.17 \\
\hline
%{Percentage}& & & & & & & & & & & & & & &  \\
{\% Difference}& & & &12.0&17.8&11.2&6.5&-16.9&-7.0&0.3&3.3&7.9&2.1&1.1&7.5 \\
\hline
{B}&150&13.2&0.126&-27.6&12.6&-10.9& &-0.161&0.150&-0.221& &5.39&1.29&3.53&  \\
%\hline
{C}&150&13.05&0.158&-16.1&9.21&-12.3&10.5&-0.145&0.215&-0.169&0.633&3.05&1.18&2.91&2.49 \\
%\hline
{D}&150&12.7&0.397&-22.6&19.6&-11.9&9.54&-0.105&0.049&-0.065&0.597&2.87&0.72&1.03&1.98 \\
%\hline
{E}&150&13.0&0.629&-11.1&5.23&-14.4&10.1&-0.215&0.217&-0.214&1.93&4.25&0.71&2.35&5.15 \\
\hline
\hline
\end{tabular}
\end{threeparttable}
\end{center}   
\end{landscape}

\normalsize
%\clearpage
%\newpage 
%\clearpage

\begin{landscape}
\begin{center} 
\scriptsize
\begin{threeparttable}[ht]
%\begin{table}[!h] \begin{center} 
\caption{\label{Marrcparam} \citet{Marr1993} Core Parameters at Closest Point of Approach of H$^{12}$CN and H$^{13}$CN Brightness Contours}
\vspace{0.5cm}
%\small

\begin{tabular}{c|c|c|c|c|c|c|d{2}|d{2}|d{2}|d{3}|d{2}|d{2}|c|c} \hline
%\begin{tabular}{|c|c|c|c|c|c|c|c|c|c|c|c|c|c|} \hline
\multicolumn{1}{c|}{Core}&\multicolumn{1}{c|}{Model}&\multicolumn{1}{c|}{Tracer}&\multicolumn{1}{c|}{Temp K}&\multicolumn{1}{c|}{Hydrogen\tnote{a}}&\multicolumn{1}{c|}{LogNMol/dV}&\multicolumn{1}{c|}{Column Density\tnote{b}} &\multicolumn{1}{c|}{Tex}&\multicolumn{1}{c|}{$\tau$}&\multicolumn{1}{c|}{Tb}&\multicolumn{1}{c|}{$\beta$}&\multicolumn{1}{c|}{x(u)}&\multicolumn{1}{c|}{x(l)}&\multicolumn{1}{c|}{Tb(H$^{12}$CN)}&\multicolumn{1}{c}{$\tau$(H$^{12}$CN)} \\
%\hline
\multicolumn{1}{c|}{ }&\multicolumn{1}{c|}{ }&\multicolumn{1}{c|}{ }& &\multicolumn{1}{c|}{Density}&\multicolumn{1}{c|}{}&\multicolumn{1}{c|}{ }& & & & & & &\multicolumn{1}{c|}{/Tb(H$^{13}$CN)}& \\
%\hline
\multicolumn{1}{c|}{ }&\multicolumn{1}{|c|}{ }& & &\multicolumn{1}{c|}{(n(H$_{2}$)$\times$10$^{4}$)}& &\multicolumn{1}{c|}{(NMol$\times10^{15}$)}& & & & & & & &  \\
\hline
%\hline
{A\tnote{c}}&{Molex}&{H$^{12}$CN$_{1-0}$}&150&1.58&13.7&2.51&6.50&7.84&3.52&0.185&0.49&0.31&4.90&2.51\\
\hline
{A}&{Radex}&{H$^{12}$CN$_{1-0}$}&150&1.58& &2.51&6.53&7.88&3.50& &0.49&0.31& & \\
\hline
%{ }&{Percentage}& & & & & & & & & & & & & \\
{ }&{\%Difference}& & & & & &0.46&0.51&-0.57& &0.0&0.0& & \\
\hline
{A}&{Molex}&{H$^{13}$CN$_{1-0}$}&150&1.58&13.7&0.228&3.76&1.62&0.718&0.586&0.460&0.462& & \\
%\hline
{A}&{Radex}&{H$^{13}$CN$_{1-0}$}&150&1.58& &0.228&3.71&1.50&0.661& &0.458&0.467& &  \\
\hline
%{ }&{Percentage}& & & & & & & & & & & & & \\
{ }&{\%Difference}& & & & & &-1.33&-7.40&-7.93& &-0.09&1.08& & \\
\hline
%\hline
{A\tnote{d}}&{Molex}&{H$^{12}$CN$_{1-0}$}&250&1.58&13.7&2.51&7.22&6.75&4.16&0.212&0.481&0.289&4.76&2.587\\
%\hline
{A}&{Radex}&{H$^{12}$CN$_{1-0}$}&250&1.58& &2.51&7.21&6.77&4.15& &0.482&0.290& &  \\
\hline
%{ }&{Percentage}& & & & & & & & & & & & & \\
{ }&{\%Difference}& & & & & &0.14&0.30&-0.24& &0.21&0.35 & & \\
\hline
{A}&{Molex}&{H$^{13}$CN$_{1-0}$}&250&1.58&13.7&0.228&4.01&1.48&0.87&0.610&0.469&0.439& & \\
%\hline
{A}&{Radex}&{H$^{13}$CN$_{1-0}$}&250&1.58& &0.228&3.96&1.37&0.81& &0.467&0.439& &  \\
\hline
%{ }&{Percentage}& & & & & & & & & & & & & \\
{ }&{\%Difference}& & & & & &-1.25&-7.43&-6.90& &-0.42&0.0& & \\
\hline
%\hline
{B\tnote{e}}&{Molex}&{H$^{12}$CN$_{1-0}$}&150&1.58&13.9&4.77&7.9&9.84&4.83& & & &5.43& \\
{B}&{Molex}&{H$^{13}$CN$_{1-0}$}&150&1.58& &0.43&3.9&2.04&0.89& & & & & \\
{C\tnote{f}}&{Molex}&{H$^{12}$CN$_{1-0}$}&150&{na}&{na}&{na}&{\mathrm{na}}&{\mathrm{na}}&{\mathrm{na}}& & & &{na}& \\
{C}&{Molex}&{H$^{13}$CN$_{1-0}$}&150&{na}&{na}&{na}&{\mathrm{na}}&{\mathrm{na}}&{\mathrm{na}}& & & &{na}& \\
{D\tnote{g}}&{Molex}&{H$^{12}$CN$_{1-0}$}&150&1.58&13.7&3.76&6.68&8.05&3.64& & & &5.87& \\
{D}&{Molex}&{H$^{13}$CN$_{1-0}$}&150&1.58& &0.34&3.69&1.39&0.62& & & & & \\
{E\tnote{h}}&{Molex}&{H$^{12}$CN$_{1-0}$}&150&0.50&13.3&1.76&7.69&2.27&4.15& & & &5.85& \\
{E}&{Molex}&{H$^{12}$CN$_{1-0}$}&150&0.50& &0.16&5.23&0.38&0.71& & & & & \\
\hline
%\hline

\end{tabular}
\begin{tablenotes}
\item[a] molecular hydrogen density, n$_{H_{2}}$, calculated as 0.5 atomic hydrogen density, n(H)
\item[b] N$_{\mathrm{molH}}^{13}$CN is derived by dividing Nmol H$^{12}$CN by 11
\item[c] see Fig.\ \ref{MarrA}
\item[d] see .\ \ref{MarrF}
\item[e] see Fig.\ \ref{MarrB}
\item[f] there is no point of closest approach see Fig.\ \ref{MarrC}
\item[g] see Fig.\ \ref{MarrD}
\item[h] contours intersect see Fig.\ \ref{MarrE}
\end{tablenotes}
\end{threeparttable}
\end{center}  
\end{landscape}
\normalsize
\paragraph*{}
Table \ref{Marrcprops} summarises the results for the five HCN cores (A to E) reported in \citet{Marr1993}. The average column density per line width and atomic hydrogen number density for each core were obtained by averaging the values of the intersection points of species pairs H$^{12}$CN(1-0)/HCN(3-2), HCO$^{+}$(1-0)/H$^{12}$CN(1-0) and H$^{13}$CN(1-0)/HCN(3-2) to provide brightness temperature values that match the plotted contour values.
 
\paragraph*{}
It is notable that the H$^{12}$CN and H$^{13}$CN brightness contours only intersect for Core E  (Fig.\ \ref{MarrE}) and do not closely approach each other in the plot for Core C (Fig.\ \ref{MarrC}). The three remaining cores, A, B and D, have close approach points listed in Table \ref{Marrcparam}.  

\subsection{Comments} \label{Marrres}
\paragraph*{}
\citet{Marr1993} used a statistical equilibrium excitation model for HCN and set boundary values for kinetic temperature (150 to 450\,K), dust temperature of 75\,K, optical depth (1 to 12) and [HCN]/[H$_{2}$] ratios (6$\times$ 10${-9}$ to 3$\times$10$_{-6}$. The [$^{12}$C]/[$^{13}$C] ratio (Z) was assumed to vary between 10 and 40 while noting that previous estimates varied from 11 (in SgrB$_{2}$ \citep{Magnum1988}) to 28 ( in the galactic centre \citep{Wannier1989}). 

\paragraph*{•}
Fig.\ \ref{MarrA} Panel (a) showed  brightness contours for Core A with alternative abundance ratios Z = 7 and Z = 28 for [H$^{12}$CN]/[H$^{13}$CN] and Z = 0.4 for [HCO$^{+}$]/[HCN] as dashed contours. The alternative ratios of Z = 28 and Z = 0.4 do not provide a definitive intersection of the contours with the H$^{12}$CN contour. The contours for Z = 7 and Z = 0.4 do intersect the H$^{12}$CN contour but at a lower hydrogen density value than for Z = 11 and Z = 0.74 which led to the adoption of Z = 11 and Z = 0.74 for the other core plots. It should also be noted that the intersection area for all tracers is not representative of the closest approach ($\Delta$logn$_{\mathrm{H}}$ = 0.1) of the H$^{12}$CN and H$^{13}$CN contours, which are closest to each other over a span of log column density per line width ($\delta$V) values from 13.6 to 14.

\paragraph*{}
The effect of changing the abundance ratio is shown in Fig.\ \ref{MarrA} Panel(a) where the relevant brightness contours shift to the left as the ratio decreases. This is especially noticeable for the [C$^{12}$]/[C$^{13}$] = 7 curve where the contour falls within the intersection points of the other three species. Similarly, a reduction in the brightness temperature shifts the contour to the left. 

\paragraph*{}
The excitation temperatures of all species and transitions are different from each other. Stimulated emission is occurring in the H$^{12}$CN, H$^{13}$CN and HCO$^{+}$ species; this was not considered by \citet{Marr1993} where maser emission was excluded from their analysis. The current results show that inverted transition levels are occurring and are optically thin not thick as suggested by \citet{Marr1993}. 

\section{Group Two}
\paragraph*{}
Sections \ref{1-0} to \ref{4-3} describe the core input data for the HCN(1-0), (3-2) and (4-3) transitions and the HCO${+}$(1-0) transition.

\subsection{HCN(1-0) and HCO$^{+}$(1-0) Input} \label{1-0}
\paragraph*{•}
The integrated intensity of each of the HCN(1-0) cores was obtained from Fig.\ 13 of \citet{Chris2005} by counting the contour levels from the lowest level to the contour closest to the core to obtain an estimated peak integrated intensity in Jy beam$^{-1}$ km s$^{-1}$. The beam quoted in the paper was converted to steradians and the FWHM spectral width for each core in km s$^{-1}$ was obtained from Table 2 of their paper. The integrated intensity was divided by the beam's area and the line width to produce a peak intensity at line centre I$_{\nu}$(0). Peak radiation temperatures at line centre were then derived using the following Rayleigh-Jeans relationship:  
\begin{equation} \label{RJTb}
\mathrm{T_{b}} = \mathrm{I}_{\nu}(0)\lambda^{2}\times 10^{-23}/2\mathrm{k} \,.
\end{equation}

\paragraph*{•}
HCO$^{+}$(1-0) intensities were obtained from the integrated intensities in Fig.\ 4 of \citet{Chris2005} and processed in the same manner as for the HCN
(1-0) data described above.
\paragraph*{•} 
Table \ref{Input1} summarises the input for both tracers. Core data, central velocities and line widths are assumed to be the same as for the HCN(1-0) data.  

\begin{landscape}
\begin{center} 

\begin{table}[ht] %\begin{center} 
\caption{\label{Input1} Input Data from \citet{Chris2005}}
\vspace{0.5cm}

%\sma)ll
%\scriptsize

\begin{tabular}{c|c|c|c|c|c|c|c} \hline
\multicolumn{1}{c|}{HCN (1-0)}&\multicolumn{1}{c|}{Diameter}&\multicolumn{1}{c|}{Area}&\multicolumn{1}{c|}{Central Vel}&\multicolumn{1}{c|}{Line Width}&\multicolumn{1}{c|}{Contour Value}&\multicolumn{1}{c|}{Integ Intensity}&\multicolumn{1}{c}{Peak Tb } \\
%\hline
\multicolumn{1}{c|}{ }&\multicolumn{1}{c|}{FWHM pc}&\multicolumn{1}{c|}{arsecs$^{2}$}&
\multicolumn{1}{c|}{km s$^{-1}$}&\multicolumn{1}{c|}{km s$^{-1}$}&\multicolumn{1}{c|}{6.675$\times$Jy beam$^{-1}$km s$^{-1}$}&\multicolumn{1}{c|}{Jy beam$^{-1}$km s$^{-1}$}&\multicolumn{1}{c}{K} \\
\hline
{Core D}&0.43&96.5&101&45.5&4.5&30.4&7.6 \\

{Core I}&0.26&35.3&-18&15&1.5&10.1&7.6 \\

{Core M}&0.26&35.3&-64&19&3.5&23.7&14.1 \\

{Core O}&0.33&56.9&-108&36.5&6.0&40.6&12.6 \\

{Core P}&0.21&23.0&-73&28.2&4.5&30.4&12.2 \\

{Core W}&0.22&25.3&56&27.9&2.5&16.9&6.8 \\
{Core Z}&0.24&25.3&58&39.6&2.5&16.9&4.8 \\
\hline
\hline
{HCO$^{+}$ (1-0)}& & & & & & &  \\ 
\hline
{Core D}& & & & &2.0&13.53&3.3 \\

{Core I}& & & & &0.8&5.412&4.0 \\

{Core M}& & & & &1.0&6.765&4.0 \\

{Core O}& & & & &1.5&10.15&3.1 \\

{Core P}& & & & &2.0&13.53&5.4 \\

{Core W}& & & & &2.0&13.53&5.4 \\
{Core Z}& & & & &1.5&10.15&2.9 \\
\hline

\end{tabular}

\end{table} 
\end{center}  
\end{landscape}

\clearpage
\normalsize
\subsection{HCN(3-2)}
\paragraph*{•}
The appropriate HCN(3-2) core to match the (1-0) core was determined in Chapter \ref{diskgeom}  by comparing x and y offsets from SgrA$^{*}$ of the (1-0) and (3-2) cores and selecting the closest match (see Table \ref{Coincores}). Peak brightness temperatures for the HCN (3-2) transition were obtained from visual estimates of the spectral values plotted in Fig.\ 1 of \citet{Jacks1993}, where the emission was plotted as Tb versus $\delta$V km s$^{-1}$. Line widths were obtained by measuring the spectral width at the half maximum T$_{b}$ level.  

\paragraph*{•}
 A core diameter of 0.25pc (6.45''), based on the average core diameter in Table
 2 of \citet{Chris2005}, was used to calculate a dilution factor of $32.7/163 =
 0.2$. The peak brightness temperature interpreted from the spectra were then
 divided by 0.2 to produce the ``resolved'' brightness temperatures used to
 plot the HCN(3-2) contours. Resolved (3-2) brightness temperatures are
 required to produce plots that are consistent with the resolved (1-0) and (4-3)
 data, where the cores fill the telescope's beam whereas the (3-2) data only fills
 0.2 of the beam (see Table \ref{scopes}). Additional brightness temperatures for
 Core D of 6.7\,K and 11.4\,K for Core O are based on the dilution factors for diameters of these two cores listed in Table 2 of \citet{Chris2005}.

\begin{center}
\begin{threeparttable}[ht]
%\begin{table}[!h] \begin{center} 
\caption{\label{Input2} Input Data from \citet{Jacks1993}}
\vspace{0.5cm}
\scriptsize

\begin{tabular}{c|c|c|c|c|c} \hline
{\citet{Jacks1993}}&{\citet{Chris2005}}&{Central Vel}&{Line Width}&{Observed}&{Resolved\tnote{a}} \\

{ }&{Equivalent Core }&km s$^{-1}$ &km s$^{-1}$ &{Peak Tb K}&Peak Tb K \\
\hline
{Core B}&{Core D\tnote{b}}&100&80&5&25 \\

{Core L}&{Core I}&-50&80&3&15 \\

{Core H}&{Core M}&-50&75&6&30 \\

{Core G}&{Core O\tnote{c}}&-25&90&5&25 \\

{Core F}&{Core P}&-40&75&4.75&23.7 \\
{Core D}&{Core W}&25&45&3.60&18 \\
{Core C}&{Core Z}&50&50&2.60&13 \\
\hline

\end{tabular}

\begin{tablenotes}
\item[a] based on a dilution factor of 0.2
\item[b] based on Core D diameter of 0.43pc dilution factor of 0.75 the resolved Tb is 6.7\,K
\item[c] based on Core O diameter of 0.33pc dilution factor of 0.44 the resolved Tb is 11.4\,K 
\end{tablenotes}
\end{threeparttable} 
\end{center}  
%\clearpage
\normalsize
\subsection{HCN (4-3)} \label{4-3}
\paragraph*{•} 
The appropriate cores corresponding to the (1-0) core were selected by comparing co-ordinates, from (Montero-Casta\~no private communication) with the HCN (1-0) core co-ordinates (see Table \ref{Coincores}). 
HCN (4-3) peak brightness temperatures were obtained from Fig.\ 7 of \citet{MMC2009} using the Rayleigh-Jeans relationship (see Eqn \ref{RJTb} with the line centre intensity,I$_{\nu}$(0) obtained by dividing the integrated intensity expressed as Jy beam$^{-1}$ by the beam area in steradians.

%\begin{equation}\label{MMCTb}
%Tb = I_{\nu}(0)\times 10^{-23} \lambda^{2}/2k
%\end{equation}

The core FWHM velocities in Table \ref{Input3} were obtained from Table 1 of \citet{MMC2009}.

\begin{landscape}
\begin{center} 
\begin{threeparttable}[ht]
\caption{\label{Input3} Input Data from \citet{MMC2009}}
\vspace{0.5cm}
\scriptsize
\begin{tabular}{c|c|c|c|c|c|c|c} \hline
\hline
\multicolumn{1}{c|}{\citet{MMC2009}}&\multicolumn{1}{|c|}{\citet{Chris2005}}&\multicolumn{1}{c|}{Core
Radius\tnote{a}}&\multicolumn{1}{c|}{Core Area}&\multicolumn{1}{c|}{Central
Vel\tnote{b}}&\multicolumn{1}{c|}{Line
Width\tnote{a}}&\multicolumn{1}{c|}{Int Emission\tnote{b}}&\multicolumn{1}{c}{Peak Tb\tnote{c}} \\
%\hline
\multicolumn{1}{c|}{ }&\multicolumn{1}{|c|}{ }&\multicolumn{1}{c|}{pc}&\multicolumn{1}{c|}{arcsecs$^{2}$}&\multicolumn{1}{c|}{km s$^{-1}$}&\multicolumn{1}{c|}{km s$^{-1}$}&\multicolumn{1}{c|}{Jy beam$^{-1}$}&\multicolumn{1}{c}{K} \\
\hline
\multicolumn{1}{c|}{Core A}&{Core D}&0.139&40.04&110&38.5&6.5&4.6 \\
\multicolumn{1}{c|}{Core AA}&{Core I}&0.145&43.9&-25&49.5&4.0&2.8 \\
%\hline
\multicolumn{1}{c|}{Core U}&{Core M}&0.110&25.3&-50&47.0&16&11.3 \\
%\hline
\multicolumn{1}{c|}{Core Q}&{Core O}&0.161&54.1&-90&51.0&14.2&10.0 \\
%\hline
\multicolumn{1}{c|}{Core N}&{Core P}&0.253&134&-40&97.0&11.0&7.8 \\
%\hline
\multicolumn{1}{c|}{Core F}&{Core W}&0.092&17.7&50&40.0&6.5&4.6 \\
\multicolumn{1}{c|}{Core E}&{Core Z}&0.098&20.1&40&55.0&8.3&5.9 \\
\hline
\hline
\end{tabular}
\begin{tablenotes}
\item[a] from \citet{MMC2009} Table 1
\item[b] estimated from spectra in \citet{MMC2009} Fig.\ 7 beam 4''.6 $\times$ 3''.0
\item[c] calculated from estimated integrated emission using Eqn \ref{RJTb}  
\end{tablenotes}
\end{threeparttable} 
\end{center}  
\end{landscape}
\normalsize

\paragraph*{•}
A kinetic gas temperature of 150\,K was again adopted for all seven cores. Additional modelling of Cores D and M was performed at a kinetic temperature of 50\,K to compare the effects on the hydrogen density value by reducing the kinetic temperature to the value used by \citet{Chris2005}.

\paragraph*{•}
Model output for the seven cores was plotted as contours of brightness temperatures (as listed in Tables \ref{Input1} to \ref{Input3}) in Figs.\ \ref{CoreD} to \ref{CoreW}.

\paragraph*{•}
\citet{Chris2005} quote a variation in dust temperatures from 20 to 80\,K and adopts T = 50\,K and an Av = 30 in their calculations compared with T = 75\,K and Av = 26.9, calculated using Eqn \ref{Av1}, and adopted for the current modelling. 75\,K was chosen as a mid value between dust temperatures of 50 to 90\,K quoted in Table 1 of \citet{Genzel1989}. Genzel mentions sub-millimetre dust emission detected by \citet{Mezger1989} which is interpreted as cold dust with a temperature of 10 to 20\,K. \citet{Mezger1989} quote \citet{Wade1987} finding a uniform  extinction of $\sim$ 27 towards the central 0$^{'}$.5 of the galactic centre.

\paragraph*{•}
Core O is modelled with and without dust to show that Molex was insensitive to dust emission at the HCN transition frequencies being analysed. Table \ref{Avval} confirms that these are very small differences, generally less than 0.1$\%$, between model outputs for the two cases, this is because the radiation effects are highly diminished at wavelengths $>$ 1000$\mu$m. For example the optical depth of the dust at 846 $\mu$m) (the wavelength of HCN(4-3)) is 4.76 $\times$ 10$^{-4}$ produces 0.03\,K (T$_{b}$ (dust) = $\tau_{\mathrm{dust}} \times$ T(dust)) as the brightness temperature for dust with a kinetic temperature of 75\,K. Comparison of Figs.\ \ref{CoreO} and \ref{CoreOa} shows that the differences are undetectable in the plots of brightness temperatures and optical depths.

\subsection{Results} \label{Chris}
\paragraph*{•}
The results for these seven cores were obtained by averaging the values of intersection points of pairs of species in Figs.\ \ref{CoreD} to \ref{CoreZ} and are summarised in Table \ref{Coincprops}.
The molecular hydrogen densities range from n$_{H\small{2}}$ \normalsize = 0.79 (Core I) to 1.58 (Core M) $\times10^{6}$cm$^{-3}$ for a kinetic temperature of T = 150\,K; these are less than a quarter to a half of the values for the optically thin scenario of \citet{Chris2005} for the HCN(1-0) transition. Modelling Cores D and M with a kinetic temperature of T - 50\,K produces a value of n$_{H\small{2}}$  \normalsize = 3.97$\times10^{6}$cm$^{-3}$  see Table \ref{Coincprops}). 

\paragraph*{•} 
Table \ref{Coincprops} summarises the results for optical depth for the seven cores where the HCN(1-0) and HCO$^{+}$(1-0) transitions indicate the presence of stimulated emission with negative excitation temperatures and opacities. The opacities  range from 1.11 to 1.96 for HCN(3-2) and 1.0 to 2.57 for HCN(4-3). 

\paragraph*{•}
The HCN column and atomic hydrogen number density values were obtained from averaging the values at the species intersection points for the brightness temperatures of HCN(1-0) with the other transitions. For example in Fig.\ \ref{CoreW} describing Core W the intersection of HCN(1-0), HCN(4-3) and HCO$^{+}$(1-0) was the lower point at logN$_{\mathrm{col}}$/$\delta$V = 13.2 and logn$_{\mathrm{h}}$ = 6.1, with the intersection of HCN(1-0) and (3-2) the upper point at logN$_{\mathrm{col}}$/$\delta$V = 13.4 and logn$_{\mathrm{H}}$ = 6.6. The average values of log Column density/$\delta$V = 13.3 and logn$_{\mathrm{H}}$ = 6.25 were the result of averaging the values of these two intersection points. It should be noted that the lower limit curve of 12.6\,K for the (3-2) transition provides a tighter intersection envelope and slightly lower values for both  HCN column and hydrogen number densities.  

\subsection{Comments} \label{Chrisres}
\paragraph*{•}
The seven HCN cores provide a good representative sample of CND conditions as they are spread throughout the ring. Core D is in the Northeast, Core I is in the East, Core M is in the South, Cores O \& P are in the Southwest and Cores W \& Z are in the North (see Fig.\ \ref{Cocores}).

\paragraph*{•}
Using the 0.2 dilution factor for the HCN(3-2) transition results in peak brightness temperature contours that have a spread of intersection points with the other HCN and HCO$^{+}$ contours; these lead to upper and lower limits for the both hydrogen number and HCN column densities.

\paragraph*{•}
The accuracy of the HCN(3-2) peak brightness temperatures is stated by \citet{Jacks1993} to be $\pm 30\%$ due to the relatively large beam size of about 12'', compared to the average core size of 6.5'', this also contributes to the large line widths for this transition compared to the other transitions of HCN and HCO$^{+}$ for this group of cores (see column 4 of Table \ref{Input2}). The effect of the large error spread can be seen in the plot of brightness temperatures for Core W in Fig.\ \ref{CoreW}, where the lower brightness temperature contour more closely fits the contours of the other transitions than the average HCN(3-2) brightness temperature.  

\begin{figure}[ht] \includegraphics[scale=0.75,bb=  0 0 580 770] 
{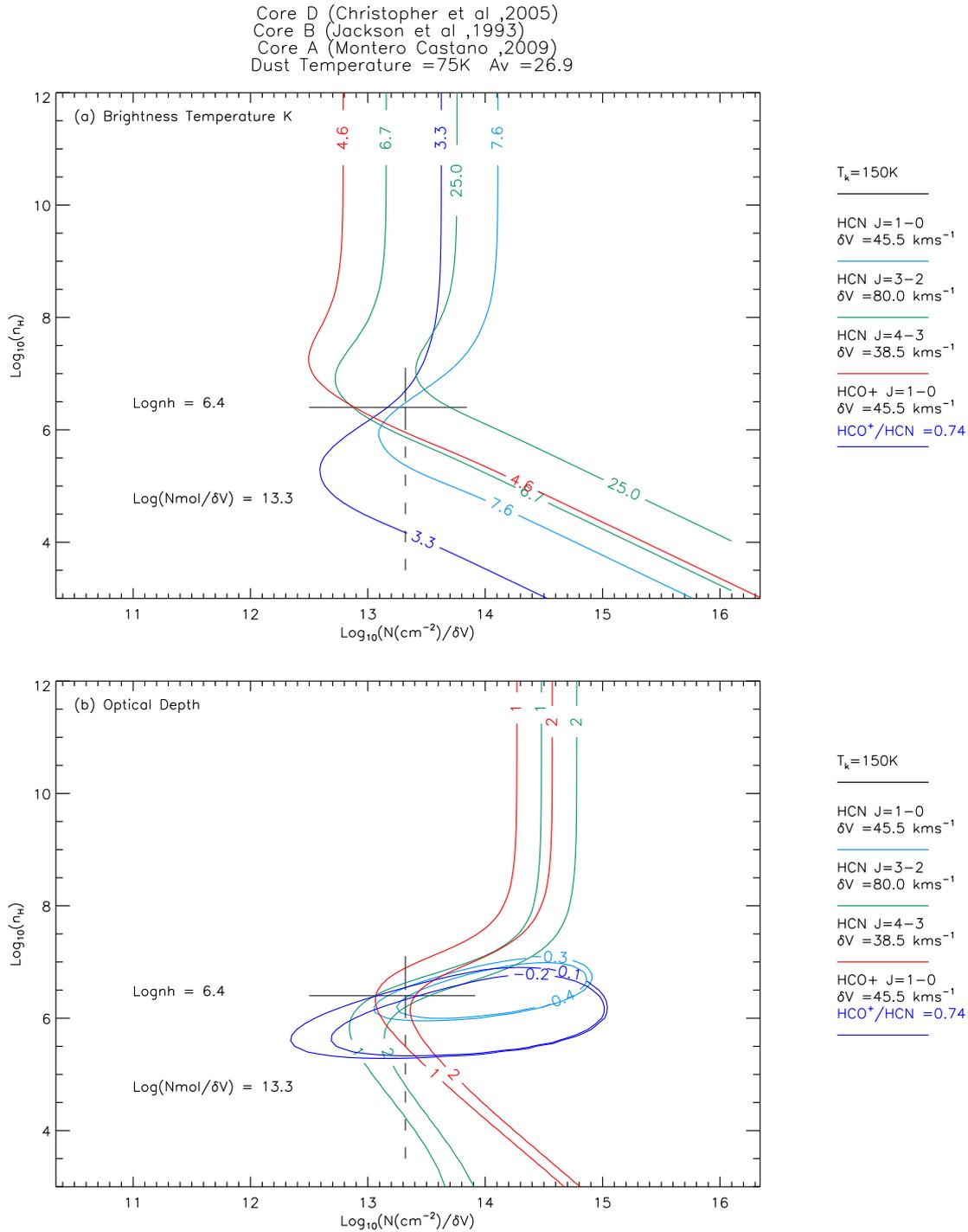} \fontsize{9} {9} \vspace{-1cm} \caption[Core D Tb and Optical Depth Plots]{\label{CoreD} Core D with kinetic temperature of 150\,K.  Panel (a) shows  brightness temperature contours for HCN (1-0), (3-2), (4-3) and HCO$^{+}$ (1-0). The average intersection point of the species is log(n$_{H}$) = 6.4, log(NMol/dV) = 13.3.
Panel (b) shows the optical depth contours for the tracers which are optically thin and shows inverted populations of HCN(1-0) and HCO$^{+}$(1-0).}
\end{figure}

\begin{figure}[ht] \includegraphics[scale=0.75,bb=  0 0 580 770] 
{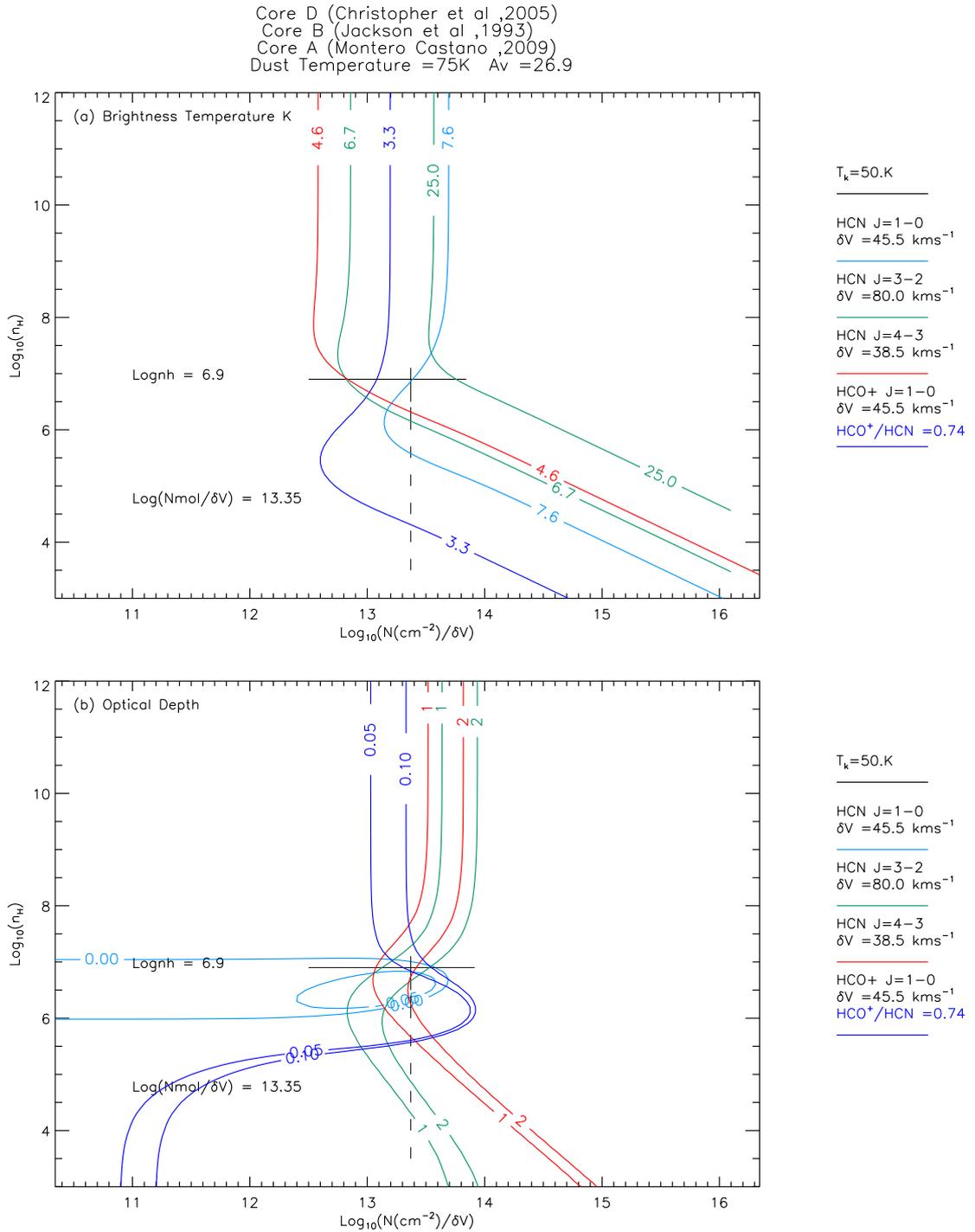} \fontsize{9} {9} \vspace{-1cm} \caption[Core D Tb and Optical Depth Plots for 50\,K]{\label{CoreD50} As for Fig.\ \ref{CoreD} but with the kinetic temperature of I = 50\,K used by \citet{Chris2005}.} 
\end{figure}

\begin{figure}[ht] \includegraphics[scale=0.75,bb=  0 0 580 770] 
{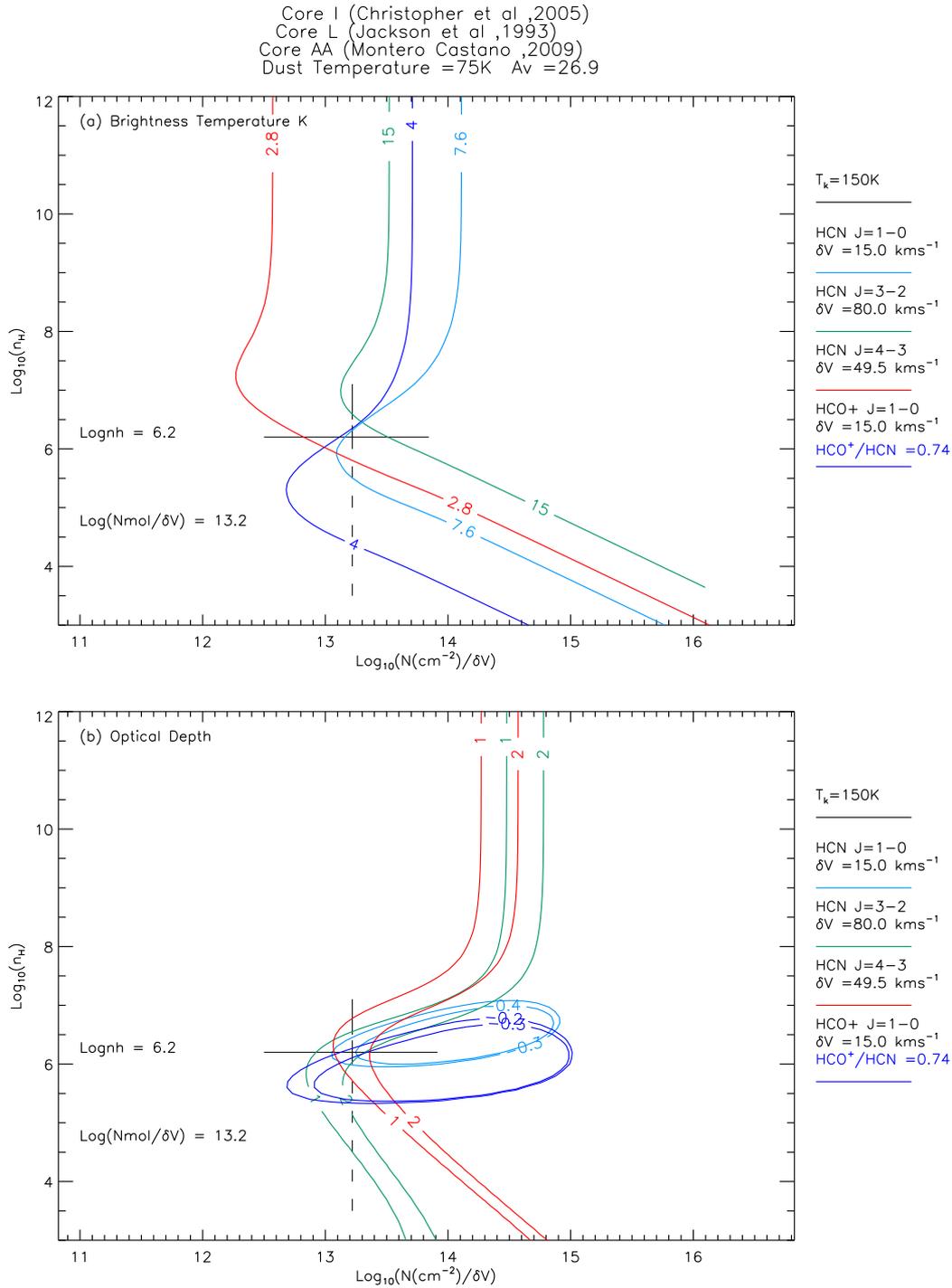} \fontsize{9} {9} \vspace{-1cm} \caption[Core I Tb and Optical Depth Plots]{\label{CoreI} As for Fig.\ \ref{CoreD} but for Core I.} 
\end{figure}

\begin{figure}[ht] \includegraphics[scale=0.75,bb=  0 0 580 770] 
{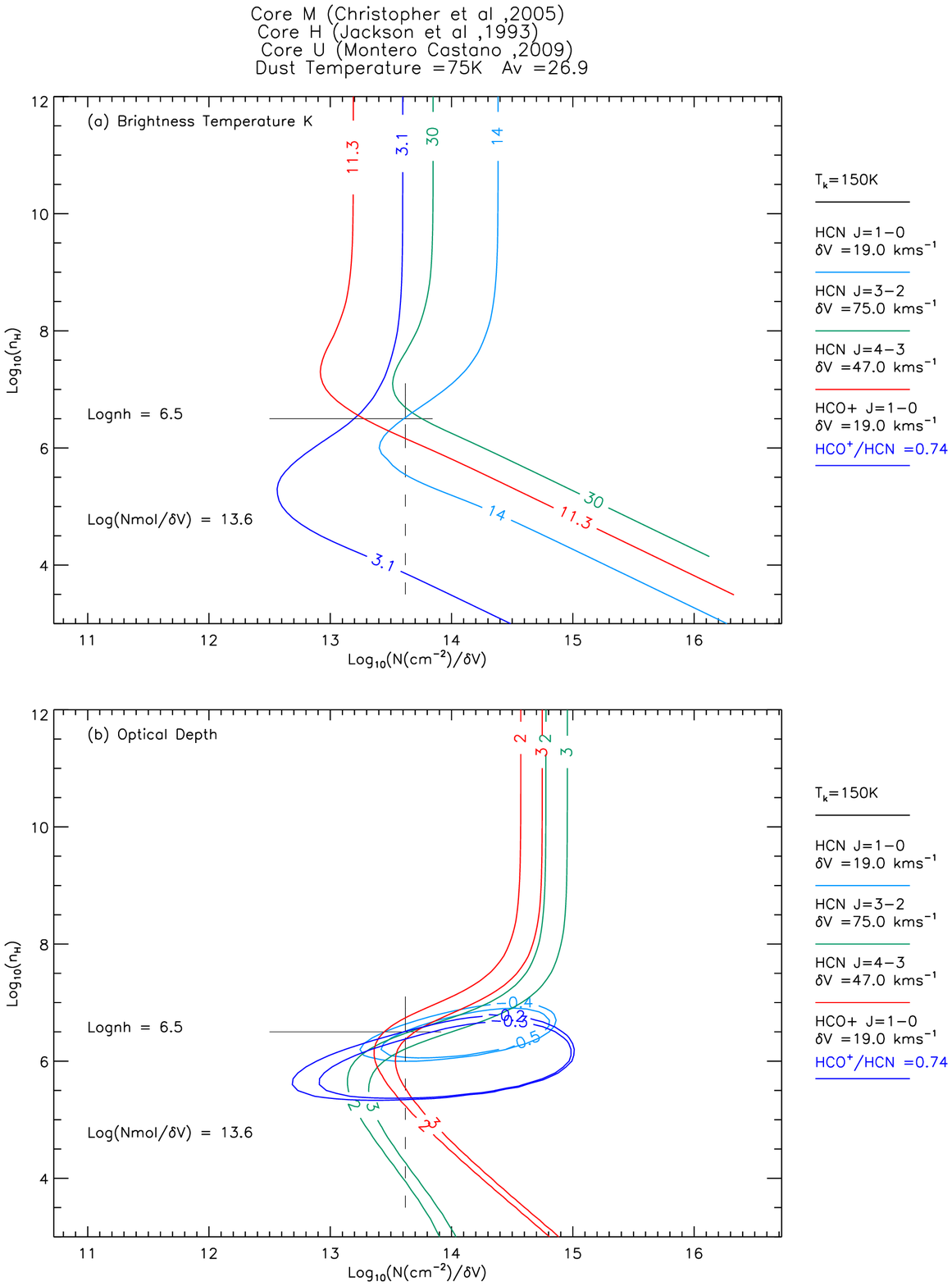} \fontsize{9} {9} \vspace{-1cm} \caption[Core M Tb and Optical Depth Plots]{\label{CoreM} As for Fig.\ \ref{CoreD} but for Core M at kinetic temperature of T = 150\,K. 
} 
\end{figure}

\begin{figure}[ht] \includegraphics[scale=0.75,bb=  0 0 580 770] 
{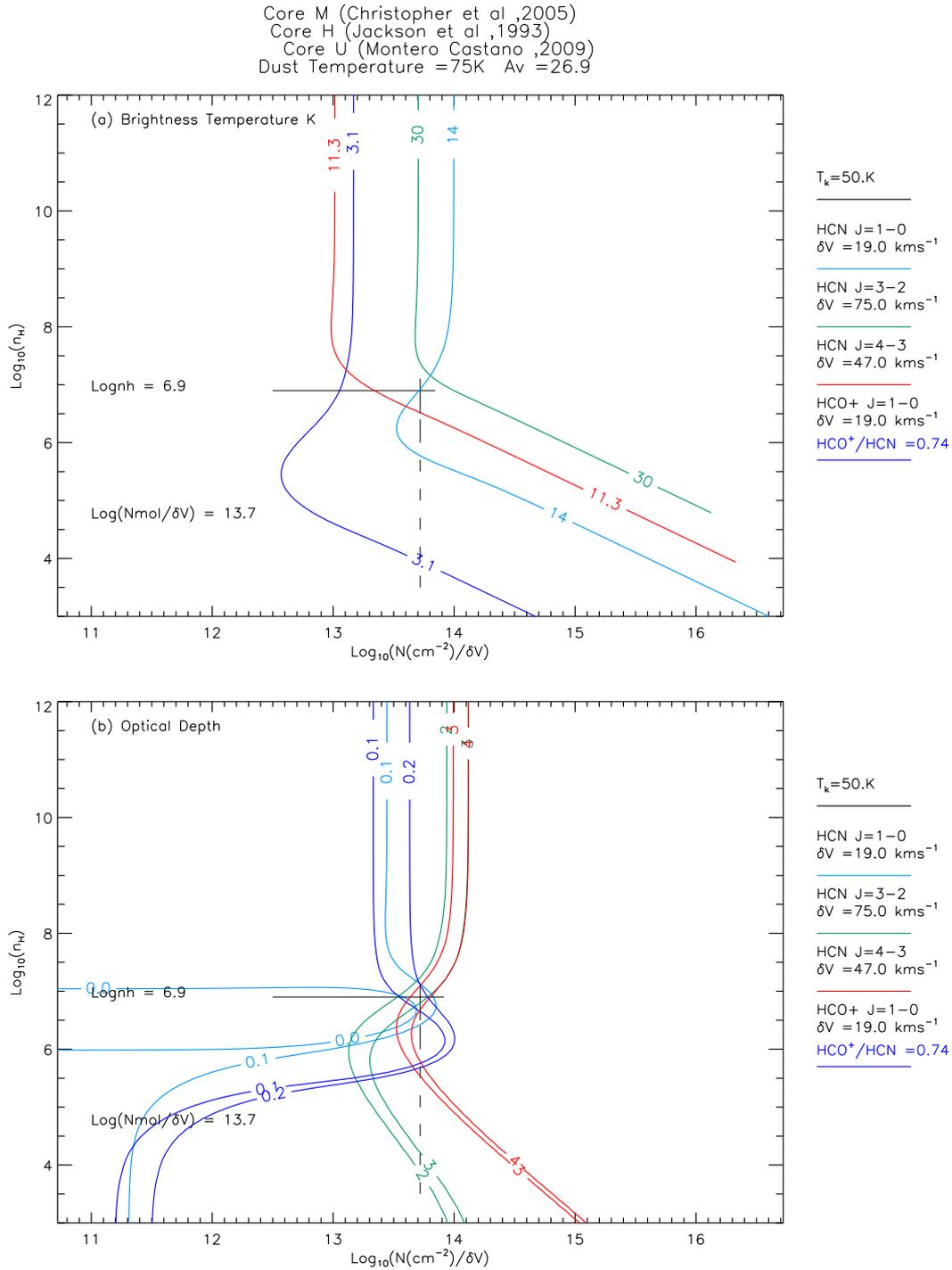} \fontsize{9} {9} \vspace{-1cm} \caption[Core M Tb and Optical Depth Plots]{\label{CoreM50}  As for Fig.\ \ref{CoreM} but with a kinetic temperature of T = 50\,K.} 
\end{figure}

\begin{figure}[ht] \includegraphics[scale=0.75,bb=  0 0 580 770] 
{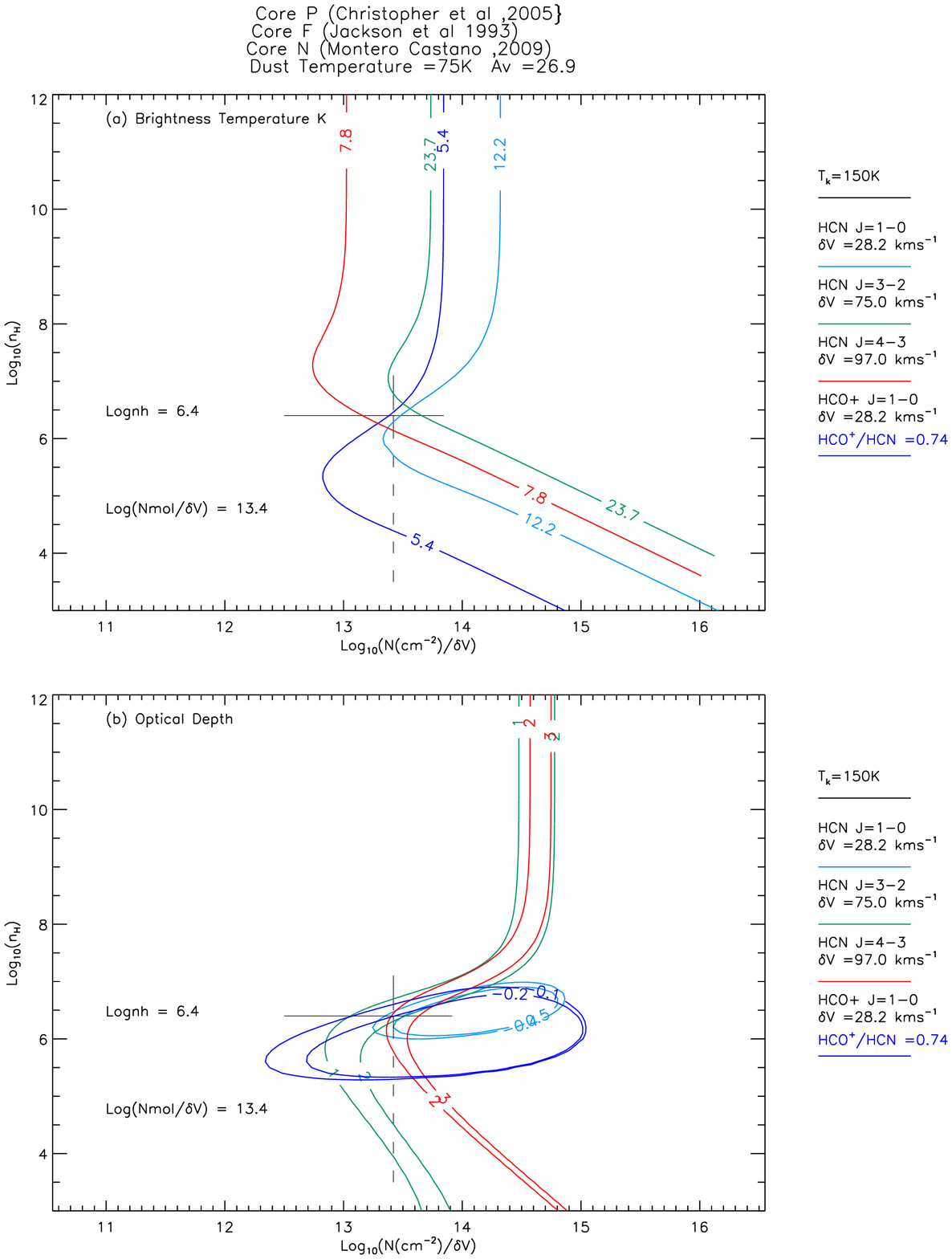} \fontsize{9} {9} \vspace{-1cm} \caption[Core P Tb and Optical Depth Plots]{\label{CoreP} As for Fig.\ \ref {CoreD} but for Core P.} 
\end{figure}

\begin{figure}[ht] \includegraphics[scale=0.75,bb=  0 0 580 770] 
{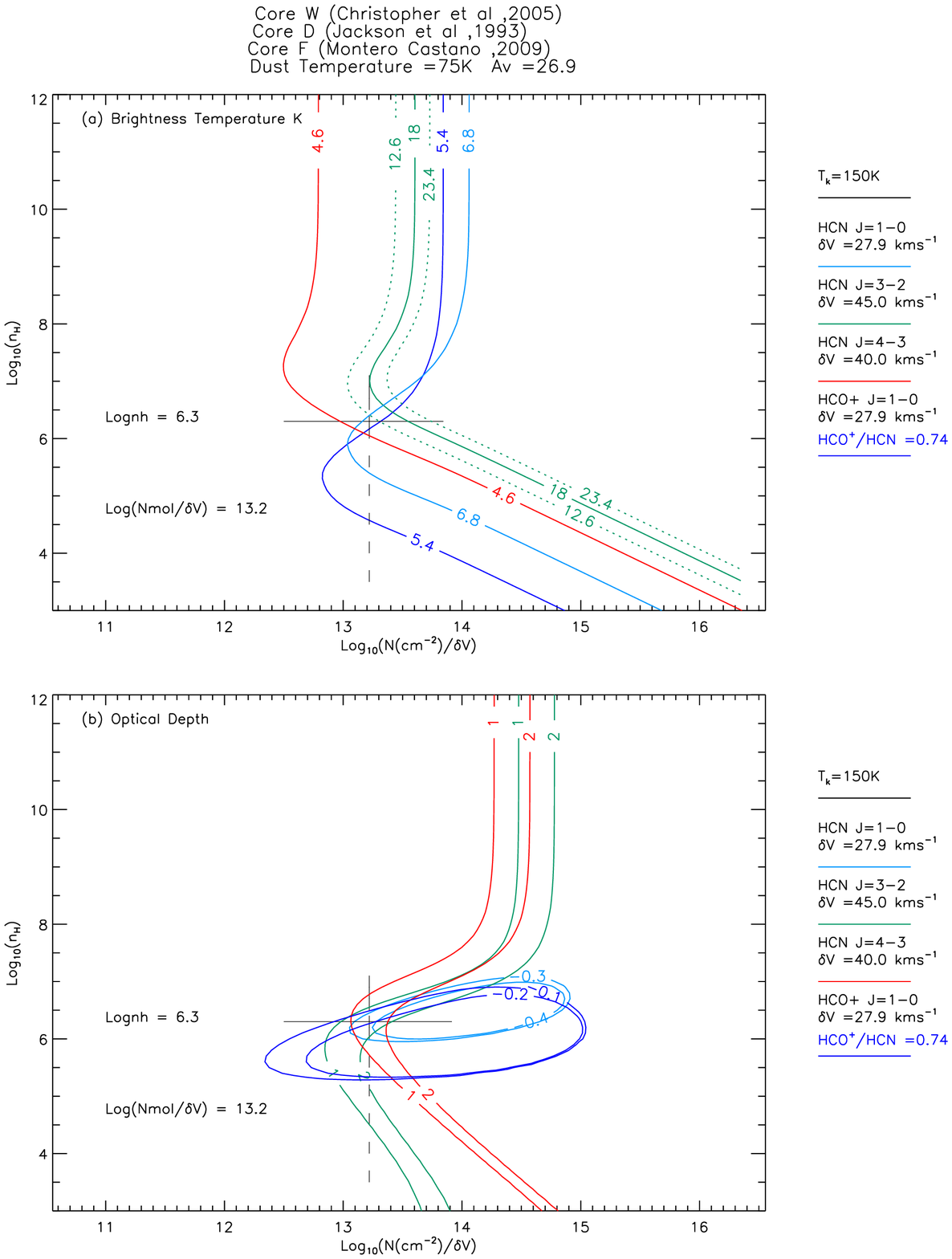} \fontsize{9} {9} \vspace{-1cm} \caption[Core W Tb and Optical Depth Plots]{\label{CoreW} As for Fig.\ \ref{CoreD} but for Core W.} 
\end{figure}

\begin{figure}[ht] \includegraphics[scale=0.75,bb=  0 0 580 770] 
{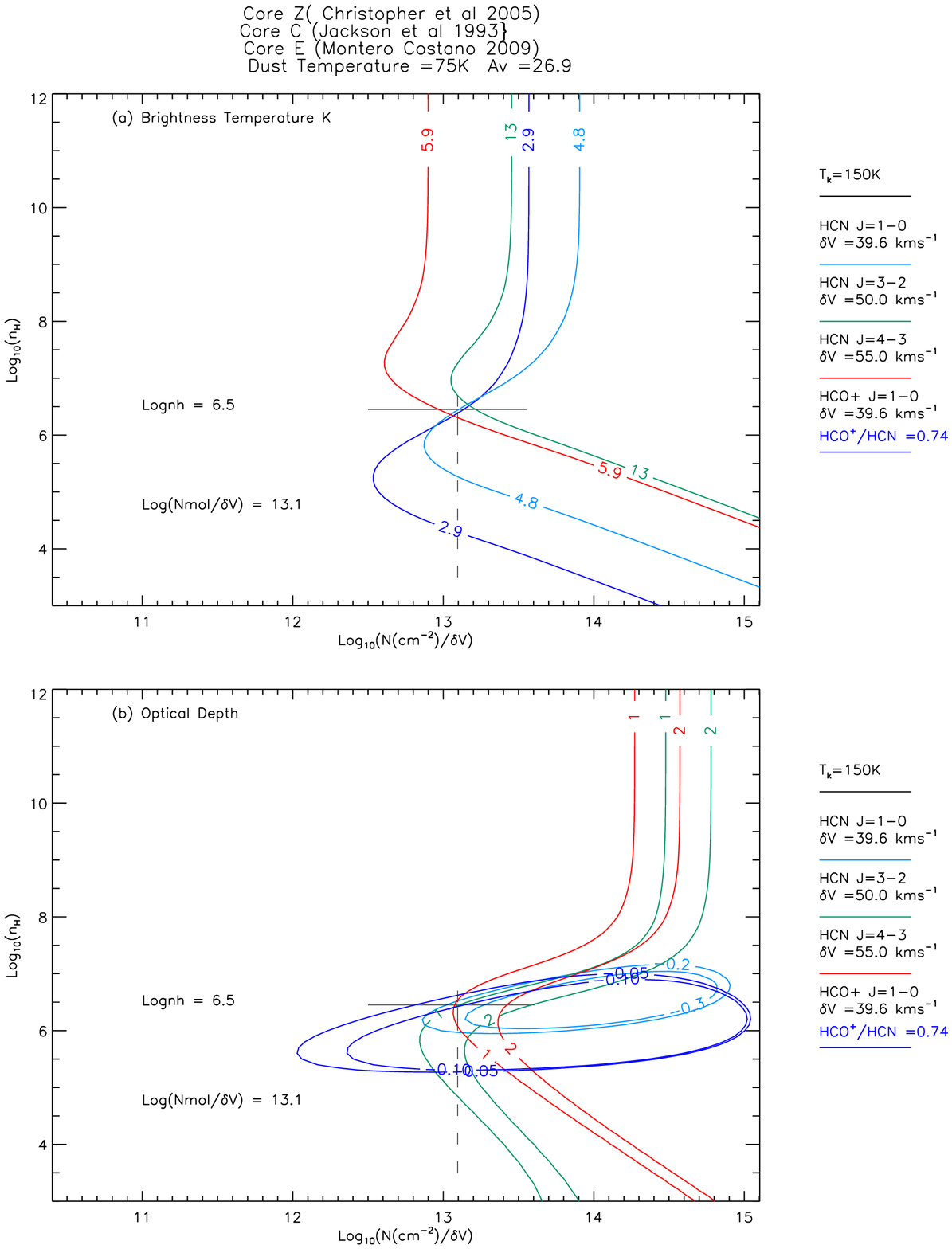} \fontsize{9} {9} \vspace{-1cm} \caption[Core Z Tb and Optical Depth Plots]{\label{CoreZ} As for Fig.\ \ref{CoreD} but for Core Z.} 
\end{figure}

\paragraph*{•}
For Core D Fig.\ \ref{CoreD} (a) shows contours for three values of the $ [\mathrm{HCO^{+}}]/[\mathrm{HCN}] $ abundance ratios, Z = 0.4, 0.74  and 1.2, (reported by \citet{Chris2005} as occurring in the CND). Lowering the abundance ratio reduces the column density per line width value. Again in modelling the second group of HCN cores the mean HCO$^{+}$ abundance value of 0.74 was used in all plots, in contrast to the value of 0.4 used by \citet{Chris2005} to outline the regions of relatively high HCN emission and low HCO$^{+}$ emission.

\paragraph*{•}
The molecular hydrogen number density for the seven HCN cores varies from 0.95 (Core I) to 1.9 (Core M) $\times10^{6}$cm$^{-3}$, which is an order of magnitude lower than the value n$_{H_{2}}$ = 10$^{7}$cm$^{-3}$ reported by \citet{Chris2005} for HCN(1-0) gas with $\large{\tau}$ \normalsize = 4 . The model results support an optically thin model as they are even lower than the results for the optically thin gas scenario reported in \citet{Chris2005} (see Table \ref{ChrisTab2}).

\paragraph*{•}
Table \ref{ChrisTab2} provides a comparison between the three core density scenarios (virial, optically thin \large{$\tau$} \normalsize $_{HCN(1-0)}$ $\leq$ 1 and optically thick \large{$\tau$} \normalsize $_{HCN(1-0)}$ = 4) discussed in  \citet{Chris2005} and those obtained from modelling the HCN cores identified in this thesis.  

\begin{figure}[ht] \includegraphics[scale=0.75,bb=  0 0 580 770] 
{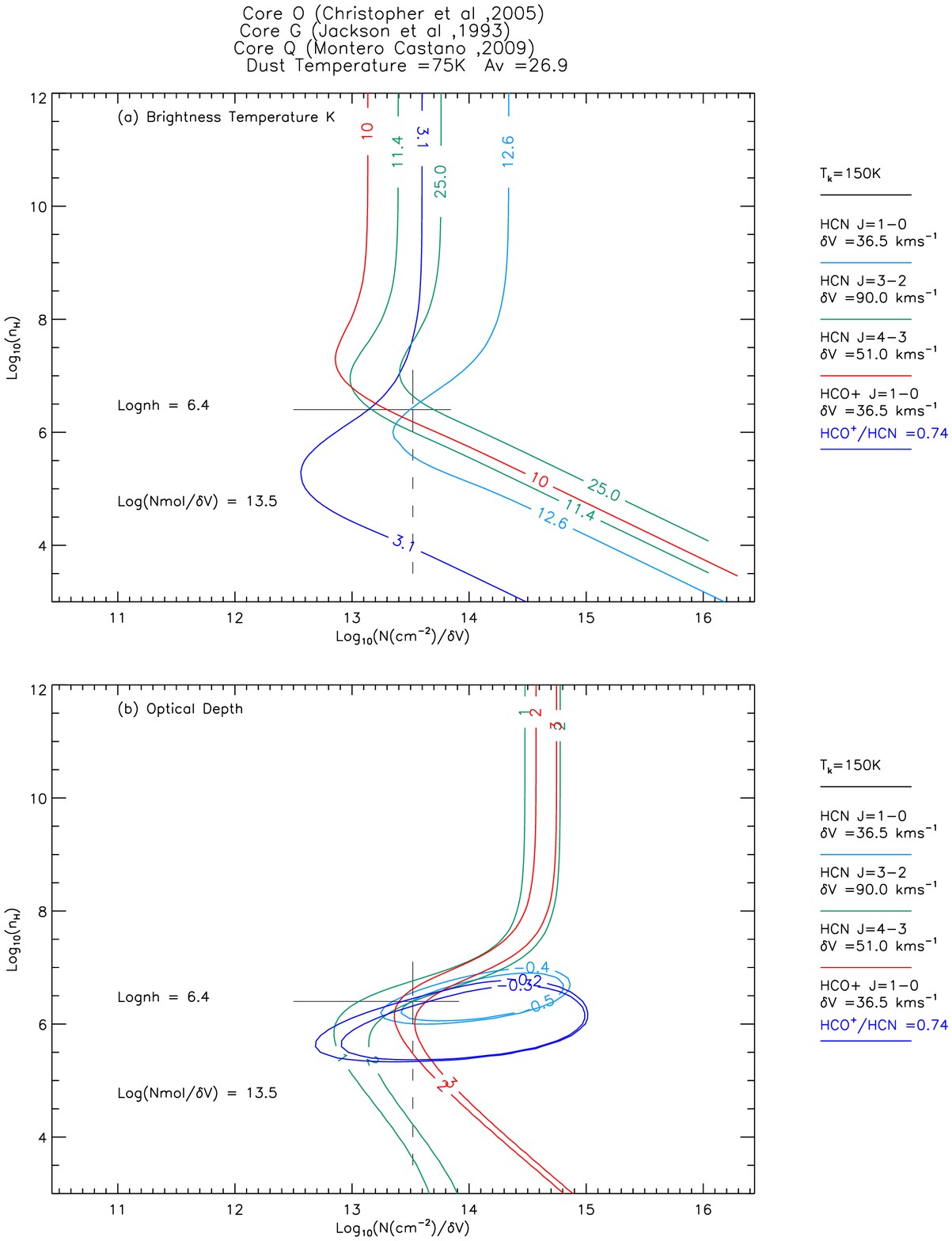} \fontsize{9} {9} \vspace{-1cm} \caption[Core O Tb and Optical Depth Plots for $Av=26.9$]{\label{CoreO} As for Fig.\ \ref{CoreD} but for Core O.} 
\end{figure}

\begin{figure}[ht] \includegraphics[scale=0.75,bb=  0 0 580 770] 
{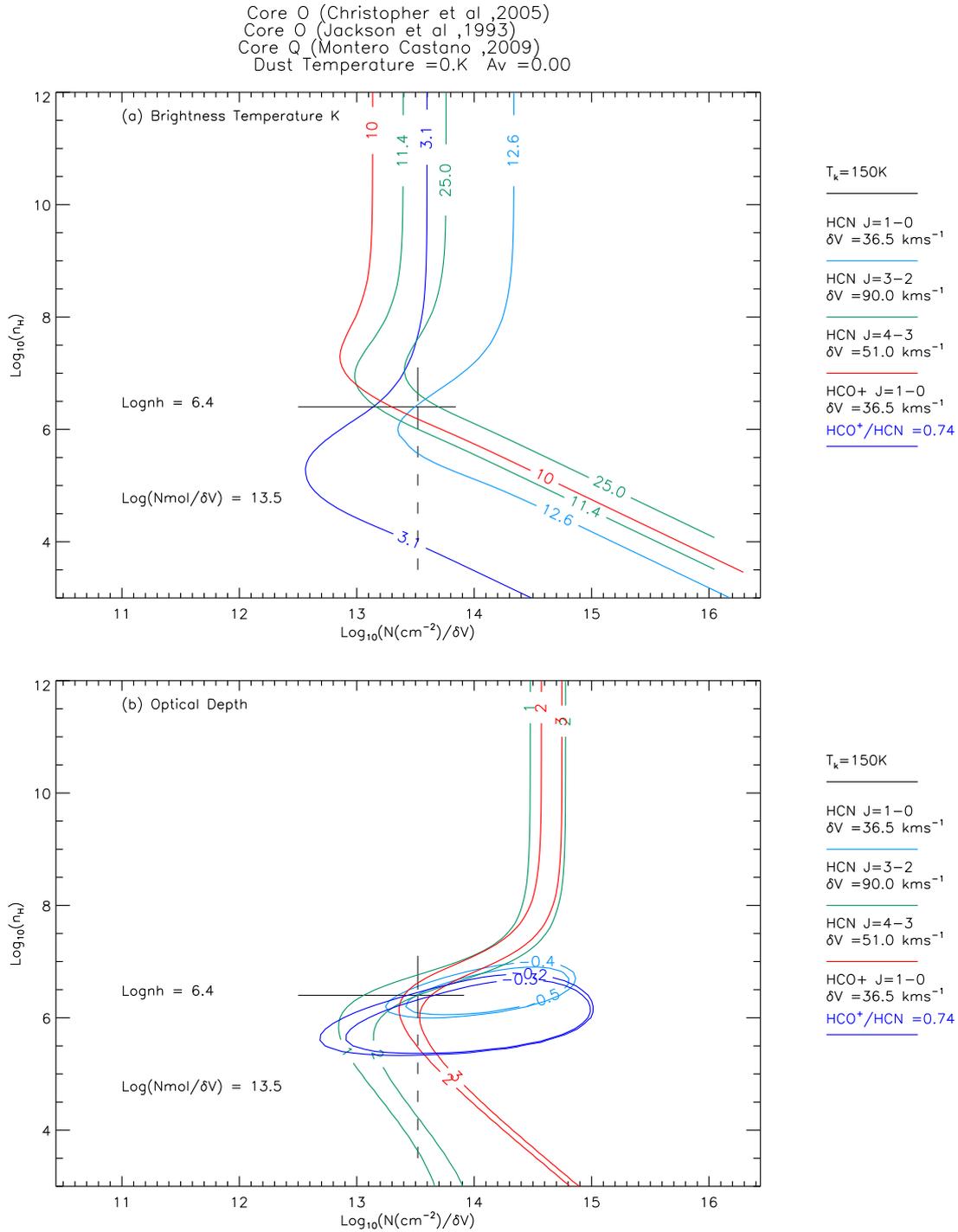} \fontsize{9} {9} \vspace{-1cm} \caption[Core O Tb and Optical Depth Plots for Td=0 $Av=0$]{\label{CoreOa} As for Fig.\ \ref{CoreO} but with no dust.} 
\end{figure}

\paragraph*{•}
Cores D and M were also modelled using a kinetic temperature of 50\,K, which  produced molecular hydrogen densities for both of n$_{\mathrm{H}_{2}}$ = 3.97$\times10^{6}$ cm$^{-3}$, which agrees closely with the \citet{Chris2005} values under optically thin conditions. The brightness temperature value of the HCN(3-2) transition contours in Cores D and O was varied to account for the different dilution factor when using the core sizes listed in Table \ref{ChrisTab2} instead of the average 0.25 pc size adopted for calculating the 0.2 dilution factor in Table \ref{Input2}. Figs.\ \ref{CoreD} and \ref{CoreO} show that the HCN(3-2) brightness contour shifts to the left as the brightness temperature decreases, so that it forms a tighter intersection area with the other HCN transitions. The resulting molecular hydrogen number density and HCN column density per line width are slightly lower than the tabulated values in Table \ref{Coincprops}   e.g.\ log(n$_{H}$) = 6.3 and log(N$_{\mathrm{mol}}$/dV) = 13.2 compared with log(n$_{H}$) = 6.4 and log(N$_{\mathrm{mol}}$/dV) = 13.5 for Core O.  

%\clearpage

\begin{landscape}
\begin{center} 
\begin{threeparttable}[ht]
\vspace{0.5cm}
\small
\scriptsize

\caption{\label{Coincprops} HCN Core Properties.}
%\begin{tabular}{c|d{2}|d{-1}|d{-1}|d{-1}|d{2}|d{2}|d{-1}|c|c|c|c|c|c|c|c} \hline
\begin{tabular}{c|d{2}|d{2}|d{2}|d{2}|d{2}|d{2}|d{2}|c|c|c|c|c|c|c|c} \hline
%\hline
%\hline
\multicolumn{1}{c|}{Group 2 Cores}&\multicolumn{1}{c|}{Kinetic}&\multicolumn{1}{c|}{Average Log}&\multicolumn{1}{c|}{n(H$_{2}$)}&\multicolumn{4}{c|}{Tex K}&\multicolumn{4}{c|}{Optical Depth $\tau$}&\multicolumn{4}{c}{Peak Brightness Temp Tb K} \\
%\hline
\multicolumn{1}{c|}{ }&\multicolumn{1}{c|}{Temp}&\multicolumn{1}{c|}{Col Density}&\multicolumn{1}{c|}{ }&\multicolumn{1}{c|}{HCN}&\multicolumn{1}{c|}{ HCN}&\multicolumn{1}{c|}{HCN}&\multicolumn{1}{c|}{HCO$^{+}$}&\multicolumn{1}{c|}{HCN}&\multicolumn{1}{c|}{HCN}&\multicolumn{1}{c|}{HCN}&\multicolumn{1}{c|}{HCO$^{+}$}&\multicolumn{1}{c|}{HCN}&\multicolumn{1}{c|}{HCN}&\multicolumn{1}{c|}{HCN}&\multicolumn{1}{c}{HCO$^{+}$} \\
%\multicolumn{1}{c|}{H$^{13}$CN}&\multicolumn{1}{c|}{HCO$^{+}$}&\multicolumn{1}{c|}(HCN}&\multicolumn{1}{c|}{H$^{12}$CN}&\multicolumn{1}{c|}{H$^{13}$CN}&\multicolumn{1}{c|}{HCO$^{+}$}&\multicolumn{1}{c|}(HCN}&\multicolumn{1}{c|}{H$^{12}$CN}&\multicolumn{1}{c|}{H$^{13}$CN}&\multicolumn{1}{c|}{HCO$^{+}$} &\multicolumn{1}{c|}(HCN} \\
%\hline
\multicolumn{1}{c|}{ }&\multicolumn{1}{c|}{ }&\multicolumn{1}{c|}{per line}&\multicolumn{1}{c|}{ }&\multicolumn{1}{c|}{{1-0}}&\multicolumn{1}{c|}{(3-2)}&\multicolumn{1}{c|}{(4-3)}&\multicolumn{1}{c|}{(1-0)}&\multicolumn{1}{c|}{(1-0)}&\multicolumn{1}{c|}{(3-2)}&\multicolumn{1}{c|}{(4-3)}&\multicolumn{1}{c|}{(1-0)}&\multicolumn{1}{c|}{(1-0)}&\multicolumn{1}{c|}{(3-2)}&\multicolumn{1}{c|}{(4-3)}&\multicolumn{1}{c}{(1-0)} \\
%\hline
\multicolumn{1}{c|}{ }&\multicolumn{1}{c|}{(K)}&\multicolumn{1}{c|}{(cm$^{-2}$/km  s$^{-1}$)}&\multicolumn{1}{c|}{(10$^{6}\times$cm$^{-3}$)}&\multicolumn{1}{c|}{(K)}&\multicolumn{1}{c|}{(K)}&\multicolumn{1}{c|}{(K)}&\multicolumn{1}{c|}{(K)}&\multicolumn{1}{c|}{ }&\multicolumn{1}{c|}{ }&\multicolumn{1}{c|}{ }&\multicolumn{1}{c|}{ }&\multicolumn{1}{c|}{(K)}&\multicolumn{1}{c|}{K}&\multicolumn{1}{c|}{(K)}&\multicolumn{1}{c}{(K)} \\
\hline
{D}&150&13.3&1.26&-21.9&44.0&13.1&-23.0&-0.30&1.11&1.06&-0.15&8.9&25.30&4.14&4.18 \\
{D}&50&13.4&3.97&-353.0&38.7&13.2&749.0&-0.04&1.25&1.45&0.01&8.45&23.2&4.9&4.49 \\
{I}&150&13.2&0.79&-15.8&28.6&11.8&-29.20&-0.33&1.29&1.66&-0.21&7.43&16.3&4.14&3.51 \\
{M}&150&13.6&1.58&-21.7&45.7&21.8&-28.7&-0.52&1.94&2.57&-0.08&17.4&32.8&12.3&2.78 \\
{M}&50&13.7&3.97&336.0&38.1&18.9&88.5&0.04&2.70&3.50&0.07&14.2&29.7&11.3&5.67 \\
{O}&150&13.5&1.26&-18.5&36.1&18.2&-28.7&-0.52&1.96&2.30&-0.09&15.0&25.8&9.8&2.89 \\
{P}&150&13.5&1.0&-18.6&35.8&16.8&-34.2&-0.53&1.93&2.23&-0.27&15.4&25.3&8.9&11.6 \\
{W}&150&13.2&1.0&-17.2&19.4&14.2&-13.4&-0.24&1.80&0.95&-0.29&6.14&11.0&4.46&5.55 \\
{Z}&150&13.1&1.58&-14.8&31.0&17.6&-21.9&-0.25&0.9&1.0&-0.1&5.1&14.6&6.5&3.1 \\
\hline
%\hline
\end{tabular}
\end{threeparttable}
\end{center}   
%\end{landscape}

%\normalsize
%\clearpage
\begin{center}
%\begin{landscape}
\begin{table}[ht] %\begin{center} 
\caption{\label{Avval} Effects of Variations in Dust Extinction for Core O.}
\vspace{0.5cm}
%\scriptsize
%\small
\scriptsize
\begin{tabular}{c|d{1}|d{1}|c|d{1}|d{3}|d{4}|d{3}|d{3}|c|d{4.2}|d{4}|d{5}} \hline

%\begin{tabular}{c|d{1}|d{-1}|c|d{-1}|d{2}|d{2}|d{2}|d{2}|c|d{4.2}|d{4}|d{5}} \hline
%\hline
\multicolumn{1}{c|}{Tracer}&\multicolumn{1}{c|}{logn$_{\mathrm{H}}$}& \multicolumn{1}{c|}{logN$_{\mathrm{mo}l}$}&\multicolumn{1}{c|}{Td}&\multicolumn{1}{c|}{Av}&\multicolumn{1}{c|}{Tex }&\multicolumn{1}{c|}{Optical}&\multicolumn{1}{c|}{Tb}&\multicolumn{1}{c|}{$\beta$}&\multicolumn{1}{c|}{Intensity}&\multicolumn{1}{c|}{Int Intensity} &\multicolumn{1}{c|}{Upper}&\multicolumn{1}{c}{Lower}  \\ %\hline
%\multicolumn{1}{|c|}{Tracer}&\multicolumn{1}{c|}{logn${H}$}&\muticolumn{1}{c|{logNmol}&\multicolumn{1}{c|}{Av}&\multicolumn{1}{c|}{Tex K}&\multicolumn{1}{c|}{Tau}&\multicolumn{1}{c|}{Tb}&\multicolumn{1}{c|}{Beta}&\multicolumn{1}{c|}{Intensity}&\multicolumn{1}{c|}{Int Itensity}&\multicolumn{1}{c|}{x(u)}&\multicolumn{1}{c|}{x(l)}\\
%\hline
\multicolumn{1}{c|}{ }&\multicolumn{1}{c|}{cm$^{-3}$}&\multicolumn{1}{c|}{(cm$^{-2}$km s$^{-1}$)}&\multicolumn{1}{c|}{(K)}&\multicolumn{1}{c|}{ }&\multicolumn{1}{c|}{(K)}&\multicolumn{1}{c|}{Depth}&\multicolumn{1}{c|}{(K)}&\multicolumn{1}{c|}{ }&\multicolumn{1}{c|}{(erg cm$^{-2}$ s$^{-1}$ sr$^{-1}$)}&\multicolumn{1}{c|}{(K km s$^{-1}$)}&\multicolumn{1}{c|}{level}&\multicolumn{1}{c}{level} \\ 
\hline
{HCN (1-0)}&6.4&13.5&0&0.0&-20.1&-0.44&13.0&1.19&3.15E-14&504.0&0.193&0.052\\
%\hline
 &6.4&13.5&75&26.9&-20.1&-0.44&13.0&1.19&3.15E-14&504.0&0.193&0.052\\
%\hline
{\% Difference}& & & &&0.03&-0.09&-0.07&-0.02&-0.07&-0.07&-0.06&-0.06 \\
\hline
%\multicolumn{1}{c|}{ }&\multicolumn{1}{c|}{ }&\multicolumn{1}{c|}{ }&\multicolumn{1}{c|}{ }&\multicolumn{1}{c|}{ }&\multicolumn{1}{c|}{ }&\multicolumn{1}{c|}{ }&\multicolumn{1}{c|}{ }&\multicolumn{1}{c|}{ }&\multicolumn{1}{c|}{ }&\multicolumn{1}{c|}{ }&\multicolumn{1}{c}{ } \\
\hline
{HCO+ (1-0)}&6.4&13.5&0&0.0&-23.9&-0.22&6.62&1.09&1.62E-14&256.0&0.101&0.028\\
%\hline
 &6.4&13.5&75&26.9&-23.9&-0.22&6.62&1.09&1.62E-14&256.0&0.101&0.028\\
%\hline
{\% Difference}& & & &0.04&-0.08&-0.04&0&-0.04&-0.04&-0.04&-0.03 \\
\hline
%\multicolumn{1}{c|}{ }&\multicolumn{1}{c|}{ }&\multicolumn{1}{c|}{ }&\multicolumn{1}{c|}{ }&\multicolumn{1}{c|}{ }&\multicolumn{1}{c|}{ }&\multicolumn{1}{c|}{ }&\multicolumn{1}{c|}{ }&\multicolumn{1}{c|}{ }&\multicolumn{1}{c|}{ }&\multicolumn{1}{c|}{ }&\multicolumn{1}{c}{ } \\
\hline
{HCN (3-2)}&6.4&13.5&0&0.0&29.4&2.06&20.4&0.518&4.42E-13&1940.0&0.241&0.266\\
%\hline
 &6.4&13.5&75&26.9&29.4&2.06&20.3&0.518&4.42E-13&1940.0&0.241&0.266\\
%\hline
{\% Difference}& & & &0.0&-0.07&0.09&0.01&-0.09&-0.09&-0.06&-0.06 \\
\hline
%\multicolumn{1}{c|}{ }&\multicolumn{1}{c|}{ }&\multicolumn{1}{c|}{ }&\multicolumn{1}{c|}{ }&\multicolumn{1}{c|}{ }&\multicolumn{1}{c|}{ }&\multicolumn{1}{c|}{ }&\multicolumn{1}{c|}{ }&\multicolumn{1}{c|}{ }&\multicolumn{1}{c|}{ }&\multicolumn{1}{c|}{ }&\multicolumn{1}{c}{ } \\
\hline
{HCN (4-3)}&6.4&13.5&0&0.0&22.7&2.36&13.7&0.48&5.31E-13&743.0&0.146&0.241\\
%\hline
 &6.4&13.5&75&26.9&22.7&2.36&13.7&0.48&5.31E-13&743.0&0.146&0.241\\
%\hline
{\% Difference}& & & &0.10&-0.13&-0.09&0.08&-0.08&-0.08&0.03&-0.05 \\
\hline
%\hline
\end{tabular}
\end{table} 
\end{center}
\end{landscape}
\normalsize
%\newpage 

\afterpage
\clearpage
\begin{landscape}
\begin{center}
\begin{table}[ht]  
\caption{\label{Radexcalcs} Radex Calculations vs Molex for Core O.}

%\vspace{0.5cm}
%\small
\scriptsize
\begin{tabular}{c|c|c|d{1}|d{1}|c|c|c|d{1}|d{3}|d{1}} \\ \hline 

%\begin{tabular}{c|c|c|d{1}|d{1}|c|c|c|d{-1}|d{-1}|d{-1}} \hline
%\begin{tabular}{|c|c|c|c|c|c|c|c|c|c|c|} \hline
%\hline
\multicolumn{1}{c|}{Model}&\multicolumn{1}{c|}{Tracer}&\multicolumn{1}{c|}{Frequency}&\multicolumn{1}{c|}{dV}&\multicolumn{1}{c|}{logn$_{\mathrm{H}}$}&\multicolumn{1}{c|}{n$_{H_{2}}$}&\multicolumn{1}{c|}{logN$_{\mathrm{Mol}}$}&\multicolumn{1}{c|}{NMol}&\multicolumn{1}{c|}{Tex}&\multicolumn{1}{c|}{Optical}&\multicolumn{1}{c}{Tb} \\
\multicolumn{1}{c|}{ }&\multicolumn{1}{c|}{ }&\multicolumn{1}{c|}{ }&\multicolumn{1}{c|}{ }&\multicolumn{1}{c|}{ }&\multicolumn{1}{c|}{ }&\multicolumn{1}{c|}{ }&\multicolumn{1}{c|}{ }&\multicolumn{1}{c|}{ }&\multicolumn{1}{c|}{Depth}&\multicolumn{1}{c}{ } \\

%\hline
\multicolumn{1}{c|}{ }&\multicolumn{1}{c|}{ }&\multicolumn{1}{c|}{(GHz)}&\multicolumn{1}{c|}{(km s$^{-1}$)}&\multicolumn{1}{c|}{ }&\multicolumn{1}{c|}{($\times10^{6}$ cm$^{-3}$)}&\multicolumn{1}{c|}{ }&\multicolumn{1}{c|}{($\times10^{15}$ cm$^{-2}$)}&\multicolumn{1}{c|}{(K)}&\multicolumn{1}{c|}{(\large{$\tau$})}&\multicolumn{1}{c}{(K)} \\
\hline
{Molex}&{HCN (1-0)} &88.6316&36.5&6.5&1.9&15.2&1.433&-21.60&-0.45&14.2 \\
%\hline
{Radex}& & & &6.5&1.9&15.2&1.433&-20.90&-0.43&13.1 \\
%\hline
{\% Difference}& & & & & & & &-3.2&-4.2&-8.0 \\
\hline
%\hline
{Molex}&{HCO+ (1-0)}&89.1885&36.5&6.7&3.0&14.9&0.794&-54.4&-0.08&4.6\\
%\hline
{Radex}& & &&6.7&3.0&14.9&0.794&-60.3&-0.07&4.5\\
%\hline
{\% Difference}& & & & & & & &-10.8&10.5&-2.2\\
\hline
%\hline
{Molex}&{HCN (3-2)}&256.8864&90&6.7&3.0&15.5&2.818&42.8&1.11&25.4\\
%\hline
{Radex}& && &6.7&3.0&15.5&2.818&43.54&1.12&25.2 \\
%\hline
{\% Difference}& & & & & & & &1.7&0.9&-0.6 \\
\hline
%\hline
{Molex}&{HCN (4-3)}&354.5055&51&6.5&1.9&15.1&1.258&23.8&1.82&13.7\\
%\hline
{Radex}& & & &6.5&1.9&15.1&1.258&23.7&1.83&13.6\\
%\hline
{\% Difference}& & & & & & & &-0.4&0.6&-1.0 \\
\hline
%\hline
\end{tabular}
\end{table} 
\end{center}  

\begin{center}
\begin{threeparttable}[!h]
%\begin{table}[!h]  

\caption{\label{ChrisTab2} HCN Cores Published Vs Modelled Properties.}

%\vspace{0.5cm}
%\small
\scriptsize
\begin{tabular}{c|c|c|c|c|c|c|c} \\ \hline
\hline
%\begin{tabular}{|c|c|c|c|c|c|c|} \hline

\multicolumn{1}{c|}{\citet{Chris2005}}&\multicolumn{1}{c|}{Deproj Dist }&\multicolumn{1}{c|}{Size FWHM}&\multicolumn{3}{c|}{\citet{Chris2005} H$_{2}$  Density Scenarios}&\multicolumn{2}{c}{Modelled\tnote{a}} \\
\multicolumn{1}{c|}{ID}&\multicolumn{1}{c|}{ }&\multicolumn{1}{c|}{  }&\multicolumn{1}{c|}{Virial}&\multicolumn{1}{c|}{Opt thick}&\multicolumn{1}{c|}{Opt thin}&\multicolumn{1}{c|}{H$_{2}$ Density}&\multicolumn{1}{c}{\large{$\tau$} \normalsize{HCN(1-0)}} \\
\multicolumn{1}{c|}{ }&\multicolumn{1}{c|}{ }&\multicolumn{1}{c|}{ }&\multicolumn{1}{c|}{ }&\multicolumn{1}{c|}{\large{$\tau$} \normalsize{=4}}&\multicolumn{1}{c|}{ }&\multicolumn{1}{c|}{ }&\multicolumn{1}{c}{ } \\
\multicolumn{1}{c|}{ }&\multicolumn{1}{c|}{(pc)}&\multicolumn{1}{c|}{(pc)}&\multicolumn{1}{c|}{($\times$10$^{6}$cm$^{-3}$)}&\multicolumn{1}{c|}{($\times$10$^{6}$cm$^{-3}$)}&\multicolumn{1}{c|}{($\times$10$^{6}$cm$^{-3}$) }&\multicolumn{1}{c|}{($\times$10$^{6}$cm$^{-3}$)}&\multicolumn{1}{c}{ } \\
\hline
{D\tnote{b}}&1.86&0.43&32.29&49.41&3.97&1.51&-0.30 \\
{I}&1.87&0.26&9.68&27.06&2.06&0.95&-0.33 \\
{M\tnote{c}}&1.75&0.26&14.78&33.43&3.56&1.90&-0.52 \\
{O}&1.60&0.33&35.04&51.48&7.36&1.51&-0.52 \\
{P}&1.39&0.21&50.82&62.00&7.06&1.20&-0.24 \\
{W}&1.47&0.22&45.28&58.32&5.44&1.20&-0.24 \\
{Z}&2.08&0.24&78.22&76.91&2.37&1.58&-0.25 \\
\hline
\hline
\end{tabular}
\begin{tablenotes}
\item[a] Values for hydrogen density and HCN(1-0) optical depth based on a kinetic temperature of 150\,K 
\item[b] Core D modelled hydrogen density for Tk=50\,K is 3.97$\times10^{6}$cm$^{-3}$ which agrees with \citet{Chris2005}
\item[c] Core M modelled hydrogen density for Tk=50\,K is 3.97$\times10^{6}$cm$^{-3}$ compared with 3.56$\times10^{6}$cm$^{-3}$ in \citet{Chris2005} \\ 
\end{tablenotes}
\end{threeparttable}
\end{center}
\end{landscape}
\normalsize
\section{Discussion} \label{discuss}
\paragraph*{}
The current modelling results have shown that molecular hydrogen densities are about 10$^{6}$cm$^{-3}$ and optical depths for the HCN(1-0) transition are $\ll$ 1. The reason for this difference from previous results is discussed in this section which highlights the modelling outcomes of both core groups. The fact that the argument of \citet{Marr1993} for equal excitation temperatures for the HCN isotopologs does not hold true for optically thin, inverted transitions as occur for HCN(1-0) and HCO$^{+}$(1-0) transitions helps explain the different results. 

\paragraph*{•}
The accuracy of the modelling was checked by comparing output parameter values with specific values for these parameters generated by Radex. Tables \ref{Marrcparam} and \ref{Radexcalcs} show the Radex results for Core A in the first group and Core O in the second group confirm that the results from LVG analysis for these cores are reasonable and that the LVG modelling give reliable results. 

\paragraph*{•}
\citet{Marr1993} used an excitation model which predicted  HCN brightness temperatures, opacities and excitation temperatures for the first 20 excitation levels. The assumption was made that based on previous estimates Z varied from 10 to 40 and core sizes that were inferred from the beam filling factor and beam size varied from 0.05 to 0.12pc. 

\paragraph*{•}
Uncertainty about the value of Z arises from the fact that \citet{Marr1993} found the value of Z from modelling was between 4 and 7 to satisfy their low boundary condition for the [HCN]/[H$_{2}$] abundance of 6$\times$10$^{-9}$ at a kinetic temperature of 150\,K and $\tau_{\mathrm{HCN}(1-0)}$ = 1. These Z values agree with those from current modelling of $\sim$ 3 to 6 from the ratios of  core T$_{b}$H$^{12}$CN to T$_{b}$H$^{13}$CN (using Eqn.\ \ref{Tb1213}). \citet{Marr1993} then reasoned that 250\,K was a more reasonable kinetic temperature based on other CND studies and assuming a Z value of 20 produced a hydrogen density of n$_{H_{2}}$ = 2$\times$10$^{6}$cm$^{-3}$ best fitted their observed brightness temperatures.       

%\paragraph*{•}
The current modelling for a kinetic temperature of 250\,K found the hydrogen density  unchanged at 0.4$\times$10$^{6}$cm$^{-3}$ (see Table \ref{Marrcprops}). \citet{Marr1993} also concluded that with their more reasonable parameters the [HCN]/[H$_{2}$] abundance was 8$\times$10$^{-8}$ an order of magnitude greater than the typical interstellar abundance of 10$^{-9}$ which was used by \citet{Chris2005}. 

\paragraph*{}
The underlying assumption of \citet{Marr1993} was that given the excitation temperatures for H$^{12}$CN and H$^{13}$CN that occupy the same space will be equal, the ratio of brightness temperatures and opacities for these HCN isotopologs was equal to Z, the ratio of $^{12}$C/$^{13}$C for small opacities. 
Their argument was developed using the following assumptions\,:

\begin{itemize}
\item Emissions from both species were assumed to have the same beam filling factors and background brightness temperatures since the spatially coincident H$^{12}$CN and H$^{13}$CN traced the same gas. 
\item Equal excitation temperatures required optical depths that were not large enough for line trapping to significantly enhance H$^{12}$CN excitation relative to that of the H$^{13}$CN. If line trapping was an issue then the optical depth estimates would be low.
\end{itemize}
\paragraph*{•}
In the limit of small opacities for both species
\begin{equation}\label{T13T12}
\frac{T_{13}}{T_{12}} = \frac{(1-\mathrm{exp}(-\tau _{13}))}{(1-\mathrm{exp}(-\tau _{12}))} = \frac{\tau _{13}}{\tau _{12}}
\end{equation}
where T =$\frac{\lambda^{\!2}}{2k}$ S$_{\nu}$ and S$_{\nu}$ is the Source function,
leads to
\begin{equation} \label{Tb1213}
\frac{T_{12}}{T_{13}} =  \frac{[H^{12}CN]}{[H^{13}CN]} = Z \,.
\end{equation}
   
\paragraph*{•}
\citet{Marr1993} then argued that as the brightness temperatures for H$^{12}$CN did not exceed 5.5\,K if H$^{13}$CN brightness temperature values were below 0.5\,K, which is the noise level and undetectable even for the lowest value of Z = 11, H$^{13}$CN emission above that level was optically thin and meant H$^{12}$CN emission was optically thick. They further argued that $(1-\mathrm{exp}(-\tau (H^{12}CN)))\approx 1$, which would be the case where $\tau$(H$^{12}$CN) = 4 which was assumed by \citet{Marr1993} and adopted by \citet{Chris2005} as the value for their optically thick scenario.

\paragraph*{•}
The above argument leads to the conclusion that \citet{Marr1993} observed H${12}$CN brightness temperatures for opaque gas i.e. (Tb(thick)), which from Eqn \ref{T13T12} for \large{$\tau$}\normalsize (12) thick and \large{$\tau$}\normalsize (13) thin, where (12) and (13) denote H$^{12}$CN and H$^{13}$CN respectively leads to
\begin{equation}
\large{\tau}\normalsize(13) = -\ln  \left[ 1-(T(13)/T(12)\right] 
\end{equation}
and
\begin{equation}
\large{\tau}\normalsize(12) = \large{\tau}\normalsize(13)\frac{T(12)}{T(13)}
\end{equation}
so that
\begin{equation} \label{H12CN}
 \large{\tau}\normalsize(12) = -Z\ln\left[ 1-T(13)/T(12) \right]  \,,
\end{equation}

where 
\begin{equation}
Z= \frac{C^{12}}{C^{13}} \,.
\end{equation}  

\paragraph*{}
The assumption of equal excitation temperatures for two HCN isotopologs occupying the same space, provided that opacities are not high enough to produce line trapping that would  enhance H$^{12}$CN excitation, is \textbf{not} supported by the LVG model results reported in Tables \ref{Marrcprops} and \ref{Marrcparam}. Instead the optical depths for both H$^{12}$CN and H$^{13}$CN are thin and T$_{\mathrm{ex}}$(12) $\neq$  T$_{\mathrm{ex}}$(13).

\paragraph*{}
Excitation temperature is very sensitive to the relative populations of the upper and lower states for weakly inverted transitions.

 The ratio of population levels is given by

\begin{equation}
\frac{n_{2}}{n_{1}} = \frac{g_{2}}{g_{1}}\exp \left(-\frac{h\nu}{\mathrm{kT}_{ex}}\right) 
\end{equation}

and differentiating this expression with respect to T$_{ex}$ gives

\begin{eqnarray}
  \frac{\mathrm{d}(n_{2}/n_{1})}{\mathrm{dT}_{ex}} &=& -\frac{g_{2}}{g_{1}}  \mathrm{exp}\left(-\frac{h\nu}{kT_{ex}}\right) \times -\frac{h\nu}{\mathrm{kT}_{ex}} \frac{1}{\mathrm{T}_{ex}}   \\ 
 &=& \frac{n_{2}}{n_{1}}\frac{h\nu}{\mathrm{kT}_{ex}} \frac{1}{\mathrm{T}_{ex}} \,. \\
\end{eqnarray}
Then 
\begin{eqnarray}
\frac{\mathrm{d}\ln(T_{ex})}{\mathrm{d}\ln(n_{2}/n_{1})} &=& \frac{kT_{ex}}{h\nu} \,.
\end{eqnarray}

For the HCN(1-0) transition $ \frac{h\nu}{k}$ = 4.25\,K   \,, \\

  so that \\

\begin{eqnarray}\label{Texrate}
\frac{\mathrm{d}\ln(T_{ex})}{\mathrm{d}\ln({n_{2}/n_{1})}} &=& \frac{T_{ex}}{4.25\,\mathrm{K}} \,.
\end{eqnarray}

\begin{figure}[ht] \includegraphics[scale=0.75,bb=  0 70 610 700] 
{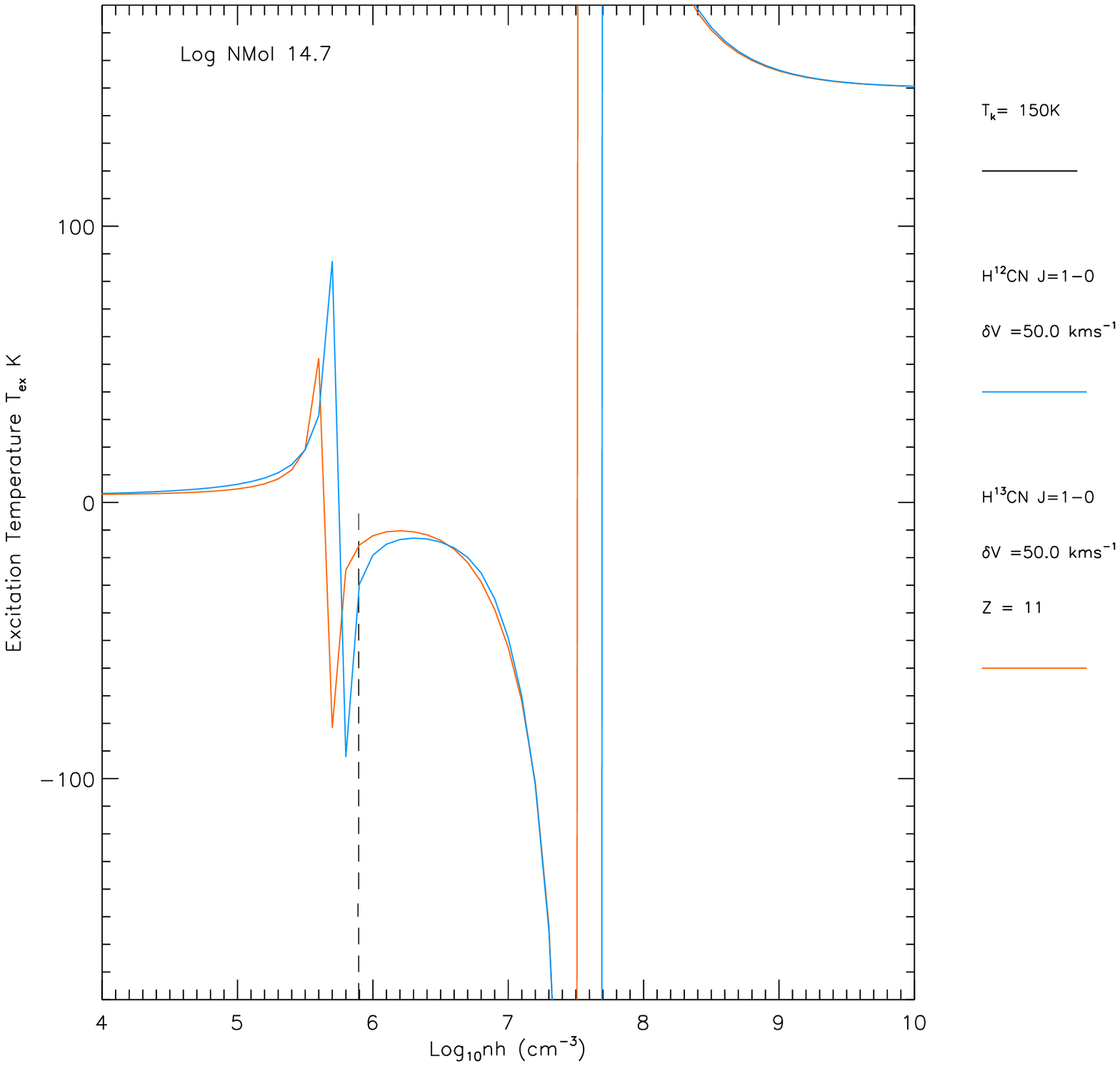} \fontsize{9} {9} \caption[Marr Core A Excitation Temperature vs Log Hydrogen Density]{\label{H12H13Tx}  Figure shows Excitation Temperature plots for H$^{12}$CN \& H$^{13}$CN (1-0) at Log Column Density of 14.7.The intersection of the plots with lognh = 5.9 indicates the values of excitation temperatures for the mean intersection of brightness temperature contours.  see Fig.\ \ref{MarrA} The plots show a clear difference  between the excitation temperatures of H$^{12}$CN \& H$^{13}$CN} 
\end{figure}

\paragraph*{}
Table \ref{Marrcprops} shows excitation temperatures for H$^{12}$CN(1-0) varying between --11 and --27.6\,K so that the ratio given by Eqn \ref{Texrate} can vary between 2.56 and 6.49. Similarly for H$^{13}$CN with an energy level of 4.14\,K for the (1-0) transition  and a range of excitation temperatures from 5.23 to 19.6\,K the rate of change can vary from 1.26 to 4.73. These rapid changes explain why it is unlikely that the excitation temperatures will be the same for both molecular species. i.e.\ small changes in n$_{2}$/n$_{1}$ lead to large changes in T$_{\mathrm{ex}}$ and Fig.\ \ref{H12H13Tx} indicates how rapidly excitation temperatures can vary. The plots show H$^{12}$CN and H$^{13}$CN excitation temperatures for a constant column density (logN$_{\mathrm{HCN}(1-0)}$ = 14.7) and changing hydrogen densities values between 10$^{3}$ and 10$^{10}$ cm$^{-3}$. A dashed line at logn$_{H}$ = 5.9 together with logN$_{HCN(1-0)}$ = 14.7 indicates the co-ordinate values of the average of the intersection points of brightness temperature lines for Core A in Fig.\ \ref{MarrA}. The intersection of the dashed line with the excitation temperature curves shows clearly that the H$^{12}$CN and  H$^{13}$CN excitation temperatures are not equal as claimed by \citet{Marr1993}.

\paragraph*{•}
\citet{Marr1993} presented a case for the abundance ratio of [HCO$^{+}$]/[HCN] to be proportional to the ratio of their optical depths times the ratio of their Einstein-A coefficients and the assumption that both were optically thick (see Eqn \ref{HCO+HCN}). They argued that while the determination of the [HCO$^{+}$]/[HCN] ratio was less certain due to their different formation processes, emission from both tracers appeared in the same regions of the CND; and because the HCO$^{+}$(1-0) transition had a similar energy to the HCN(1-0) transition, the HCO$^{+}$ emission should therefore have the same beam filling factor, excitation temperature and background temperature. If this was the case the optically thick brightness temperature for HCO$^{+}$ was the same as for HCN:

\begin{equation}
T(\mathrm{thick}, \mathrm{HCO}^{+}) \sim  T(\mathrm{thick}, \mathrm{HCN}) =  T(\mathrm{H^{12}CN}) \,.
\end{equation}

The optical depth for HCO$^{+}$ was then given by the expression
\begin{equation}
\large{\tau}\normalsize_{\mathrm{HCO}^{+}} = -\ln[1-(\mathrm{T_{HCO^{+}}}/\mathrm{T_{H^{12}CN}})] \,,
\end{equation}

and \citet{Marr1993} also note that if \large{$\tau$}\normalsize(HCO$^{+}$)  $ > 5 $  then 
\begin{equation}
\large{\tau}\normalsize(\mathrm{HCO}^{+}) \approx \large{\tau}\normalsize(\mathrm{H^{12}CN}) \,,
\end{equation}
 
and as \large{$\tau$}\normalsize(HCO$^{+}$) cannot be measured with any confidence, it can be inferred that it was optically thick from the fact that \large{$\tau$}\normalsize(H$^{12}$CN) was optically thick. Given that emission from both molecules trace the same gas the ratio of their abundances was related to the ratio of their optical depths via:

\begin{equation}
\frac{[\mathrm{HCO}^{+}]}{[\mathrm{H^{12}CN}]} =  \frac{[\large{\tau}\normalsize(\mathrm{HCO}^{+})]}{[\large{\tau}\normalsize(\mathrm{H^{12}CN})]} \times \frac{\mathrm{A}_{21}(\mathrm{H^{12}CN})}{\mathrm{A}_{21}(\mathrm{HCO}^{+})}  
\end{equation}

where A is the molecule's Einstein A coefficient (2.4 $\times 10^{-5}$ for HCN(1-0) and 5.5 $\times 10^{-5}$ for HCO$^{+}$(1-0)) so that the equation becomes

\begin{equation}\label{HCO+HCN}
\frac{[HCO^{+}]}{[{H^{12}CN}]} = 0.44 \times \frac{[\large{\tau}\normalsize(\mathrm{HCO}^{+})]}{[\large{\tau}\normalsize(\mathrm{H^{12}CN})]} \,. 
\end{equation}

Using Eqn \ref{HCO+HCN} and the values modelled in Molex for \large{$\tau$}\normalsize (HCO$^{+}$) and \large{$\tau$}\normalsize (H$^{12}$CN) in Table \ref{Marrcparam}, the [HCO$^{+}$]/[HCN] ratio ranges between 0.4 (Core A) to 0.6 (Core B). 

\paragraph*{}
\citet{Marr1993} summarised their results for the [HCO$^{+}$]/[{H$^{12}$CN}] ratio as varying between 0.1 down to 0.06  with a [C$^{12}$]/[C$^{13}$] value of Z = 30. These [HCO$^{+}$]/[{H$^{12}$CN}] values were significantly lower than the rates determined from higher resolution observations as between 0.4 and 1.1 \citep{Chris2005}. The current model used values [HCO$^{+}$]/[H$^{12}$CN] of Z = 0.74 and [H$^{12}$CN]/[H$^{13}$CN] of Z = 11  and produced more credible intersection points for the peak brightness temperature contour plots of the respective molecules than the abundance values adopted in papers by \citet{Marr1993} and \citet{Chris2005}.

\paragraph*{•}
Modelling the first core group produced molecular hydrogen densities, between 0.13 and 0.63 $\times 10^{6}$cm$^{-3}$ which are lower than for the second core group where the densities are between 0.95 and 1.9 $\times 10^{6}$cm$^{-3}$ at a kinetic temperature of T = 150\,K and 3.97 $\times 10^{6}$cm$^{-3}$ at T = 50\,K.  

\paragraph*{•}
Table \ref{[HCN]/[H2]} presents values for the [HCN]/[H$_{2}$] ratio for HCN(1-0) in cores A and E in the first group and in cores D and O in the second group from the current modelling results. The effects of varying the core diameters and kinetic temperatures are shown. The ratio ranges between a high of 5$\times10^{-7}$ and a low of 5$\times10^{-10}$. The high [HCN]/[H$_{2}$] ratios occur for HCN number densities of $20\leq [HCN] \leq 80 \times10^{-4}$cm$^{-2}$ arising from either a thick HCN slab or a smaller core size. More typical ratios of of $0.2 \leq [HCN]/[H_{2}] \leq 1.2 \times10^{-9}$ result from the second core group analysis where observations have resolved the core sizes. 

\paragraph*{}
It should be noted that in their conclusions \citet{Chris2005} concede that ``Some of the HCN emission may be enhanced by shock chemistry implying that estimates of optically thick and optically thin core masses and densities (assuming standard HCN) are overestimates''.

\paragraph*{•}
\citet{Chris2005} mention H$^{13}$CN observations for the NE and SW lobes of the CND and comparison of these with H$^{12}$CN observations support their optically thick scenario, however there are no details provided in their paper to check this scenario.   

\paragraph*{•}
Greater confidence in the current results would require resolved observation of the CND cores in the HCN (3-2) transition. In spite of this shortcoming, the model analysis of the \citet{Marr1993} cores gives considerable confidence that the HCN gas is optically thin.

\paragraph*{•}
The following chapter summarises the work done to present the case for a low hydrogen number density scenario in the CND base on LVG modelling and then lists the conclusions arising from this work. 

\begin{landscape}

\begin{center}
\begin{threeparttable}[ht]

\caption{\label{[HCN]/[H2]} [HCN]/[H$_{2}]$ Abundance Ratios from Modelled Cores}

%\vspace{0.5cm}
%\small
\scriptsize
%\begin{tabular}{c|c|c|c|d{2}|d{2}|d{3}|d{2}} \hline

\begin{tabular}{c|c|c|c|c|c|c|c} \hline
\hline
\multicolumn{1}{c|}{Core}&\multicolumn{1}{c|}{Kinetic}&\multicolumn{1}{c|}{FWHM}&\multicolumn{1}{c|}{Log HCN}&\multicolumn{1}{c|}{HCN}&\multicolumn{1}{c|}{HCN}&\multicolumn{1}{c|}{Hydrogen}&\multicolumn{1}{c}{[HCN]/[H$_{2}$]} \\
\multicolumn{1}{c|}{Group}&\multicolumn{1}{c|}{Temp}&\multicolumn{1}{c|}{Diameter}&\multicolumn{1}{c|}{Column Density per}&\multicolumn{1}{c|}{Column}&\multicolumn{1}{c|}{Density}&\multicolumn{1}{c|}{Density}&\multicolumn{1}{c}{Abundance} \\
\multicolumn{1}{c|}{ID}&\multicolumn{1}{c|}{ }&\multicolumn{1}{c|}{}&\multicolumn{1}{c|}{line width}&\multicolumn{1}{c|}{Density}&\multicolumn{1}{c|}{}&\multicolumn{1}{c|}{}&\multicolumn{1}{c}{Ratio} \\
\multicolumn{1}{c|}{Pairs}&\multicolumn{1}{c|}{(K)}&\multicolumn{1}{c|}{(pc)}&\multicolumn{1}{c|}{}&\multicolumn{1}{c|}{($\times10^{14}$cm$^{-2}$ km s$^{-1}$)}&\multicolumn{1}{c|}{($\times10^{-4}$cm$^{-3}$)}&\multicolumn{1}{c|}{($\times$10$^{6}$cm$^{-3}$)}&\multicolumn{1}{c}{($\times10^{-9}$)} \\
\hline
{First\tnote{g}}& & & & & & & \\
{A\tnote{a}}&150&{0.43\tnote{d}}&13.0&4.45&3.36&0.397&0.85 \\
{B\tnote{b}}&150&0.43&13.7&25.06&18.90&0.016&120.00 \\
{A\tnote{a}}&150&{0.10\tnote{e}}&13.0&4.45&14.40&0.397&3.60 \\
{B\tnote{b}}&150&0.10&13.7&25.06&81.20&0.016&510.00 \\
{E\tnote{a}}&150&{0.33\tnote{d}}&13.0&8.80&8.64&0.629&1.40 \\
{G\tnote{b}}&150&0.10&13.0&8.80&28.50&0.63&450.00 \\
{Second\tnote{h}}& & & & & & & \\
{D\tnote{c}}&150&0.43&13.3&9.08&6.84&1.26&0.54 \\
{B\tnote{b}}&150& & & & & & \\
{D\tnote{c}}&50&0.43&13.4&11.40&8.61&3.96&0.21 \\
{B\tnote{b}}&50& & & & & & \\
{O\tnote{c}}&150&{0.33\tnote{d}}&13.5&8.80&11.54&1.26&0.91 \\
{G\tnote{b}}&150& & & & & & \\
{O\tnote{c}}&150&{0.25\tnote{f}}&13.5&11.54&15.50&1.26&1.20 \\
{G\tnote{b}}&150& & & & & & \\
\hline
\hline
\end{tabular}
\begin{tablenotes}
\item[a] H$^{12}$CN(1-0) \citet{Marr1993}
\item[b] HCN(3-2) \citet{Jacks1993}
\item[c] HCN(1-0) \citet{Chris2005}  
\item[d] actual core diameter from Table 2 \citet{Chris2005}
\item[e] average core diameter from \citet{Marr1993}
\item[f] average core diameter from Table 2 \citet{Chris2005}
\item[g] core pairs A, H$^{12}$(1-0) \& B, HCN(3-2); E, H$^{12}$(1-0) \& G, HCN(3-2).
\item[h] core pairs D, HCN(1-0) \& B, HCN(3-2); O, HCN(1-0)\& G, HCN(3-2).
\end{tablenotes}
\end{threeparttable}
\end{center}
\end{landscape}

\normalsize
\cleardoublepage 

\chapter{Summary and Conclusions}\label{concls}

\paragraph*{•}
The Circumnuclear Disk (CND) is a torus of molecular dust and gas rotating about the galactic centre which has a massive black hole (4$\times$10$^{6}$M$_{\odot}$), SgrA$^{*}$. Early observations using different molecules e.g.\ CO, HCN, NH$_{3}$ and CS inferred a low molecular hydrogen density n$_{H_{2}} \leq$ 10$^{6}$ cm$^{-3}$. More recent observations using HCN, HCO$^{+}$, NH$_{3}$ and CS \citep{Chris2005,MMC2009} infer denser gas n$_{H_{2}} \geq$ 10$^{7}$ cm$^{-3}$ which has led \citet{Genzel2010} to discuss two scenarios and present evidence that supports the lower hydrogen density scenario.

The two scenarios \citet{Genzel2010} used to describe the cores are \,:
\begin{enumerate}
\item Scenario one considers the gas cores to be cool and sufficiently dense n$_{H_{2}} \sim$ 10$^{7}$ cm$^{-3}$ to be tidally stable leading to an estimated life for the CND of some M yr with the ensuing possibility for star formation within the disk. Evidence supporting this claim is an apparent core stability and the presence of methanol masers near three of the cores \citep{FYZ2008} that could indicate hot gas flaring from a condensing core in the very early stages of star formation. 
\item Scenario two proposes warm, less dense nH$_{2} \sim$ 10$^{6}$ cm$^{-3}$ gas which is tidally unstable leading to a comparatively short disk lifetime of $<$ 10$^{5}$ years and no core condensation. Evidence presented in support of this scenario suggests that the mass of the CND is insufficient (3 to 4  $\times$ 10$^{6}$ M$_{\odot}$) to disrupt the orbits of the stellar cluster in the central cavity and there is no evidence of individual dense cores (dark spots) in IR CND images.
\end{enumerate}
\paragraph*{•} 
The present study sought to reconcile the observations in different HCN transitions in order to identify cores observed in multiple transitions by comparing locations  in the CND and kinematic properties (radial velocities). This work was described in Chapter \ref{diskgeom} and resulted in the selection of two groups of cores for analysis which is reported in Chapter \ref{anal}. The LVG model produced hydrogen densities in the CND of n(H$_{2}$) = (1--4)$\times$10$^{6}$ cm$^{-3}$ with an optically thin, inverted ($\tau$ negative)HCN(1-0) transition that supports the lower hydrogen density scenario. 

\paragraph*{•}
Data from the twenty-six HCN (1-0) cores listed in \citet{Chris2005} provided the information for calculating the true (deprojected) core offsets from SgrA$^{*}$, using trial and error to select the disk's tilt angle, $\alpha$ = 60$^{\circ}$ out of the plane of the sky. The expression for core model radial velocities was based on the disk's orientation relative to the plane of the sky and a  CND rotational velocity of 110 km s$^{-1}$ was adopted to provide a comparison with observed core radial velocities. This comparison illustrated the warped nature of the disk and its composition of circulating gas streamers that rotate at varying velocities. Eighteen of the twenty-six cores fit within the accepted range of disk parameters, while seven cores have radial velocities that are significantly different from the model disk. Variations in the rotational velocity of the disk were discussed in Chapter \ref{CNDintro}.

\paragraph*{•}
The following conclusions are derived from the results presented in this thesis.\\
 
The figure of the deprojected HCN (1-0) cores clearly shows the circular structure of the CND with an inner cavity of $\sim$ 1.6 pc. The attempt to fit the HCN(1-0) core radial velocities to the model radial velocity distribution curves  demonstrates the warped nature of the disk's rotating streamers and that the rotational velocity of some streamers vary significantly from the accepted 110 km s$^{-1}$. 

\paragraph*{•} 
The LVG modelling of the core group based on \citet{Marr1993} data and the the core group based on more recent data \citep{Chris2005,MMC2009} in which the core groups were resolved gave consistent results. Both core groups gave molecular hydrogen number densities that show that the optically thin scenario outlined in \citet{Genzel2010} is more likely than the optically thick scenario proposed by \citet{Chris2005} and supported by \citet{MMC2009}. The low molecular hydrogen number densities imply that the cores are not tidally stable and that the disk is a transient structure with a kinematic lifespan of $\sim$ 10$^{5}$ to 10$^{6}$ years which would be too short for star formation and also means that the estimated total mass of the CND could be up to an order of magnitude less than the previous estimate of 10$^{6}$ M$_{\odot}$. 

\paragraph*{•}
The \citet{Marr1993} proposition that excitation temperatures for molecules occupying the same location in the CND is incorrect, the excitation temperatures are indeed different for the prevailing CND conditions and this is why their hypothesis fails. 

\paragraph*{•}
The discovery of methanol masers by \citet{FYZ2008} in the vicinity of Cores F, G and V may indicate higher hydrogen densities for these cores as a result of a shock mechanism. Those authors proposed that these cores may have condensed to the stage of forming protostars and that the masers are a result of jet outflows from these cores some $10^{4}$ yr after core collapse.

\paragraph*{•}
The possibility of a few localised higher density pockets within the CND could be due to either shocks emanating from one or more of the following \,: outflows from the young UV stars in the cavity inside the CND, the shock front of the SgrA East SNR and core-core collisions within the CND. Core-core collisions were suggested by \citet{Sjman2008} and \citet{FYZ2001} as the mechanism, which was was formulated by \citet{Elitz1976} for producing the shocks needed for the formation of 1720 MHz OH masers, which are found in the CND and might be a trigger for star formation. 

\paragraph*{•}
Greater certainty for  the hydrogen density results could be attained if resolved HCN (3-2) transition observations were available for modelling in addition to the currently  available (1-0) and (4-3) transition data. The sub millimetre array (SMA) operates at frequencies from 180 to 700 GHz which includes the HCN(3-2) transition frequency of 265.886 GHz. Plans for increasing the sensitivity of this instrument are now in progress (see the SMA Newsletter Number 12, August 2011).   

\paragraph*{•}
A wider CND survey of the H$^{13}$CN(1-0) transition  might clarify the optically thick/thin question and hence the hydrogen density value. An analysis of the NH$_{3}$ (1-1), (2-2), (3-3) and (6-6)  transitions \citep{HandH2002} would also be useful to check hydrogen densities, particularly as the (6-6) transition traces denser and hotter regions more effectively than HCN. Such a study should be done in the future to reinforce the results of this study. 

\paragraph*{•}
A possible test of  maser effects of negative optical depths in the (1-0) transitions of HCN and HCO$^{+}$ would be to search for a bright (T$_{b} \sim$ 1000\,K) source of continuum emission (such as a quasar or radio galaxy ?) aligned with one of the disk's cores. A comparison of off HCN line frequency, on line frequency of core observations would then detect an increase in the background intensity caused by amplification of HCN or HCO$^{+}$(1-0) emission as it passes through the core. 

\paragraph*{•}
The results from this study clearly favour the warm, low density hydrogen gas scenario with a CND mass of  3 to 4 $\times$ 10$^{5}$ M$_{\odot}$ and tidal forces pulling the cores apart.    
%\include{chapter_6}
%\include{chapter_7}

%%%%% Other chapters in here
%\input{chap_other}

%%%%% Conclusion
%\input{chap_conclusion}

%\appendix

%%%%% Appendix
%\input{chap_appendix}
 
%\backmatter

%%%%% List of symbols
% your thesis may not need this, so comment out or delete the following line
%\input{listofsymbols}

%%%%% Bibliography, in BibTeX format (the references.bib file)

\bibliography{references}{}

\begin{thebibliography}{40}
\providecommand{\natexlab}[1]{#1}
\providecommand{\url}[1]{\texttt{#1}}
\expandafter\ifx\csname urlstyle\endcsname\relax
  \providecommand{\doi}[1]{doi: #1}\else
  \providecommand{\doi}{doi: \begingroup \urlstyle{rm}\Url}\fi

\bibitem[{Becklin} et~al.(1982){Becklin}, {Gatley}, and {Werner}]{Becklin1982}
E.~E. {Becklin}, I.~{Gatley}, and M.~W. {Werner}.
\newblock {Far-infrared observations of Sagittarius A - The luminosity and dust
  density in the central parsec of the Galaxy}.
\newblock \emph{\apj}, 258:\penalty0 135--142, July 1982.

\bibitem[{Bohlin} et~al.(1978){Bohlin}, {Savage}, and {Drake}]{Bohlin1978}
R.~C. {Bohlin}, B.~D. {Savage}, and J.~F. {Drake}.
\newblock {A survey of interstellar H I from L-alpha absorption measurements.
  II}.
\newblock \emph{\apj}, 224:\penalty0 132--142, Aug. 1978.

\bibitem[{Christopher} et~al.(2005){Christopher}, {Scoville}, {Stolovy}, and
  {Yun}]{Chris2005}
M.~H. {Christopher}, N.~Z. {Scoville}, S.~R. {Stolovy}, and M.~S. {Yun}.
\newblock {HCN and HCO$^{+}$ Observations of the Galactic Circumnuclear Disk}.
\newblock \emph{\apj}, 622:\penalty0 346--365, Mar. 2005.

\bibitem[{Draine} and {Lee}(1984)]{Draine1984}
B.~T. {Draine} and H.~M. {Lee}.
\newblock {Optical properties of interstellar graphite and silicate grains}.
\newblock \emph{\apj}, 285:\penalty0 89--108, Oct. 1984.

\bibitem[{Elitzur}(1976)]{Elitz1976}
M.~{Elitzur}.
\newblock {Inversion of the OH 1720-MHz Line}.
\newblock \emph{\apj}, 203:\penalty0 124--131, Jan. 1976.

\bibitem[{Etxaluze} et~al.(2011){Etxaluze}, {Smith}, {Tolls}, {Stark}, and
  {Gonzalez-Alfonso}]{Etxa2011}
M.~{Etxaluze}, H.~A. {Smith}, V.~{Tolls}, A.~A. {Stark}, and
  E.~{Gonzalez-Alfonso}.
\newblock {The Galactic Centre in the Far Infrared}.
\newblock \emph{ArXiv e-prints 1108.0313}, Aug. 2011.

\bibitem[{Genzel}(1989)]{Genzel1989}
R.~{Genzel}.
\newblock {The Circumnuclear Disk (review)}.
\newblock In {M.~Morris}, editor, \emph{The Center of the Galaxy}, volume 136
  of \emph{IAU Symposium}, pages 393--405, 1989.

\bibitem[{Genzel} et~al.(2010){Genzel}, {Eisenhauer}, and
  {Gillessen}]{Genzel2010}
R.~{Genzel}, F.~{Eisenhauer}, and S.~{Gillessen}.
\newblock {The Galactic Center massive black hole and nuclear star cluster}.
\newblock \emph{Reviews of Modern Physics}, 82:\penalty0 3121--3195, Oct. 2010.

\bibitem[{Guesten} et~al.(1987){Guesten}, {Genzel}, {Wright}, {Jaffe},
  {Stutzki}, and {Harris}]{Guesten1987}
R.~{Guesten}, R.~{Genzel}, M.~C.~H. {Wright}, D.~T. {Jaffe}, J.~{Stutzki}, and
  A.~I. {Harris}.
\newblock {Aperture synthesis observations of the circumnuclear ring in the
  Galactic center}.
\newblock \emph{\apj}, 318:\penalty0 124--138, July 1987.

\bibitem[{Harris} et~al.(1985){Harris}, {Jaffe}, {Silber}, and
  {Genzel}]{Harris1985}
A.~I. {Harris}, D.~T. {Jaffe}, M.~{Silber}, and R.~{Genzel}.
\newblock {CO 7-6 submillimeter emission from the galactic center - Warm
  molecular gas and the rotation curve in the central 10 parsecs}.
\newblock \emph{\apjl}, 294:\penalty0 L93--L97, July 1985.

\bibitem[{Herrnstein} and {Ho}(2002)]{HandH2002}
R.~M. {Herrnstein} and P.~T.~P. {Ho}.
\newblock {Hot Molecular Gas in the Galactic Center}.
\newblock \emph{\apjl}, 579:\penalty0 L83--L86, Nov. 2002.

\bibitem[{Jackson} et~al.(1993){Jackson}, {Geis}, {Genzel}, {Harris}, {Madden},
  {Poglitsch}, {Stacey}, and {Townes}]{Jacks1993}
J.~M. {Jackson}, N.~{Geis}, R.~{Genzel}, A.~I. {Harris}, S.~{Madden},
  A.~{Poglitsch}, G.~J. {Stacey}, and C.~H. {Townes}.
\newblock {Neutral gas in the central 2 parsecs of the Galaxy}.
\newblock \emph{\apj}, 402:\penalty0 173--184, Jan. 1993.

\bibitem[{Karlsson} et~al.(2003){Karlsson}, {Sjouwerman}, {Sandqvist}, and
  {Whiteoak}]{Karlsson2003}
R.~{Karlsson}, L.~O. {Sjouwerman}, A.~{Sandqvist}, and J.~B. {Whiteoak}.
\newblock {18-cm VLA observations of OH towards the Galactic Centre. Absorption
  and emission in the four ground-state OH lines}.
\newblock \emph{\aap}, 403:\penalty0 1011--1021, June 2003.

\bibitem[{Liszt} et~al.(1985){Liszt}, {Burton}, and {van der Hulst}]{Liszt1985}
H.~S. {Liszt}, W.~B. {Burton}, and J.~M. {van der Hulst}.
\newblock {Associations between neutral and ionized gas in SGR A}.
\newblock \emph{\aap}, 142:\penalty0 237--244, Jan. 1985.

\bibitem[{Lockett} and {Elitzur}(1989)]{Lockett1989}
P.~{Lockett} and M.~{Elitzur}.
\newblock {An escape probability treatment of line fluorescence and overlap in
  astrophysics}.
\newblock \emph{\apj}, 344:\penalty0 525--530, Sept. 1989.

\bibitem[{Mangum} et~al.(1988){Mangum}, {Rood}, {Wadiak}, and
  {Wilson}]{Magnum1988}
J.~G. {Mangum}, R.~T. {Rood}, E.~J. {Wadiak}, and T.~L. {Wilson}.
\newblock {Observations of the C-13 isomers of cyanoacetylene - Implications
  for carbon isotope studies in the Milky Way}.
\newblock \emph{\apj}, 334:\penalty0 182--190, Nov. 1988.

\bibitem[{Marr} et~al.(1993){Marr}, {Wright}, and {Backer}]{Marr1993}
J.~M. {Marr}, M.~C.~H. {Wright}, and D.~C. {Backer}.
\newblock {HCO$^{+}$, H$^{13}$CN and H$^{12}$CN aperture synthesis observations
  of the circumnuclear ring in the Galactic center}.
\newblock \emph{\apj}, 411:\penalty0 667--673, July 1993.

\bibitem[{Marshall} et~al.(1995){Marshall}, {Lasenby}, and
  {Harris}]{Marshall1995}
J.~{Marshall}, A.~N. {Lasenby}, and A.~I. {Harris}.
\newblock {HCN observations of the circumnuclear disc in the Galactic Centre}.
\newblock \emph{\mnras}, 277:\penalty0 594--608, Nov. 1995.

\bibitem[{Mehringer} and {Menten}(1997)]{Mehringer1997}
D.~M. {Mehringer} and K.~M. {Menten}.
\newblock {44 GHz Methanol Masers and Quasi-thermal Emission in Sagittarius
  B2}.
\newblock \emph{\apj}, 474:\penalty0 346--361, Jan. 1997.

\bibitem[{Mezger} et~al.(1989){Mezger}, {Zylka}, {Salter}, {Wink}, {Chini},
  {Kreysa}, and {Tuffs}]{Mezger1989}
P.~G. {Mezger}, R.~{Zylka}, C.~J. {Salter}, J.~E. {Wink}, R.~{Chini},
  E.~{Kreysa}, and R.~{Tuffs}.
\newblock {Continuum observations of SGR A at mm/submm wavelengths}.
\newblock \emph{\aap}, 209:\penalty0 337--348, Jan. 1989.

\bibitem[{Mezger} et~al.(1996){Mezger}, {Duschl}, and {Zylka}]{Mezger1996}
P.~G. {Mezger}, W.~J. {Duschl}, and R.~{Zylka}.
\newblock {The Galactic Center: a laboratory for AGN?}
\newblock \emph{\aapr}, 7:\penalty0 289--388, 1996.

\bibitem[{Montero-Casta{\~n}o} et~al.(2009){Montero-Casta{\~n}o}, {Herrnstein},
  and {Ho}]{MMC2009}
M.~{Montero-Casta{\~n}o}, R.~M. {Herrnstein}, and P.~T.~P. {Ho}.
\newblock {Gas Infall Toward Sgr A* from the Clumpy Circumnuclear Disk}.
\newblock \emph{\apj}, 695:\penalty0 1477--1494, Apr. 2009.

\bibitem[{Oka} et~al.(2007){Oka}, {Nagai}, {Kamegai}, {Tanaka}, and
  {Kuboi}]{Oka2007}
T.~{Oka}, M.~{Nagai}, K.~{Kamegai}, K.~{Tanaka}, and N.~{Kuboi}.
\newblock {A CO J = 3-2 Survey of the Galactic Center}.
\newblock \emph{\pasj}, 59:\penalty0 15--23, Feb. 2007.

\bibitem[{Oka} et~al.(2011){Oka}, {Nagai}, {Kamegai}, and {Tanaka}]{Oka2011}
T.~{Oka}, M.~{Nagai}, K.~{Kamegai}, and K.~{Tanaka}.
\newblock {A New Look at the Galactic Circumnuclear Disk}.
\newblock \emph{\apj}, 732:\penalty0 120,1--10, May 2011.

\bibitem[{Rieke} and {Lebofsky}(1985)]{Rieke1985}
G.~H. {Rieke} and M.~J. {Lebofsky}.
\newblock {The interstellar extinction law from 1 to 13 microns}.
\newblock \emph{\apj}, 288:\penalty0 618--621, Jan. 1985.

\bibitem[{Rohlfs} and {Wilson}(2006)]{Rohlfs2006}
K.~{Rohlfs} and T.~{Wilson}.
\newblock \emph{{Molecules in Interstellar Space}}, pages 360--426.
\newblock Astronomy and Astrophysics Library. 2006.

\bibitem[{Serabyn} and {Lacy}(1985)]{Serabyn1985}
E.~{Serabyn} and J.~H. {Lacy}.
\newblock {Forbidden NE II observations of the galactic center - Evidence for a
  massive block hole}.
\newblock \emph{\apj}, 293:\penalty0 445--458, June 1985.

\bibitem[{Serabyn} et~al.(1986){Serabyn}, {Guesten}, {Walmsley}, {Wink}, and
  {Zylka}]{SGW1986}
E.~{Serabyn}, R.~{Guesten}, J.~E. {Walmsley}, J.~E. {Wink}, and R.~{Zylka}.
\newblock {CO 1 - 0 and CS 2 - 1 observations of the neutral disk around the
  galactic center}.
\newblock \emph{\aap}, 169:\penalty0 85--94, Nov. 1986.

\bibitem[{Serabyn} et~al.(1989){Serabyn}, {G{\"u}sten}, and {Evans}]{SGE1989}
E.~{Serabyn}, R.~{G{\"u}sten}, and N.~J. {Evans}, II.
\newblock {CS Multitransition Observations of the Circumnuclear Disk}.
\newblock In {M.~Morris}, editor, \emph{The Center of the Galaxy}, volume 136
  of \emph{IAU Symposium}, pages 417--419, 1989.

\bibitem[{Sjouwerman} and {Pihlstr{\"o}m}(2008)]{Sjman2008}
L.~O. {Sjouwerman} and Y.~M. {Pihlstr{\"o}m}.
\newblock {Very Large Array Observations of Galactic Center OH 1720 MHz Masers
  in Sagittarius A East and in the Circumnuclear Disk}.
\newblock \emph{\apj}, 681:\penalty0 1287--1295, July 2008.

\bibitem[{Sutton} et~al.(1990){Sutton}, {Danchi}, {Jaminet}, and
  {Masson}]{Sutton1990}
E.~C. {Sutton}, W.~C. {Danchi}, P.~A. {Jaminet}, and C.~R. {Masson}.
\newblock {CO J = 3-2 observations of the neutral disk in Sagittarius A West}.
\newblock \emph{\apj}, 348:\penalty0 503--514, Jan. 1990.

\bibitem[{{\v S}ubr} et~al.(2009){{\v S}ubr}, {Schovancov{\'a}}, and
  {Kroupa}]{Subr2009}
L.~{{\v S}ubr}, J.~{Schovancov{\'a}}, and P.~{Kroupa}.
\newblock {The warped young stellar disc in the Galactic centre}.
\newblock \emph{\aap}, 496:\penalty0 695--699, Mar. 2009.

\bibitem[{van der Tak} et~al.(2007){van der Tak}, {Black}, {Sch{\"o}ier},
  {Jansen}, and {van Dishoeck}]{VderT2007}
F.~F.~S. {van der Tak}, J.~H. {Black}, F.~L. {Sch{\"o}ier}, D.~J. {Jansen}, and
  E.~F. {van Dishoeck}.
\newblock {A computer program for fast non-LTE analysis of interstellar line
  spectra. With diagnostic plots to interpret observed line intensity ratios}.
\newblock \emph{\aap}, 468:\penalty0 627--635, June 2007.

\bibitem[{van Langevelde} and {van de Tak}(2004)]{vdeTak2004}
H.~J. {van Langevelde} and F.~{van de Tak}.
\newblock {Radiative Bookkeeping : a guide to astronomical molecular
  spectroscopy and radiative transfer problems with an emphasis on RADEX}.
\newblock pages 1--18, 2004.

\bibitem[{Wade} et~al.(1987){Wade}, {Geballe}, {Krisciunas}, {Gatley}, and
  {Bird}]{Wade1987}
R.~{Wade}, T.~R. {Geballe}, K.~{Krisciunas}, I.~{Gatley}, and M.~C. {Bird}.
\newblock {Ionization state in and reddening to the center of the Galaxy}.
\newblock \emph{\apj}, 320:\penalty0 570--572, Sept. 1987.

\bibitem[{Wannier}(1989)]{Wannier1989}
P.~G. {Wannier}.
\newblock {Abundances in the Galactic Center}.
\newblock In {M.~Morris}, editor, \emph{The Center of the Galaxy}, volume 136
  of \emph{IAU Symposium}, pages 107--119, 1989.

\bibitem[{Yusef-Zadeh} et~al.(1999){Yusef-Zadeh}, {Roberts}, {Goss}, {Frail},
  and {Green}]{FYZ1999}
F.~{Yusef-Zadeh}, D.~A. {Roberts}, W.~M. {Goss}, D.~A. {Frail}, and A.~J.
  {Green}.
\newblock {High-Resolution Observations of OH (1720 MHZ) Masers toward the
  Galactic Center}.
\newblock \emph{\apj}, 512:\penalty0 230--236, Feb. 1999.

\bibitem[{Yusef-Zadeh} et~al.(2001){Yusef-Zadeh}, {Stolovy}, {Burton},
  {Wardle}, and {Ashley}]{FYZ2001}
F.~{Yusef-Zadeh}, S.~R. {Stolovy}, M.~{Burton}, M.~{Wardle}, and M.~C.~B.
  {Ashley}.
\newblock {High Spectral and Spatial Resolution Observations of Shocked
  Molecular Hydrogen at the Galactic Center}.
\newblock \emph{\apj}, 560:\penalty0 749--762, Oct. 2001.

\bibitem[{Yusef-Zadeh} et~al.(2008){Yusef-Zadeh}, {Braatz}, {Wardle}, and
  {Roberts}]{FYZ2008}
F.~{Yusef-Zadeh}, J.~{Braatz}, M.~{Wardle}, and D.~{Roberts}.
\newblock {Massive Star Formation in the Molecular Ring Orbiting the Black Hole
  at the Galactic Center}.
\newblock \emph{\apj}, 683:\penalty0 L147--L150, Aug. 2008.

\bibitem[{Zhao} et~al.(2009){Zhao}, {Morris}, {Goss}, and {An}]{Zhao2009}
J.~{Zhao}, M.~R. {Morris}, W.~M. {Goss}, and T.~{An}.
\newblock {Dynamics of Ionized Gas at the Galactic Center: Very Large Array
  Observations of the Three-dimensional Velocity Field and Location of the
  Ionized Streams in Sagittarius A West}.
\newblock \emph{\apj}, 699:\penalty0 186--214, July 2009.

\end{thebibliography}
\bibliographystyle{abbrvnat}

\end{document}